\documentclass[11pt,oneside,fleqn]{book}

\usepackage{mm}
\usepackage{psfig,epsfig}
\usepackage{latexsym}
\usepackage{cite}
\usepackage{axodraw}
\usepackage{abox}

\renewcommand{\baselinestretch}{1.5}
\renewcommand{\d}{{\rm d}}
\def\OMIT#1{{}}

\newcommand{\be}{\begin{equation}}
\newcommand{\ee}{\end{equation}}
\newcommand{\bea}{\begin{eqnarray}}
\newcommand{\eea}{\end{eqnarray}}
\newcommand{\beq}{\begin{equation}}
\newcommand{\eeq}{\end{equation}} 
\newcommand{\beqa}{\begin{eqnarray}}
\newcommand{\eeqa}{\end{eqnarray}}
\newcommand{\bey}{\begin{eqnarray}}
\newcommand{\eey}{\end{eqnarray}}

\def\ifmath#1{\relax\ifmmode#1\else$#1$\fi}

\def\D0{D\O}

\def\vereq#1#2{\lower4pt\vbox{\baselineskip1pt\lineskip1pt
    \ialign{\\$#1\hfill##\hfil\\$\crcr#2\crcr\sim\crcr}}}

\def\openone{\leavevmode\hbox{\small1\kern-3.8pt\normalsize1}}%

\newcommand{\gev}{\,\mbox{GeV}}
\newcommand{\mev}{\,\mbox{MeV}}
\newcommand{\keV}{\ensuremath{\mathrm{ke\kern -0.1em V}}}
\newcommand{\eV}{\ensuremath{\mathrm{e\kern -0.1em V}}}
\newcommand{\MeV}{\ensuremath{\mathrm{Me\kern -0.1em V}}}
\newcommand{\GeV}{\ensuremath{\mathrm{Ge\kern -0.1em V}}}
\newcommand{\TeV}{\ensuremath{\mathrm{Te\kern -0.1em V}}}

\newcommand{\Bbar}{\,\overline{\!B}}

\def\B0bar{\Bbar{}^0}

\def\lqcd{\Lambda_{\rm QCD}}

\def\bsg{\ifmmode B\to X_s\gamma\else $B\to X_s\gamma$\fi}

\def\bd{${B^0_d}$}

\newcommand{\taller}[2][.3ex]{\setbox\@tempboxa\hbox{#2}%
  \@tempdima #1 \advance\@tempdima\ht\@tempboxa \ht\@tempboxa\@tempdima
  \box\@tempboxa}
\newcommand{\deeper}[2][.3ex]{\setbox\@tempboxa\hbox{#2}%
  \@tempdima #1 \advance\@tempdima\dp\@tempboxa \dp\@tempboxa\@tempdima
  \box\@tempboxa}
\newcommand{\mrm}[1]{\ensuremathbox{\mathrm{#1}}}

\def\d0{D^0}
\def\d0b{\bar{D^0}}

\def\be{\begin{equation}}
\def\ee{\end{equation}}
\def\bea{\begin{eqnarray}}
\def\eea{\end{eqnarray}}

\def\bd{${B^0_d}$}
\def\lb{${\Lambda^0_b}$}

\def\lbdec{${\Lambda^0_b \rightarrow J/\psi \Lambda}$}
\def\lbbardec{${\overline{\Lambda^0_b} \rightarrow J/\psi \overline{\Lambda}}$}

\def\bddecks{${B_d^0 \rightarrow J/\psi K^0_{S}}$}

\makeindex

\begin{document}
\ifx\href\undefined\else\hypersetup{linktocpage=true}\fi 
\beforepreface
\afterpreface


\newcommand{\kevcc}{\!\mathrm{keV}/c^2}
\newcommand{\kevc}{\!\mathrm{keV}/c}
\newcommand{\kev}{\!\mathrm{keV}}
\newcommand{\ev}{\!\mathrm{eV}}
\newcommand{\evcc}{\!\mathrm{eV}/c^2}
\newcommand{\mevcc}{\!\mathrm{MeV}/c^2}
\newcommand{\mevc}{\!\mathrm{MeV}/c}
\newcommand{\gevcc}{\!\mathrm{GeV}/c^2}
\newcommand{\gevc}{\!\mathrm{GeV}/c}
\newcommand{\gevb}{\!\mathrm{GeV}^{-2}c^2}
\newcommand{\gevt}{\!\mathrm{GeV}^{2}/c^2}
\newcommand{\micron}{\!\mathrm{\mu m}}
\newcommand{\cm}{\!\mathrm{cm}}
\newcommand{\mm}{\!\mathrm{mm}}
\newcommand{\ns}{\!\mathrm{ns}}
\newcommand{\chisquare}{\chi^2}


\newcommand{\kshort}{K_S}
\newcommand{\klong}{K_L}
\newcommand{\kz}{K^0}
\newcommand{\kzbar}{\overline{K}^0}
\newcommand{\kp}{K^+}
\newcommand{\km}{K^-}
\newcommand{\kpm}{K^{\pm}}
\newcommand{\kmp}{K^{\mp}}
\newcommand{\piz}{\pi^0}
\newcommand{\pip}{\pi^+}
\newcommand{\pim}{\pi^-}
\newcommand{\pipm}{\pi^{\pm}}
\newcommand{\pimp}{\pi^{\mp}}
\newcommand{\ep}{e^+}
\newcommand{\eepm}{e^{\pm}}
\newcommand{\ppm}{p^{\pm}}
\newcommand{\kstar}{K^*}
\newcommand{\kstarz}{K^{*0}}
\newcommand{\kstarzbar}{\overline{K}^{*0}}
\newcommand{\kstarp}{K^{*+}}
\newcommand{\kstarm}{K^{*-}}
\newcommand{\kstarpm}{K^{*\pm}}
\newcommand{\kstarmp}{K^{*\mp}}
\newcommand{\kkx}{K^+K^-(1750)}
\newcommand{\etaa}{\eta(550)}
\newcommand{\phia}{\phi(1020)}
\newcommand{\phib}{\phi(1680)}
\newcommand{\phipb}{\phi'(1680)}
\newcommand{\rhoa}{\rho(770)}
\newcommand{\rhob}{\rho(1450)}
\newcommand{\rhoc}{\rho(1700)}
\newcommand{\omegaa}{\omega(782)}
\newcommand{\omegab}{\omega(1650)}
\newcommand{\fza}{f_0(1370)}
\newcommand{\fzb}{f_0(1500)}
\newcommand{\fzc}{f_0(1710)}
\newcommand{\rhopa}{\rho'(1250)}
\newcommand{\rhopb}{\rho'(1600)}
\newcommand{\rhoppb}{\rho''(1600)}
\newcommand{\omegapiz}{\omega\piz(1250)}
\newcommand{\bone}{b_1(1235)}
\newcommand{\Jpsi}{J/\psi}
\newcommand{\jpsi}{J/\psi}


\newcommand{\kpkm}{K^+K^-}
\newcommand{\ksks}{K_SK_S}
\newcommand{\ksk}{K^*K}
\newcommand{\kskpi}{K_SK^{\pm}\pi^{\mp}}
\newcommand{\kstark}{K^{*\pm}K^{\mp}}
\newcommand{\kskstar}{K^{*0}K_S}
\newcommand{\epem}{e^+e^-}
\newcommand{\twopi}{\pip\pim}
\newcommand{\threepi}{\pip\pim\piz}
\newcommand{\fourpi}{\pip\pim\pip\pim}
\newcommand{\fourpiz}{\pip\pim\piz\piz}


\newcommand{\qbar}{\overline{q}}
\newcommand{\ubar}{\overline{u}}
\newcommand{\dbar}{\overline{d}}
\newcommand{\sbar}{\overline{s}}
\newcommand{\cbar}{\overline{c}}
\newcommand{\bbar}{\overline{b}}
\newcommand{\qqbar}{q\overline{q}}
\newcommand{\nnbar}{n\overline{n}}
\newcommand{\uubar}{u\overline{u}}
\newcommand{\ddbar}{d\overline{d}}
\newcommand{\ssbar}{s\overline{s}}
\newcommand{\ccbar}{c\overline{c}}
\newcommand{\bbbar}{b\overline{b}}
\newcommand{\qqbarg}{\qqbar g}
\newcommand{\qqq}{qqq}


\newcommand{\photox}{\gamma N \rightarrow X N}
\newcommand{\photokkx}{\gamma N \rightarrow \kkx N}
\newcommand{\photokkxtokpkm}{\gamma N \rightarrow \kkx N \rightarrow \kpkm N}
\newcommand{\photovp}{\gamma p \rightarrow V p}
\newcommand{\epemtov}{\epem \rightarrow V}
\newcommand{\photoxtokpkm}{\gamma N \rightarrow X N \rightarrow \kpkm N}
\newcommand{\photoxtoksktokskpi}{\gamma N \rightarrow X N \rightarrow \ksk N \rightarrow \kskpi N}
\newcommand{\photoxtokstarktokskpi}{\gamma N \rightarrow X N \rightarrow \kstark N \rightarrow \kskpi N}
\newcommand{\photoxtokskstartokskpi}{\gamma N \rightarrow X N \rightarrow \kskstar N \rightarrow \kskpi N}
\newcommand{\kkxtokpkm}{\kkx \rightarrow \kpkm}
\newcommand{\kkxtoksks}{\kkx \rightarrow \ksks}
\newcommand{\kkxtoksk}{\kkx \rightarrow \ksk}
\newcommand{\kkxtokskpi}{\kkx \rightarrow \kskpi}
\newcommand{\kkxtokskstarfull}{\kkx \rightarrow \kstarzbar\kz \rightarrow \km\pip\kshort}
\newcommand{\kkxtokstarkfull}{\kkx \rightarrow  \kstarp\km \rightarrow \kshort\pip\km}
\newcommand{\kkxtokskoverkpkm}{\Gamma(\kkxtoksk)/\Gamma(\kkxtokpkm)}


\newcommand{\vdm}{\mathrm{VDM}}
\newcommand{\vmd}{\mathrm{VDM}}
\newcommand{\gvdm}{\mathrm{GVDM}}
\newcommand{\gvmd}{\mathrm{GVDM}}
\newcommand{\vdmratio}{\rho  \!:\!\omega \!:\!\phi  = 9\!:\!1\!:\!2}
\newcommand{\vmdratio}{\rho  \!:\!\omega \!:\!\phi  = 9\!:\!1\!:\!2}
\newcommand{\gvdmratio}{\rho'\!:\!\omega'\!:\!\phi' = 9\!:\!1\!:\!2}
\newcommand{\gvmdratio}{\rho'\!:\!\omega'\!:\!\phi' = 9\!:\!1\!:\!2}


\newcommand{\hxv}{\mathrm{H \! \times \! V}}


\newcommand{\jpc}{J^{PC}}
\newcommand{\zeromm}{0^{--}}
\newcommand{\zeromp}{0^{-+}}
\newcommand{\zeropm}{0^{+-}}
\newcommand{\zeropp}{0^{++}}
\newcommand{\onemm}{1^{--}}
\newcommand{\onemp}{1^{-+}}
\newcommand{\onepm}{1^{+-}}
\newcommand{\onepp}{1^{++}}
\newcommand{\twomm}{2^{--}}
\newcommand{\twomp}{2^{-+}}
\newcommand{\twopm}{2^{+-}}
\newcommand{\twopp}{2^{++}}
\newcommand{\threemm}{3^{--}}
\newcommand{\threemp}{3^{-+}}
\newcommand{\threepm}{3^{+-}}
\newcommand{\threepp}{3^{++}}
\newcommand{\jp}{J^{P}}
\newcommand{\zerom}{0^{-}}
\newcommand{\zerop}{0^{+}}
\newcommand{\onem}{1^{-}}
\newcommand{\onep}{1^{+}}
\newcommand{\twom}{2^{-}}
\newcommand{\twop}{2^{+}}
\newcommand{\onehalfp}{{\frac{1}{2}}^{+}}
\newcommand{\onehalfm}{{\frac{1}{2}}^{-}}
\newcommand{\threehalfp}{{\frac{3}{2}}^{+}}
\newcommand{\threehalfm}{{\frac{3}{2}}^{-}}
\newcommand{\spinzero}{\rm{spin}\!\!-\!\!0}
\newcommand{\spinhalf}{\rm{spin}\!\!-\!\!\frac{1}{2}}
\newcommand{\spinone}{\rm{spin}\!\!-\!\!1}
\newcommand{\spinthreehalf}{\rm{spin}\!\!-\!\!\frac{3}{2}}
\newcommand{\threepzero}{{^{3}P_0}}
\newcommand{\slj}{{^{2S+1}L_J}}


\newcommand{\kkxevents}{11,\!700 \pm 480}
\newcommand{\kkxmass}{1753.5 \pm 1.5 \pm 2.3~\mevcc}
\newcommand{\kkxwidth}{122.2\pm 6.2\pm 8.0~\mevcc}
\newcommand{\kkxtpsteep}{69.2 \pm 2.1~\gevb}
\newcommand{\kkxtpshallow}{4.17 \pm 0.21~\gevb}
\newcommand{\phiatpsteep}{77.71 \pm 0.59~\gevb}
\newcommand{\phiatpshallow}{1.71 \pm 0.14~\gevb}


\newcommand{\pt}{p_T}
\newcommand{\tp}{t'}
\newcommand{\ebeam}{E_{BEAM}}
\newcommand{\sqrts}{\sqrt{s}}
\newcommand{\sqrtsdef}{\sqrts = (M_N^2 + 2M_N\ebeam)^{\frac{1}{2}}}
\newcommand{\sinsint}{\sin^2\theta}
\newcommand{\coscost}{\cos^2\theta}
\newcommand{\sint}{\sin\theta}
\newcommand{\cost}{\cos\theta}
\newcommand{\twoexp}{a_1e^{-b_1\tp}+a_2e^{-b_2\tp}} 
\newcommand{\tpwtsteep}{\frac{a_1e^{-b_1\tp}}{a_1e^{-b_1\tp}+a_2e^{-b_2\tp}}}
\newcommand{\tpwtshallow}{\frac{a_2e^{-b_2\tp}}{a_1e^{-b_1\tp}+a_2e^{-b_2\tp}}}
\newcommand{\oneexp}{ae^{-b\tp}}
\newcommand{\Pbeam}{P_{BEAM}}
\newcommand{\Pkk}{P_{KK}}
\newcommand{\Pres}{P_{res}}
\newcommand{\tpdef}{\tp \equiv |t| - |t|_{min} \approx \pt^2}
\newcommand{\tdef}{t \equiv (\Pbeam - \Pkk)^2}
\newcommand{\breitwigner}{\frac{\Gamma/2\pi}{(M-M_0)^2 + \Gamma^2/4}}


\newcommand{\ebeamcuta}{10 < \ebeam < 40~\gev}
\newcommand{\ebeamcutb}{40 < \ebeam < 70~\gev}
\newcommand{\ebeamcutc}{70 < \ebeam < 100~\gev}
\newcommand{\ebeamcutd}{100 < \ebeam < 130~\gev}
\newcommand{\ebeamcute}{130 < \ebeam < 160~\gev}
\newcommand{\ebeamcutf}{10 < \ebeam < 160~\gev}
\newcommand{\ebeamcutaa}{50 < \ebeam < 60~\gev}
\newcommand{\ebeamcutbb}{60 < \ebeam < 70~\gev}
\newcommand{\ebeamcutcc}{70 < \ebeam < 80~\gev}
\newcommand{\ebeamcutdd}{80 < \ebeam < 90~\gev}
\newcommand{\ebeamcutee}{90 < \ebeam < 100~\gev}
\newcommand{\ebeamcutff}{50 < \ebeam < 100~\gev}
\newcommand{\ebeamcutaaa}{50 < \ebeam < 75~\gev}
\newcommand{\ebeamcutbbb}{75 < \ebeam < 100~\gev}
\newcommand{\ptcuta}{\pt < 0.15~\gevc}
\newcommand{\ptcutb}{\pt > 0.15~\gevc}
\newcommand{\tpcuta}{\tp < 0.0225~\gevt}
\newcommand{\phiacut}{1010 < M(\kpkm) < 1035~\mevcc}
\newcommand{\kkxcutlsb}{1500 < M(\kpkm) < 1600~\mevcc}
\newcommand{\kkxcut}{1640 < M(\kpkm) < 1860~\mevcc}
\newcommand{\kkxcutrsb}{1900 < M(\kpkm) < 2050~\mevcc}


\newcommand{\tlm}{t_{lm}}
\newcommand{\Io}{I(\Omega)}
\newcommand{\effo}{\eta(\Omega)}
\newcommand{\Ylm}{Y_l^m}
\newcommand{\Ylmo}{Y_l^m(\Omega)}
\newcommand{\YLM}{Y_L^M}
\newcommand{\YLMo}{Y_L^M(\Omega)}
\newcommand{\YJM}{Y_J^M}
\newcommand{\Alm}{A_{lm}}
\newcommand{\Almo}{A_{lm}(\Omega)}
\newcommand{\Vlm}{V_{lm}}
\newcommand{\Dlm}{D^l_{m0}}
\newcommand{\Dlmo}{D^l_{m0}(\Omega)}
\newcommand{\reDlmo}{\mathrm{Re}{\lbrace \Dlmo \rbrace}}
\newcommand{\reDlmoi}{\mathrm{Re}{\lbrace \Dlm(\Omega_i) \rbrace}}
\newcommand{\eAlm}{{^{\epsilon}A_{lm}}}
\newcommand{\eAlmo}{{^{\epsilon}A_{lm}}(\Omega)}
\newcommand{\eVlm}{{^{\epsilon}V_{lm}}}
\newcommand{\eVlmo}{{^{\epsilon}V_{lm}}(\Omega)}
\newcommand{\dOmega}{d\Omega}
\newcommand{\SeffIo}{\int{\effo \Io \dOmega}}
\newcommand{\tlmdef}{\int{\Io \Dlmo \dOmega}}


\newcommand{\lqed}{{\cal L}_{QED} = \overline{\psi}
                    (i\gamma^{\mu}\partial_{\mu} -
                     qA_{\mu}\gamma^{\mu} - m) \psi -
                    \frac{1}{4}F_{\mu\nu}F^{\mu\nu}}
\newcommand{\fqed}{F^{\mu\nu} \equiv \partial^{\mu}A^{\nu}
                                     - \partial^{\nu}A^{\mu}}
\renewcommand{\lqcd}{{\cal L}_{QCD} = \overline{q_i}
                   (i\partial_{\mu}\gamma^{\mu}\delta_{ij} -
                    g\frac{\lambda_{ij}^a}{2}A_{\mu}^a\gamma^{\mu} -
                    m\delta_{ij}) q_j -
                    \frac{1}{4}F_{\mu\nu}^aF^{a\mu\nu}}
\newcommand{\fqcd}{F_a^{\mu\nu} \equiv \partial^{\mu}A_a^{\nu}
                                     - \partial^{\nu}A_a^{\mu}
                                     - gf_{abc}A_b^{\mu}A_c^{\nu}}

\chapter{The Standard Model of Particle Physics}
\label{ch:SM}

The Standard Model~\cite{weinberg} is the currently accepted theory of the fundamental particles and their interactions. It has proven to be a successful theory. Even precision measurements have found no deviations so far from its predictions, with the exception of the neutrino masses~\cite{neutrino}. 
However a number of problems remain, The Standard Model makes no room for including gravity and has the unattractive feature of 18 free parameters or 21 when incorporating massive neutrinos. And we have not measured all the parameters to the same precision. Also we have yet to discover the Higgs which is the cornerstone of the Standard Model. 
This short chapter will simply list its main components.

Within the Standard Model, there are two broad categories of particles.  The fundamental fermions (fermions are particles with fractional spin) are considered to be the matter, the stuff of the universe.  For example, the quarks in the protons and neutrons in an atomic nucleus and the electrons surrounding it belong to this category.  The fundamental bosons (bosons are particles with integer spin), on the other hand, are responsible for the forces between the matter particles.  For example, the quarks in protons and neutrons of the nucleus are held together by gluons and electrons are bound to the nucleus by the exchange of ``virtual'' photons.

The interactions between matter particles take place through the exchange of the fundamental bosons.  These interactions are described by the Lagrangian of the Standard Model, leading to equations that specify rules for calculating quantities such as the probabilities for certain reactions to occur, referred to as cross sections.

\section{The Fundamental Particles}

According to the Standard Model, there are 24 fundamental matter particles (see Tables~\ref{tab_quarks} and~\ref{tab_leptons}) -- six quarks (the up, down, strange, charm, bottom and top quarks) coming in three different colors~\cite{breadbutter} and six leptons (the electron, muon, tau, and a neutrino)~\cite{pdg2004}. All of the elements of the periodic table can be built from combinations of only three of these 24:  the up quark, the down quark, and the electron.  An oxygen atom, for example,  has eight electrons surrounding a nucleus of eight protons and eight neutrons.  Protons are built from two up quarks and one down quark; and neutrons are two down quarks and one up quark.

\begin{table}[p]
\centering
\caption[Properties of the quarks]{Properties of the quarks.
\label{tab_quarks}}
\begin{tabular}{|c|c|c|c|}
  \hline
    Name                &     Symbol     &    Charge    \\
  \hline
  \hline
    Up Quark            &      $u$        &  $+2/3$  \\
  \hline
    Down Quark          &      $d$        &  $-1/3$  \\
  \hline
    Strange Quark       &      $s$         &  $-1/3$  \\
  \hline
    Charm Quark         &      $c$       &  $+2/3$  \\
  \hline
    Bottom Quark        &      $b$      &  $-1/3$  \\
  \hline
    Top Quark           &      $t$       &  $+2/3$  \\
  \hline
\end{tabular}
\vskip 0.25in
\end{table}

\begin{table}[p]
\centering
\caption[Properties of the leptons]{Properties of the leptons.
\label{tab_leptons}}
\begin{tabular}{|c|c|c|c|}
  \hline
    Name                &     Symbol    & Mass   &    Charge    \\
  \hline
  \hline
    Electron            &      $e$       &  $0.511~\mevcc$   &  $-1$  \\
  \hline
    Muon                &      $\mu$     &  $105.6~\mevcc$   &  $-1$  \\
  \hline
    Tau                 &      $\tau$    &  $1.777~\gevcc$   &  $-1$  \\
  \hline
    Electron            &    $\nu_e$     &    $<3~\evcc$     &  0   \\
    Neutrino            &                &                   &      \\
  \hline
    Muon                &   $\nu_{\mu}$  &  $<0.17~\mevcc$   &  0   \\
    Neutrino            &                &                   &      \\
  \hline
    Tau                 &  $\nu_{\tau}$  &  $<18~\mevcc$     &  0   \\
    Neutrino            &                &                   &      \\
  \hline
\end{tabular}
\vskip 0.25in
\end{table}

Because of the strength of the forces between them (the strong force), quarks are confined to exist in composites.  A combination of a quark and an antiquark is a meson (e.g., a pion or kaon), and a combination of three quarks is a baryon (e.g., a proton or neutron). Mesons and baryons are collectively termed hadrons.  While the vast majority of the matter we encounter in everyday life consists only of up and down quarks, the other four quarks are equally as fundamental.  The essential difference is only in their greater masses.

The lepton category of the fundamental particles consists of three negatively charged particles, of which the electron is prototypical, and their three very weakly interacting neutral partners, the neutrinos.  The muon and tau differ from the electron only in mass.  Neutrinos are only emitted during weak processes and only interact weakly, and are therefore extraordinarily difficult to detect. All of these fundamental fermions have anti-matter partners.

\section{The Fundamental Forces}

\begin{table}[t]
\centering
\caption[Properties of the gauge bosons]{Properties of the gauge bosons.
\label{tab_bosons}}
\begin{tabular}{|c|c|c|c|}
  \hline
    Name                &     Symbol     &     Mass          &    Charge    \\
  \hline
  \hline
    Photon              &   $\gamma$     &     0             & 0 \\
  \hline
    {\it W} Boson             &   $W^{\pm}$    & $80.4~\gevcc$     & $\pm 1$ \\
  \hline
    {\it Z} Boson             &   $Z^0$        & $91.19~\gevcc$    & 0 \\
  \hline
    Gluon               &    $g$         &     0             & 0 \\
  \hline
\end{tabular}
\vskip 0.25in
\end{table}

The fundamental particles interact through four different forces.  Probably the most familiar of the forces is gravity, the force of attraction between massive bodies. However due to the smallness of the masses of the fundamental particles, the force of gravity between any of them is negligible.  It is hoped that gravity will someday be described by a theory unifying it to the other three forces, but the Standard Model does not incorporate the gravitational force.  The forces of the Standard Model are all described by the exchange of force-carrying particles, the gauge bosons (see Table~\ref{tab_bosons}).

Electromagnetism is the force responsible for the repulsion between like charges, the attraction between unlike charges, the deflection of charged particles in magnetic fields, etc.  In the Standard Model, the force is due to the exchange of ``virtual'' photons, which interact with any charged body.  In addition, in Quantum Electrodynamics (QED), the Standard Model theory of electromagnetism, photons can convert to electrons and positrons, an electron can emit a photon, electrons and positrons can annihilate into photons, etc.

The strong force acts only on quarks, and is due to the exchange of gluons.  Similiar to the electric charge for electromagnetism, the strong force proceeds through a charge of its own, the ``color'' charge.  But unlike electromagnetism, where the photon has no electric charge of its own, gluons do carry color charge, allowing them to interact among themselves and thus creating a much more complex situation.  The strong force binds quarks tightly into hadrons, so tightly that the quarks never appear unbound.  The Standard Model theory of the strong force is Quantum Chromodynamics (QCD).

Finally, the weak force affects all of the fundamental particles.  It is carried by the $W^{\pm}$ and $Z^0$ bosons.  Nuclear beta decay is the most familiar example of this force, where one of the down quarks of a neutron converts to an up quark by emitting a $W^-$, which subsequently decays to an electron and an electron antineutrino.  The Standard Model theory of the weak force and QED are united into a single theory, the electroweak theory~\cite{electroweak}, by introducing a Higgs Boson. The search for the Higgs Boson, the last of the Standard Model particles to be experimentally undiscovered, is one of the major efforts of contemporary high energy physics.

\section{Lagrangians}

The mathematical structure of the Standard Model is contained in a series of Lagrangians. 
\begin{equation}
\label{eqn_langra_sum}
{\cal L}_{SM} = {\cal L}_{QED} +  {\cal L}_{QCD} + {\cal L}_{Weak}.
\end{equation}

 For example, the QED Lagrangian can be written as:
      \begin{equation}
        \label{eqn_lqed}
        \lqed ,
      \end{equation}
where
      \begin{equation}
        \label{eqn_fqed}
        \fqed .
      \end{equation}
Here, $\psi$ represents a particle with charge $q$, and $A_{\mu}$ represents the photonic vector field. 
The first term describes the kinetic energy of the particle, and together with the mass term, they constitue the Lagrangian density of a free particle.
 The second term describes the interaction of the particle with the electromagnetic field. The last term describes the free electromagnetic field.


\null\vfill
\begin{center}
\Huge Part I
\end{center}
\null\vfill

\chapter{CPT Formalism}

The combined symmetry of charge conjugation (C), parity (P), and 
time reversal (T) is believed to be respected
by all local, point-like, Lorentz covariant field 
theories, such as the Standard Model we outlined in Chapter~\ref{ch:SM}.  
However, extensions to the Standard Model based on string 
theories do not necessarily require CPT invariance,
and observable effects at low-energies may be within
reach of experiments studying flavor 
oscillations~\cite{kostelecky-95}.
A parametrization~\cite{kostelecky} in which 
CPT and T violating parameters
appear has been developed, which allows
experimental investigation in many physical systems including 
atomic systems, Penning traps, and neutral meson 
systems~\cite{alanbook}. Using this parameterization we
present the first experimental search for CPT violation in the charm
meson system.

\section{Mixing Formalism}
Before we introduce CPT formalism, we want to outline mixing formalism, since much of the notation from standard mixing in charm mesons is used there.

Assuming CP conservation in the charm meson system, the CP eigenstates of the neutral $D$ meson can be written as,
\begin{equation}
\vert D_1\rangle = \frac{1}{\sqrt{2}} \left(\vert D^0 \rangle + \vert \overline{D^0}\rangle \right), \qquad {\rm and} \qquad \vert D_2\rangle = \frac{1}{\sqrt{2}} \left(\vert D^0 \rangle - \vert \overline{D^0}\rangle\right).
\label{eq:cpstates1}
\end{equation}
If we define $CP|D^0\rangle =| \overline{D^0}\rangle $, it then follows that $|D_1\rangle $ is a CP-even state and $|D_2\rangle $ is CP-odd. The time evolution of the $|D_1\rangle $ and $|D_2\rangle $ states is given by
\begin{eqnarray}
|D_i(t)\rangle  = e^{-i(M_i - i\frac{\Gamma_i}{2})t}|D_i(0)\rangle ,
\label{eq:timeevolution}
\end{eqnarray}
where $M_i$ and $\Gamma_i$ are the mass and the width for state $i$. Rearranging Eq.~\ref{eq:cpstates1}, we find in terms of $D_1$ and $D_2$, that a pure $D^0$ state produced at time $t=0$ is
\begin{eqnarray}
|D^0(t=0)\rangle  =\frac{1}{\sqrt{2}} \left(|D_1\rangle  + |D_2\rangle \right).
\label{eq:d0pure}
\end{eqnarray}
We obtain the time evolution of $D^0$ by plugging in the time evolution of the $D_1$ and $D_2$ states as given by Equation~\ref{eq:timeevolution}:
\begin{eqnarray}
|D^0(t)\rangle  = \frac{1}{\sqrt{2}}\left(e^{-i(M_1 -i\frac{\Gamma_1}{2})t}|D_1(0)\rangle  + e^{-i(M_2 -i\frac{\Gamma_2}{2})t}|D_2(0)\rangle \right).
\label{eq:d0_func_d1_d2}
\end{eqnarray}
This can be expressed in terms of the $|D^0\rangle $ and $|\overline{D^0}\rangle $ by using the relations in the Equations~\ref{eq:cpstates1} and combining like terms:
\begin{eqnarray}
|D^0(t)\rangle  = \frac{1}{2}\left(A_+|D^0\rangle  + A_-|\overline{D^0}\rangle \right),
\label{eq:with_a+_a-}
\end{eqnarray}
with 
\begin{eqnarray}
A_{\pm} = e^{-i(M_1 - i \frac{\Gamma_1}{2})t} \pm e^{-i(M_2 - i \frac{\Gamma_2}{2})t}.
\label{eq:a+a-}
\end{eqnarray}
The terms $A_{\pm}$ can be arranged in more convenient forms by using the definitions:
\begin{eqnarray}
\Gamma = \frac{\Gamma_1 + \Gamma_2}{2}, ~~M = \frac{M_1 + M_2}{2}, ~~x = \frac{-(M_1 - M_2)}{\Gamma},~{\rm and}~~y = \frac{\Gamma_1 - \Gamma_2}{2\Gamma}.
\label{eq:gamma}
\end{eqnarray}
The expressions for $A_{\pm}$ with these definitions are
\begin{eqnarray}
A_+ = 2 e^{-(iM + \frac{\Gamma}{2})t} {\rm cosh} \Biggl[ (y + ix) \frac{\Gamma t}{2}\Biggr]
\label{eq:A+}
\end{eqnarray}
and 
\begin{eqnarray}
A_- = -2 e^{-(iM + \frac{\Gamma}{2})t} {\rm sinh}\Biggl[ (y + ix) \frac{\Gamma t}{2}\Biggr].
\label{eq:A-}
\end{eqnarray}
Now we pose the question: what is the probability of an originally pure $D^0$ state to decay to $K^+\pi^-$? Define $\langle f|$ as the vector representing the final state $K^+\pi^-$. The amplitude for this decay process is:
\begin{eqnarray}
&&\langle f|D^0(t)\rangle  =
\nonumber \\
 && e^{-(iM + \frac{\Gamma}{2})t} \Bigg\{ {\rm cosh} \bigg[(y + ix) \frac{\Gamma t} {2}\bigg] \langle f|D^0\rangle  - {\rm sinh} \bigg[(y + ix) \frac{\Gamma t} {2}\bigg] \langle f|\overline{D^0}\rangle \Bigg\},
\label{eq:amplitude1}
\end{eqnarray}
where $\langle f|D^0\rangle $ is the Double Cabbibo Suppressed (DCS) decay amplitude and $\langle f|\overline{D^0}\rangle $ is the Cabbibo Favored (CF) amplitude.  
The DCS to CF amplitude ratio is written as 
\begin{eqnarray}
\frac{\langle f|D^0\rangle }{\langle f|\overline{D^0}\rangle } = -\sqrt{R_{DCS}} e^{-i\delta},
\label{eq:dcstocf}
\end{eqnarray}
where $R_{DCS}$ is the DCS to CF branching ratio and $\delta$ is a strong force phase between DCS and CF amplitudes. Plugging this in and approximating the hyperbolic functions with the first term of their Taylor series expansions we obtain
\begin{eqnarray}
\langle f|D^0(t)\rangle  =  e^{-(iM + \frac{\Gamma}{2})t} \langle f|\overline{D^0}\rangle \bigg[ -\sqrt{R_{DCS}} e^{-i\delta} - (y + ix) \frac{\Gamma t} {2} \bigg].
\label{eq:amplitude2}
\end{eqnarray}
Finally the probability is the absolute value square of the amplitude:
\begin{eqnarray}
&&|\langle f|D^0(t)\rangle |^2 = 
\nonumber \\
&& e^{-\Gamma t} |\langle f|\overline{D^0}\rangle |^2 \Bigg[ R_{DCS} + \sqrt{R_{DCS}} (y\, {\rm cos} \delta -x\, {\rm sin} \delta) \Gamma t + (\frac{x^2 + y^2}{4}) \Gamma^2 t^2\Bigg].
\label{eq:probability1}
\end{eqnarray}
We use the soft pion from the decay $D^{*+}\rightarrow D^0\pi^+$
to tag the flavor of the $D$ at production, and
the kaon charge in the decay $D^0\rightarrow K^- \pi^+$
to tag the $D$ flavor at the time of decay. Right-sign signal~(RS) is obtained by requring that the soft pion charge is equal the opposite of kaon charge. 
Wrong-sign signal~(WS) is obtained by requring that the soft pion charge is equal the kaon charge.   
Define a quantity $R(t)$, which is the time-dependent rate for the WS process relative to CF~(RS) branching fraction or 
\begin{eqnarray}
R(t) = \frac{|\langle f|D^0(t)\rangle |^2}{|\langle f|\overline{D^0}\rangle |^2},
\label{eq:rt}
\end{eqnarray}
and define the parameters $x'$ and $y'$ that are related to the mixing parameters, $x$ and $y$, by a strong phase rotation:
\begin{eqnarray}
x' = x {\rm cos} \delta + y {\rm sin} \delta, \qquad y' = y {\rm cos} \delta - x {\rm sin} \delta.
\end{eqnarray}
Redefine $t$ in units of $D^0$ lifetime ( where $\Gamma t = t/\tau_{D^0}$ ) to obtain an expression for the lifetime evolution of the decay $ D^0 \rightarrow K^+ \pi^-$:
\begin{eqnarray}
R(t) = \Biggl[ R_{DCS} + \sqrt{R_{DCS}} y' t + (\frac{x'^2 + y'^2}{4}) t^2\Biggr] e^{-t/\tau_{D^0}}.
\label{eq:Rfinal}
\end{eqnarray}
The first term in Eq.~\ref{eq:Rfinal} is a pure DCS decay amplitude, the second term is the interference of DCS and mixing, and the third term is a pure mixing term.

\section{Proper Time Asymmetry}

The time evolution of a neutral-meson state is governed by a $2\times2$
effective Hamiltonian matrix $\Lambda$ in  the Schr\"odinger
equation. For a complex 2$\times $2 matrix, it is possible to write the two diagonal elements as the sum and difference of two complex numbers. It is also possible to write the off-diagonal elements as the product and ratio of two complex numbers. Using these two facts, which ultimately permit the clean representation of T and CPT-violating quantities, a general expression for $\Lambda$ can be taken as~\cite{kostelecky}:
\begin{eqnarray}
\Lambda = \frac{1}{2}\Delta \lambda \left ( \matrix{ U + \xi & VW^{-1} \cr VW & U-\xi}\right ),
\label{eq:matrix}
\end{eqnarray}
where the parameters $U,V,W$, and $\xi$ are complex. The requirements that the trace of the matrix is $tr \Lambda =\lambda$ and that the determinant is $det \Lambda =\lambda_1 \lambda_2$ impose the identifications $U=\lambda/\Delta \lambda$, $V=\sqrt{1-\xi^2}$ on the complex parameters $U$ and $V$. The free parameters in Eq.~\ref{eq:matrix} are therefore $W$ and $\xi$. These can be regarded as four independent real quantities: $W=we^{i\omega}$, and $\xi = {\rm Re}\,\xi + i {\rm Im}\,\xi$. One of these four real numbers, the argument $\omega$, is arbitrary and physically irrelevant. The other three are physical. The modulus of $W$ controls T violation with $w=1$ if and only if T is preserved. The two remaining real numbers, ${\rm Re}\,\xi$ and ${\rm Im}\,\xi$, control CPT violation and both are zero if and only if CPT is preserved. The quantities $w$ and $\xi$ can be expressed in terms of the components of $\Lambda$ as $w = \sqrt{|\Lambda_{21}/\Lambda_{12}|}$, $\xi=\Delta\Lambda/\Delta\lambda$, where $\Delta\Lambda
=\Lambda_{11}-\Lambda_{22}$ and $\Delta\lambda$ is the difference in
the eigenvalues. $\lambda_1 = M_1 -\frac{1}{2}i\Gamma_1$, $\lambda_2 = M_2 -\frac{1}{2}i\Gamma_2$ and $\Delta \lambda =\Gamma(x-i~y)$.
 $\xi$ is phenomenologically introduced and therefore
independent of the model.     
Indirect CPT violation occurs if and
only if the difference of diagonal elements of $\Lambda$ is nonzero.
To determine the time-dependent decay amplitudes and probabilities, it is useful to obtain an explicit expression for the time evolution of the neutral D meson. Doing the same exercise as in the mixing formalism (with $\Lambda$ as the matrix) we obtain,
\begin{eqnarray}
\left(\matrix{D^0(t,\hat{t},p) \cr \overline{D^0}(t,\hat{t},p)}\right) = \left (\matrix{C + S\xi & SVW \cr SVW^{-1} & C-S\xi}\right) \left(\matrix{D^0 \cr \overline{D^0} }\right ).
\label{eq:cptmatrix}
\end{eqnarray}
The functions $C$ and $S$ depend on the meson proper time $t$, sidereal time $\hat{t}$ and are given by:
\begin{eqnarray}
C={\rm cos} \left(\frac{1}{2}\Delta \lambda t\right ) e^{-\frac{1}{2}i\lambda t} \qquad 
S = -i {\rm sin} \left(\frac{1}{2}\Delta \lambda t\right ) e^{-\frac{1}{2}i\lambda t}.
\label{eq:CS}
\end{eqnarray}

One can easily extract time-dependent decay probabibilities by manipulating Eq.~\ref{eq:cptmatrix}.    
For the decay of $D^0$ to a right-sign final state $f$ (which could
be a semileptonic mode, or a Cabibbo favored hadronic mode (with DCS negligible)), 
the time-dependent decay probability is
\begin{eqnarray}
P_f(t) & \equiv & | \langle f | T | D^0(t) \rangle |^2                         \nonumber  \\
       &=& {1\over{2}} |F|^2 {\rm exp}({-{\gamma \over{2}}t})  
           \times [ ( 1 + |\xi|^2 ) {\rm cosh} \Delta\gamma t/2 + 
           (1 - |\xi|^2 ) {\rm cos} \Delta m t                                   \nonumber \\
       &&  - 2~{\rm Re}~\xi~{\rm sinh} \Delta\gamma t/2 
           - 2~{\rm Im}~\xi~{\rm sin}\Delta m t              ].
\label{eq:rsdecay}
\end{eqnarray}

\noindent
The time-dependent probability for the  decay of $\overline{D}^0$
to a right-sign final state $\overline f$,  
$\overline{P}_{\overline f}(t)$, may be obtained by replacing in the above equation 
$\xi \rightarrow - \xi$ and
$F \rightarrow \overline{F}$.  
In the formula, $F$ represents the basic
transition amplitude for the decay $D^0\rightarrow f$,
$\Delta\gamma$ and $\Delta m$ are
the differences in physical decay widths and masses for the propagating eigenstates
and can be related to the usual mixing parameters $x  = \Delta M/\Gamma$
and $y = \Delta \Gamma/{2 \Gamma}$.  
The complex parameter $\xi$ controls the CPT violation and is seen to modify
the shape of the time dependent decay probabilities.
Expressions for wrong-sign decay probabilities involve both CPT and T violation
parameters that scale the probabilities, leaving the shape unchanged.
Using only right-sign decay modes, the following asymmetry can be formed,
\begin{eqnarray}
A_{CPT}(t)  =  {{\overline{P}_{\overline f}(t) - P(t)}\over{\overline{P}_{\overline f}(t) + P(t)}},
\label{eq:asym}
\end{eqnarray}

\noindent
which is sensitive to the CPT violating parameter $\xi$:
\begin{eqnarray}
A_{CPT} = {{ 2~{\rm Re}~\xi~{\rm sinh}\Delta\gamma t/2 + 2~{\rm Im}~\xi~{\rm sin} \Delta m t} \over {
 ( 1 + |\xi|^2 ) {\rm cosh} \Delta\gamma t/2 + 
           (1 - |\xi|^2 ) {\rm cos} \Delta m t 
}}.
\label{eqn:acpt}
\end{eqnarray}

\begin{figure}[tbh]
\begin{centering}
\psfig{figure=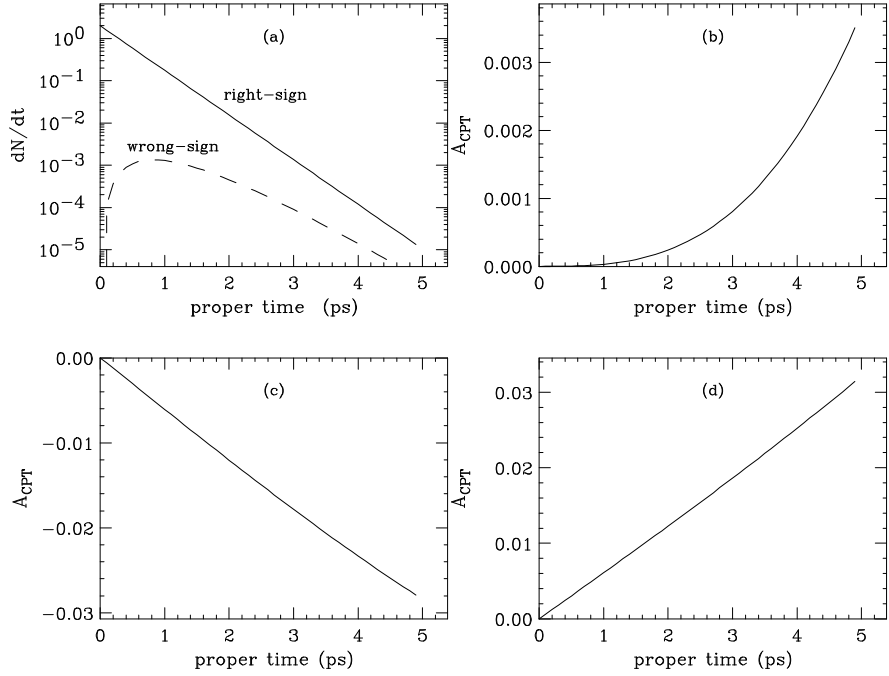,height=3.5in}
\caption{
(a) Proper time decay probabilities for
    $D^0$ decay, right-sign (solid) and wrong-sign (dashed) and
(b) $A_{CPT}$ with  Re~$\xi$ = 5\%, Im~$\xi$ = 5\%.
(c) $A_{CPT}$ with  Re~$\xi$ = 0,   Im~$\xi$ = 5\%.
(d) $A_{CPT}$ with  Re~$\xi$ = 5\%, Im~$\xi$ = 0.
 }  
\label{fig:asym}
\end{centering}
\end{figure}

We can gain insight into the anticipated experimental sensitivity by plotting
these functions with some reasonable assumptions.  We use  95\% confidence level~(C.L.) 
upper bounds on the mixing parameters $x$ and  $y$ of 5\%,  which is at the 
upper range of the current experimental sensitivity, as discussed previously.  
In Fig.~\ref{fig:asym}(a) we plot the proper time decay probabilities for
$D^0$ decay under the assumption of CPT violation at the level of 
Re~$\xi$ = 5\% and Im~$\xi$ = 5\%, which are independent parameters in the framework.  
One sees a CPT violation-induced wrong-sign contribution that vanishes both at
zero proper time and at long proper times.  This causes a distortion
from a purely exponential decay of a $D^0$ (and $\overline{D}^0$), which is then visible in the
asymmetry plot, $A_{CPT}$ as shown in Fig.~\ref{fig:asym}(b).  
Because of the small oscillation frequency
and short lifetime, one sees only the start of the oscillation, growing 
beyond 0.3\% at long proper times.  
Evident from Eqn.~\ref{eqn:acpt} is that positive values of
 Re~$\xi$ and Im~$\xi$  work to oppose one another in the
asymmetry in a linear fashion.  
This is shown in the nearly linear behavior of $A_{CPT}$
in Figs.~\ref{fig:asym}(c,d) with parameters 
Im~$\xi = 5\%$, Re~$\xi = 0$  and
Im~$\xi = 0$, Re~$\xi = 5\%$ respectively, and consequently CPT
asymmetries larger by a factor of 10 at long proper times.  
In practice, experiments will be sensitive to either Re~$\xi$ or
Im~$\xi$, but not both simultaneously.

\section{Double Cabbibo Suppressed Interference}
In the previous section, we generated the asymmetry by assuming the basic transition amplitudes to be $\langle  f|T|D^0\rangle =F$, $\langle f|T|\d0b\rangle =0$, 
$\langle \overline{f}|T|\d0b\rangle =\overline{F}$, $\langle \overline{f}|T|D^0\rangle =0$. This is fine as long as we are dealing with a semileptonic decay or if we neglect the Double Cabbibo Suppressed decays~(DCS) of hadronic modes. The general framework developed in Ref.~\cite{kostelecky} assumes that the DCS effects are negligible.  However in our forming the asymmetry we had to deal with DCS that were comparable to Cabbibo favored decays. We therefore will report the asymmetry in the previous section with DCS effects included. 
For the decay $D^0 \rightarrow K^-\pi^+ $ the basic transition
amplitudes are then  $\langle f|T|D^0\rangle =F$, $\langle f|T|\d0b\rangle =F_{\rm DCS}$, 
$\langle \overline{f}|T|\d0b\rangle =\overline{F}$, and  $\langle \overline{f}|T|D^0\rangle =\overline{F_{\rm DCS}}$.


For us to see if we can
neglect DCS, we have to start with the more general framework where the
DCS interference with the Cabbibo favored decays and the DCS term are not
neglected. Now we assume that $\langle f|T|\d0b\rangle =F_{\rm DCS}$ and 
$\langle \overline{f}|T|D^0\rangle =\overline{F_{\rm DCS}}$ are not neglected.
To touch base with the formalism already developed for mixing, 
we will take the ratio $\langle \overline{f}|T|D^0\rangle /\langle \overline{f}|T|\d0b\rangle =-\sqrt{R_{\rm
DCS}}~e^{-i\delta}$ and $\langle f|T|\d0b\rangle /\langle f|T|D^0\rangle =-\sqrt{R_{\rm DCS}}~e^{i\delta}$.  
With these in mind, the time dependent probability into a right-sign decay as in Eq.~\ref{eq:rsdecay} including DCS presence takes this more general form:

\begin{eqnarray}
P_f(t) & \equiv & | \langle  f | T | D^0(t) \rangle |^2                         \nonumber  \\
       &=& {1\over{2}} |F|^2 {\rm exp}({-{\gamma \over{2}}t})  
           \times [ ( 1 + |\xi|^2 ) {\rm cosh} \Delta\gamma t/2 + 
           (1 - |\xi|^2 ) {\rm cos} \Delta m t                                   \nonumber \\
       &&  - 2~{\rm Re}~\xi~{\rm sinh} \Delta\gamma t/2 
           - 2~{\rm Im}~\xi~{\rm sin}\Delta m t  -2~\sqrt {R_{DCS}}\times
  \nonumber \\
       &&   [(-{\rm sinh}\Delta\gamma t/2~{\rm cos}\delta -{\rm
           sin}\Delta m t~{\rm sin}\delta)
  \nonumber \\
       &&   +({\rm cosh} \Delta\gamma t/2 -{\rm cos}\Delta m t)\times
       ({\rm Re}\xi~{\rm cos}\delta +{\rm Im}\xi~{\rm sin}\delta)]
  \nonumber \\
      &&   +R_{DCS}\times ({\rm cosh} \Delta\gamma t/2-{\rm cos}\Delta m t)].
\label{eq:rsdecay_g}
\end{eqnarray}

\noindent
The time-dependent probability for the  decay of $\overline{D}^0$
to a right-sign final state $\overline f$,  
$\overline{D}_{\overline f}(t)$, may be obtained by replacing
$\xi \rightarrow - \xi$ and
$F \rightarrow \overline{F}$ in the above equation.  

 $R_{DCS}$ is the ratio of DCS decay to the Cabbibo favored decay. $\delta$ is the strong mixing phase between DCS decay and
Cabbibo favored.  
Using only right-sign decay modes, we form the asymmetry as in Eq.~\ref{eq:asym}
~that is sensitive to the CPT-violating parameter $\xi$.
In the case of negligible contributions
from DCS decay, $ |\langle  f | T | \overline{D^0(t)} \rangle| =0$, we
can form an identical asymmetry contribution as given by Eq.~\ref{eqn:acpt}:
\begin{eqnarray}
A_{CPT}^{0}(t) = {{ 2~{\rm Re}~\xi~{\rm sinh}\Delta\gamma t/2 + 2~{\rm Im}~\xi~{\rm sin} \Delta m t} \over {
 ( 1 + |\xi|^2 ) {\rm cosh} \Delta\gamma t/2 + 
           (1 - |\xi|^2 ) {\rm cos} \Delta m t 
}}.
\label{eqn:acpt_dcs}
\end{eqnarray}

We can form the same asymmetry as in Eq.~\ref{eqn:acpt} by using,
instead of probabilities given by Eq.~\ref{eq:rsdecay}, the
probabilities given by Eq.~\ref{eq:rsdecay_g} \footnote{We
have assumed that  T is not violated and thus the phase related to T
is zero, i.e., the only phase that enters in the probabilities due to
interference is the strong phase.}. 
We have this additional interference term in the $A_{CPT}$ expression, which ignores the small contributions in the denominator:
\begin{eqnarray}
A_{CPT}^{int}(t) = {{2~\sqrt {R_{DCS}}~\times({\rm cosh}\Delta\gamma
 t/2-{\rm cos} \Delta m t)\times ({\rm Re}\xi~{\rm cos}\delta +{\rm Im}\xi~{\rm sin}\delta)}\over {
 ( 1 + |\xi|^2 ) {\rm cosh} \Delta\gamma t/2 + 
           (1 - |\xi|^2 ) {\rm cos} \Delta m t 
}}.
\label{eqn:acpt_int}
\end{eqnarray}

With these new modifications, the total $A_{CPT}$ is the sum of the
two contributions, $A_{CPT}=A_{CPT}^{0}(t)+A_{CPT}^{int}(t)$. 
From this general expression of $A_{CPT}$, two different approaches can
be taken. 
In the first approach we assume that Eq.~15
of~\cite{kostelecky} does not hold but Eq.~21 of~\cite{kostelecky}
is valid since it is phenomenologically introduced and is not
dependent on the model. Equation~15 of~\cite{kostelecky} is  ${\rm Re}~\xi=x~{\rm Im}\xi/y$ in terms of $x,y$ mixing values. 
With small values of mixing, and the fact that $D^0$ has
a relatively short lifetime, the following approximation is valid:   
\begin{eqnarray}
A_{CPT}=Re\xi~y~t/\tau-Im\xi~x~t/\tau + O(\sqrt(R_{\rm DCS})(x^2+y^2)).
\label{eqn:acpt_sim}
\end{eqnarray}

In the second approach, we assume that Eq.~15
of~\cite{kostelecky} does hold, so we have ${\rm
Re}~\xi=x~{\rm Im}\xi/y$. Under this constraint, and with small values
of $x,y$ mixing, our expression of $A_{CPT}$ takes this form

\begin{eqnarray}
A_{\rm CPT} &=& \frac{{\rm Re}\,\xi (x^2 + y^2) (t/\tau)^2}{2x}
\left[ \frac{xy}{3} (t/\tau) + \sqrt{R_{\rm DCS}}\left(
x\,\cos{\delta}+ y\,\sin{\delta} \right) \right]. 
\label{eq:acpt_sim2}
\end{eqnarray}
  
This scenario is more strict since it requires that Eq.~15 of
\cite{kostelecky} is valid. While in the first approach we neglected the
DCS decay interference term as small, we can not neglect it in the
second approach because DCS interference term plays a comparable role
in the $A_{CPT}$. 
In this thesis and the resulting journal result~\cite{ref:abaz}, we consider both these
scenarios since the question of the constraint given by Eq.~15
of~\cite{kostelecky} is an open question.  

\section{Lorentz Violating Parameters}
In the CPT and Lorentz-violating extension~(SME) to the  Standard 
Model~\cite{colladay-kostelecky},
the CPT violating parameters may depend on lab momentum, orientation, and 
sidereal time~\cite{kostelecky-prl,kostelecky}. It can be shown that~\cite{kostelecky-prl}
\begin{equation}
 \Delta \Lambda \approx \beta^{\mu}\Delta a_{\mu},
\label{eq:lorentz}
\end{equation}
 where $\beta^{\mu} =\gamma(1,\vec{\beta})$ is the four-velocity of the $D$ meson in the observer frame. The effect of Lorentz and CPT violation in the SME appears in Eq.~\ref{eq:lorentz} via the factor $\Delta a_{\mu} =r_{q_1} a^{q_1}_{\mu} - r_{q_2} a^{q_2}_{\mu}$, where $a^{q_1}_{\mu}$ and $a^{q_2}_{\mu}$ are CPT- and Lorentz-violating coupling coefficients for the two valence quarks in the $D$ meson, and where $r_{q_1}$ and $r_{q_2}$ are quantities resulting from quark-binding and normalization effects. The coefficients $a^{q_1}_{\mu}$ and $a^{q_2}_{\mu}$ for Lorentz and CPT-violation have mass dimension one and emerge from terms in the Lagrangian for the SME of the form $-a^q_{\mu}\bar{q}\gamma^{\mu}q$, where $q$ specifies the quark flavor.    
A significant consequence of the four-momentum dependence arises from the rotation of the Earth relative to the constant vector $\Delta \vec{a}$. This leades to sidereal variations for CPT violating parameter $\xi$. 
In the case of FOCUS, 
a forward, fixed-target spectrometer, the $\xi$ parameter assumes the 
following form:
\begin{eqnarray}
\xi(\hat{t}, p)  & = & {\gamma(p)\over{\Delta\lambda}}[ \Delta a_0 + \beta \Delta a_Z {\rm cos} \chi 
+ \beta {\rm sin} \chi (\Delta a_Y {\rm sin} \Omega \hat{t} + \Delta a_X {\rm cos} \Omega \hat{t}) ],
\label{eq:xi}
\end{eqnarray}
\noindent
where $\Omega$ and $\hat{t}$ are the sidereal frequency and time respectively, 
and $X, Y, Z$ are non-rotating coordinates with $Z$ aligned with the
Earth's rotation axis. $\Delta a_X$, $\Delta a_Y$, $\Delta a_0$ and $\Delta a_Z$ are the differences in the Lorentz-violating coupling coefficients between valence quarks. $\chi$ is the angle between the $D^0$ momentum and the Z axis. The parameter $\Delta\lambda$ in terms of {\it x} and {\it y} is:
\begin{eqnarray}
\Delta\lambda =-\Delta m - {{i\Delta\gamma}\over {2}}=x\Gamma-iy\Gamma=\Gamma(x-i~y)={{(x-iy)}\over{\tau}},
\label{eq:dellam}
\end{eqnarray}
where $\tau$ is the mean lifetime of the $D^0$ meson.
\noindent

\section{Sidereal Time}

Since CPT-violating parameters depend on sidereal time, 
we outline what sidereal time is and how we calculate it. 
Sidereal time is time
according to the stars and not the sun.
Careful observation of the sky will show that any specific star will cross directly overhead (on the meridian) about four minutes earlier every day. In other words, the day according to the stars (the sidereal day) is about four minutes shorter than the day according to the sun (the solar day). 
 If we measure a day from noon to noon -- from when the sun crosses the
meridian (directly overhead) to when the sun crosses the meridian --
again we will find the average solar day is about 24 hours. If we
measure the day according to a particular star from when that star
crosses the meridian to when that star crosses the meridian again we
will find the average sidereal day is 23 hours and 56 minutes long. The 4
minute lag is explained by the fact that the earth not only rotates about
its axis but also proceeds along its orbit around the sun, while with
respect to the stars the earth's motion around the sun can be
neglected.
Denote $\hat{t}$ as Greenwich Mean Sidereal Time~(GMST), and 
$d$ the number of days that have passed or have to be passed since the
epoch. In most astronomy books, the epoch starts on
January 1st, 2000 AD, 12:00 noon Greenwich London time.
There is an algorithm that finds the total number of days
since that epoch~\cite{JeanMeeus}. Let $d$ be the total number of full
days that have passed or are to pass since that epoch, $y$ the number of years, and $m$ the number of months:
\begin{eqnarray}
d=367y-{\rm int}({{7}\over{4}}(y+{\rm int}({{m+9}\over{12}})))+{\rm int}({{275m}\over{9}})+day-730531.5.
\label{eq:days}
\end{eqnarray}
Now we want to have the total number of full days plus the fraction of a day, so we have $d_{tot}=d+(h+{min}/60)/24$ in order to find sidereal time at
that particular time of the day. $h$ is the Greenwich UT hour and
$min$ is Greenwich UT minute.
Now that we have the total number of days including the fractional part, the GMST angle\cite{JeanMeeus} is given by:
\begin{eqnarray}
 GMST_{angle}= 280.46061837 +360.98564736629\times d_{tot}.
\label{eq:gmsa}
\end{eqnarray}
From this number we remove
multiples of 360, and what is left is the hour angle. In order to
convert to hours we take $\hat{t}=GMST_{hour}=GMST_{angle}24/360$. In our experiment, spill number comes with a time stamp, so we had to map spills with time
stamps known to within 1 minute. A spill results in collisions and events. These time stamps were in Chicago time,
and had to be converted to Universal Time~(UT). During the year that
data was taken, two important dates had to be included, Daylight Saving 
Time ending October 26, 1996 00:00 and Daylight Saving Time begining April
 6, 1997 00:00. During Daylight Saving Time, we have to add 5 hours to 
Chicago local time to find Universal Time, but when there is no Daylight
Saving Time, we add 6 hours to Chicago Local Time. After they are converted to UT, Eq.~\ref{eq:days}
and Eq.~\ref{eq:gmsa} were used to find the Greenwich Mean 
Sidereal Hour $\hat{t}$.

\section{Previous Searches}

Searches for CPT violation have been
made in the neutral kaon system. Using an earlier CPT
formalism~\cite{kostelecky99}, KTeV
reported a bound on the CPT figure of merit
$r_K \equiv |m_{K^0} - m_{\overline{K}\!\,^0}|/m_{K^0} < (4.5 \pm 3) \times
10^{-19}$~\cite{ktev}. A more recent analysis, using
the framework described in reference~\cite{kostelecky} and more data extracted limits
on the coefficients for Lorentz violation of 
$\Delta a_{X}, \Delta a_{Y}<9.2\times 10^{-22}$~GeV~\cite{ktevlv}.  
CPT tests in $B^0$ meson decay have been 
made by OPAL at LEP~\cite{opal}, and by
Belle at KEK which has recently reported 
$r_B \equiv |m_{B^0} - m_{\overline{B}\!\,^0}|/m_{B^0} < 1.6 \times 10^{-14}$~\cite{belle}.

To date, no experimental search for CPT violation has been
made in the charm quark sector.  This is due in part to 
the expected suppression of $D^0-\d0b$  oscillations
in the ``Standard Model'', and the lack of a strong mixing
signal in the experimental data.


Recent mixing searches include
a study of lifetime differences between charge-parity (CP) eigenstates
from FOCUS, which  reported\cite{focusycp, belleycp, babary} a value for 
the parameter $y_{CP} = (3.42 \pm 1.39 \pm 0.74)\%$.  
The CLEO Collaboration has reported 
95\% confidence level bounds  on mixing parameters $x'$ and $y'$ 
(related to the usual parameters $x$ and $y$ by a strong phase 
shift):\cite{cleo}
$(1/2)x'^2 < 0.041\%$ and $-5.8\% < y' < 1\%$.  
FOCUS has reported\cite{linkrd} a study of the doubly Cabbibo 
suppressed ratio ($R_{DCS}$)  for the decay $D^0\rightarrow K^+\pi^-$ and has extracted
a contour limit on $y'$ (of order few \%) under varying assumptions of
$R_{DCS}$ and $x'$.
The question
arises -- what can be learned about indirect CPT violation 
given the apparent smallness of mixing in the charm
system? It turns out that even in the absence of a strong mixing 
signal one can still infer the level of CPT violation 
sensitivity through study of the time dependence of 
$D^0$ decays, which we show in this thesis.

\chapter{
The E831/FOCUS Experiment at Fermilab
}

FOCUS is a high-energy photoproduction experiment that took data during the Fermilab 1996--1997 fixed-target run.  A bremsstrahlung-generated photon beam with energies ranging from approximately 20 to 300~$\gev$ was incident on a BeO target.  While the primary purpose of the FOCUS experiment is to study the photoproduction of charm and the properties of charmed mesons and baryons, the experiment has also been able to collect an impressive sample of light quark events.  This chapter will give a brief overview of the FOCUS experiment and its detector.

\section{Physics Overview}

The FOCUS experiment has been at the forefront of charm physics since its analysis efforts began around 1998.  Improving on its predecessor, E687~\cite{expE687}, FOCUS has been able to reconstruct over one million $D$ mesons (see Fig.~\ref{fig_goldenmode}).  Over thirty papers have been published on topics such as semileptonic charm decays, charmed baryon lifetimes, CP violation in the charm sector, and the spectroscopy of charmed meson and baryon excited states\footnote{A list of publications and more detail concerning ongoing physics analyses can be found at http://www-focus.fnal.gov/}.

{
\begin{figure}[p]
  \begin{center}
  \leavevmode
  \epsfxsize=4truein
  \epsfbox{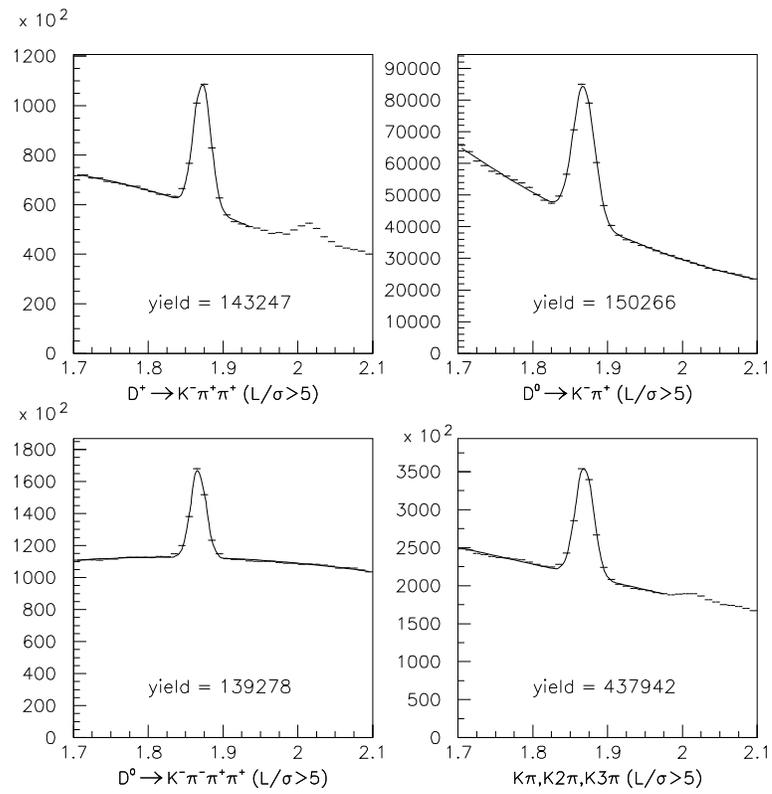}
  \end{center}
  \caption[The FOCUS $D$ meson signal in three different final states]
          {The FOCUS $D$ meson signal in three different final states. $L/\sigma>5$ was applied, where $L/\sigma$ is the $D$ meson decay length over its error.
  \label{fig_goldenmode}}
\end{figure}
}

\section{The Accelerator}

During the Fermilab 1996--1997 fixed-target run, 800$\gev$ protons from the Fermilab Tevatron were used to feed an array of fixed-target experiments.  In the Main Switchyard, the proton beam extracted from the Tevatron was split into a meson beam line, a neutrino beam line, and a proton beam line. The FOCUS photon beam originated from the proton beam line.  Figure~\ref{fig_tevatron} shows the general layout of Fermilab and the fixed target lines.

{
\begin{figure}[p]
  \begin{center}
  \leavevmode
  \epsfxsize=5truein
  \epsfbox{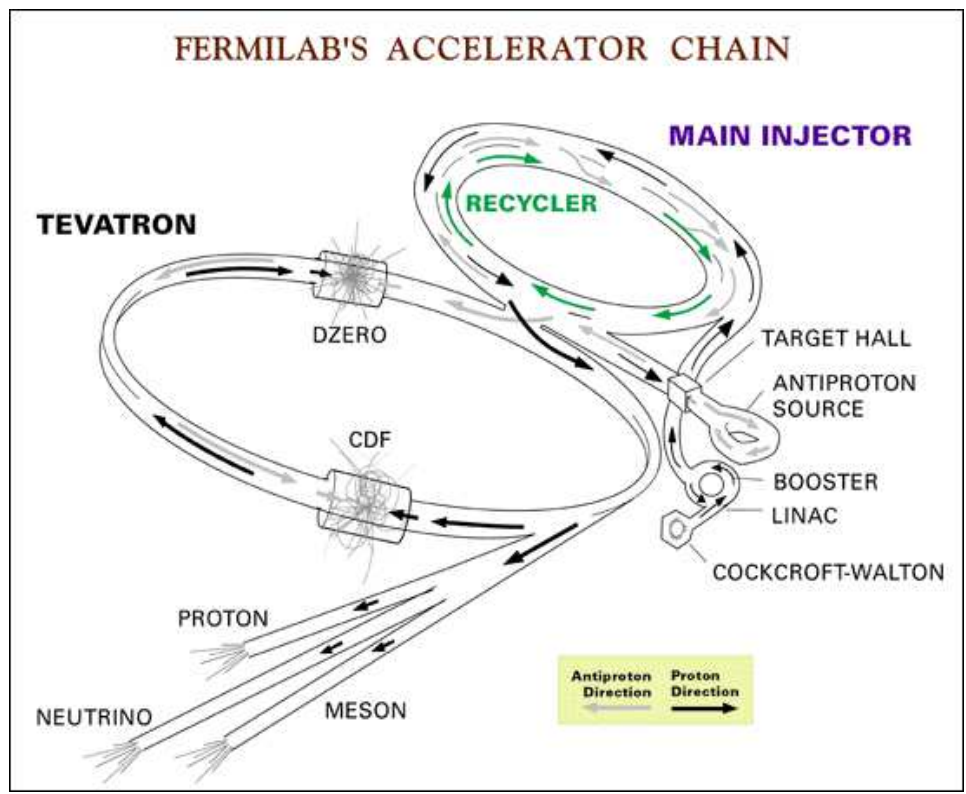}
  \end{center}
  \caption[A schematic of the layout of the Fermilab accelerator complex]
          {A schematic of the layout of the Fermilab accelerator complex.
  \label{fig_tevatron}}
\end{figure}
}

The 800~$\gev$ protons of the Tevatron are generated in a series of five stages, each stage increasing the energy of the beam. The process begins with the Cockcroft-Walton, where electrons are added to hydrogen atoms to form negatively charged ions.  The negative electric charge allows the $H^-$ ions to be accelerated across an electrostatic gap to an energy of 750~$\kev$. Next, the $H^-$ ions are fed into a linear accelerator (Linac).  The Linac accelerates the $H^-$ ions from 750~$\kev$ to 400~$\mev$ using a series of RF cavities.  Once at the end of the accelerator, the ions are stripped of their electrons in a thin carbon foil, the result of which is a 400~$\mev$ proton beam. From the Linac, the proton beam is picked up by the Booster synchrotron.  With a relatively small diameter of 500 feet, the Booster accelerates the protons from 400~$\mev$ to 8~$\gev$. By way of the Main Injector, the protons are now ready to enter the much larger Main Ring, a synchrotron with a 4 mile circumference housed in the same tunnel as the Tevatron.  The Main Ring brings the energy of the protons from 8~$\gev$ up to 150~$\gev$. In the final stage of acceleration, the protons are transferred from the Main Ring to the Tevatron.  Using 1000 superconducting magnets, the Tevatron boosts the proton energy from 150~$\gev$ to its final energy of 800~$\gev$ for fixed target experiments.

During the fixed-target run period, the Tevatron held 1000~proton bunches separated by 20~$\ns$.  The acceleration process went through a one minute cycle:  40~seconds were spent filling the Tevatron with 800~$\gev$ protons, and then during the remaining 20~seconds the protons were extracted from the Tevatron and routed through the Main Switchyard.  The FOCUS experiment was located in Wideband Hall at the end of the proton fixed target line.  Data collection within the FOCUS experiment was divided into separate ``runs,'' periods of roughly one hour of running.

\section{The Photon Beam}

{
\begin{figure}[p]
  \begin{center}
  \leavevmode
  \epsfxsize=4truein
  \epsfbox{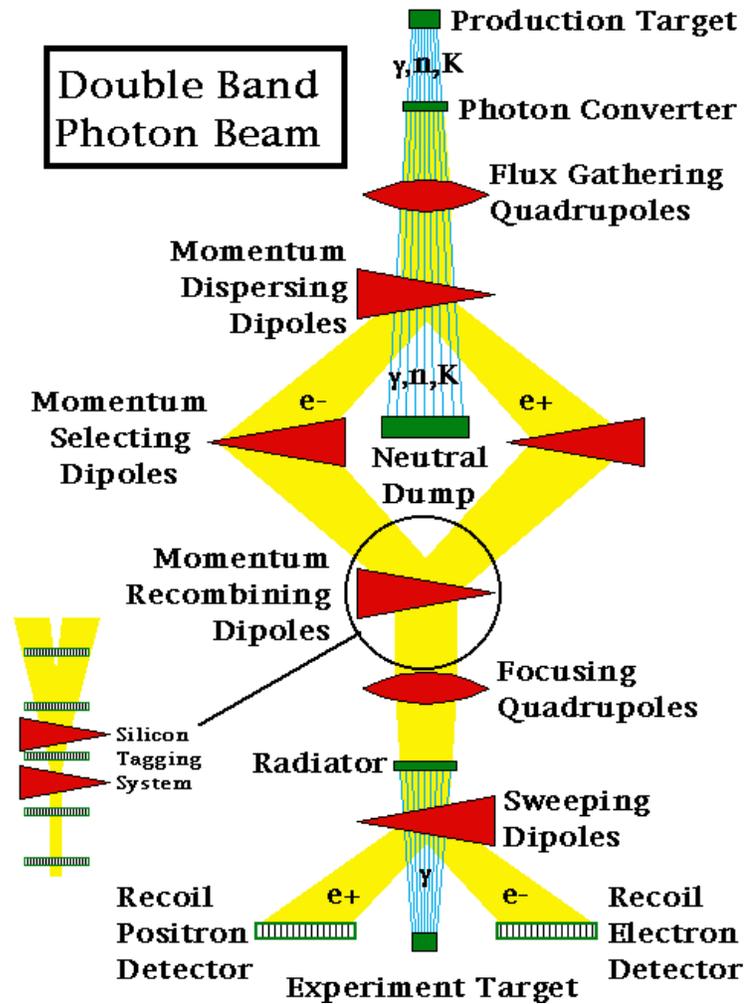}
  \end{center}
  \caption[Generating the high energy photon beam used by FOCUS]
          {Generating the high energy photon beam used by FOCUS.
  \label{fig_beam}}
\end{figure}
}

Once the 800~$\gev$ protons have been extracted from the Tevatron and sent down the proton fixed target line, the proton beam is converted to a photon beam~\cite{E831beam} through a series of stages (see Fig.~\ref{fig_beam}).  The process begins 365~m upstream of the FOCUS experimental target where the 800~$\gev$ proton beam interacts with the 3.6~meter long liquid deuterium production target.  This interaction results in a spray of all varieties of charged and neutral particles.  The charged particles are swept away by dipole magnets and collimators, leaving only neutral particles, primarily photons, neutrons, and $\klong$'s.  These neutral particles are sent through a lead converter that converts most of the photons in the neutral beam to $\epem$ pairs. The $\epem$ pairs are guided around a thick beam dump using a series of dipole magnets, and the remaining neutral particles in the beam are absorbed by the dump.  The series of dipole magnets leading the $\epem$ pairs around the neutral particle dump consists of (1) the Momentum Dispersing Dipoles, the magnets that initially cause electrons to bend one way and positrons the other; (2) the Momentum Selecting Dipoles, which are optimized to select electrons and positrons with momenta around 300~$\gev$; and (3) the Momentum Recombining Dipoles, the magnets that recombine the electrons and positrons into a single beam.  Once around the neutral beam dump, the $\epem$ beam is further focused by the Focusing Quadrupoles.

The $\epem$ beam can now be used to generate a photon beam using the bremsstrahlung process.  About 40~m upstream from the FOCUS experimental target, the $\epem$ beam is sent through a lead radiator. The individual electrons and positrons radiate photons through bremsstrahlung. Because of the extremely high energy of the $\epem$ beam of around 300$\gev$, the radiated photons travel in a direction nearly identical to the original direction of the $\epem$ beam.  
After radiating, the electrons and positrons are swept into instrumented beam dumps (the Recoil Positron and Recoil Electron detectors) by the Sweeping Dipoles, and only a high energy photon beam remains.  The mean photon beam energy is around 150$\gev$, but in addition there is a long low energy tail reaching down to around 20$\gev$.

The energy of each photon in the beam nominally is measured by the beam tagging system.  Before entering the Radiator, the energies of the electrons and positrons are measured by a set of five silicon planes interspersed between the Recombining Dipoles.  After passing through the Radiator, when the electrons and positrons are swept to opposite sides of the experimental target, their energies are again measured, this time by lead-glass calorimeters, the Recoil Electron and Recoil Positron detectors.  The energy of the radiated beam photon is then just the difference in energy of the electron (or positron) before and after the Radiator.  In the case of a multiple bremsstrahlung event, the energy of the extra noninteracting photons is measured by a small central calorimeter, the Beam Gamma Monitor, and this energy is subtracted from the original measurement.  In other words, the tagged photon beam energy ($\ebeam$) is calculated from:
    \begin{equation}
          \ebeam = E_{INC} - E_{OUT} - E_{BGM} ,
    \label{eqn_ebeamtag}
    \end{equation}
where $E_{INC}$ is the incident electron (positron) energy before radiating, $E_{OUT}$ is the electron (positron) energy after radiating, and $E_{BGM}$ is the energy of any additional photons produced in a multiple bremsstrahlung event.  The energy resolution of the beam tagging system is approximately 16 GeV.

\section{The Spectrometer}

The FOCUS detector, building upon the previous E687 photoproduction experiment~\cite{expE687}, is a forward multi-particle spectrometer designed to measure the interactions of high energy photons on a segmented BeO target (see Fig.~\ref{fig_spectrometer}).  BeO was chosen as the target material to maximize the ratio of hadronic to electromagnetic interactions. The target was segmented into four sections to allow for a majority of charmed particles to decay outside of the target material.

{
\begin{figure}[p]
  \begin{center}
  \leavevmode
  \epsfxsize=5truein
  \epsfbox{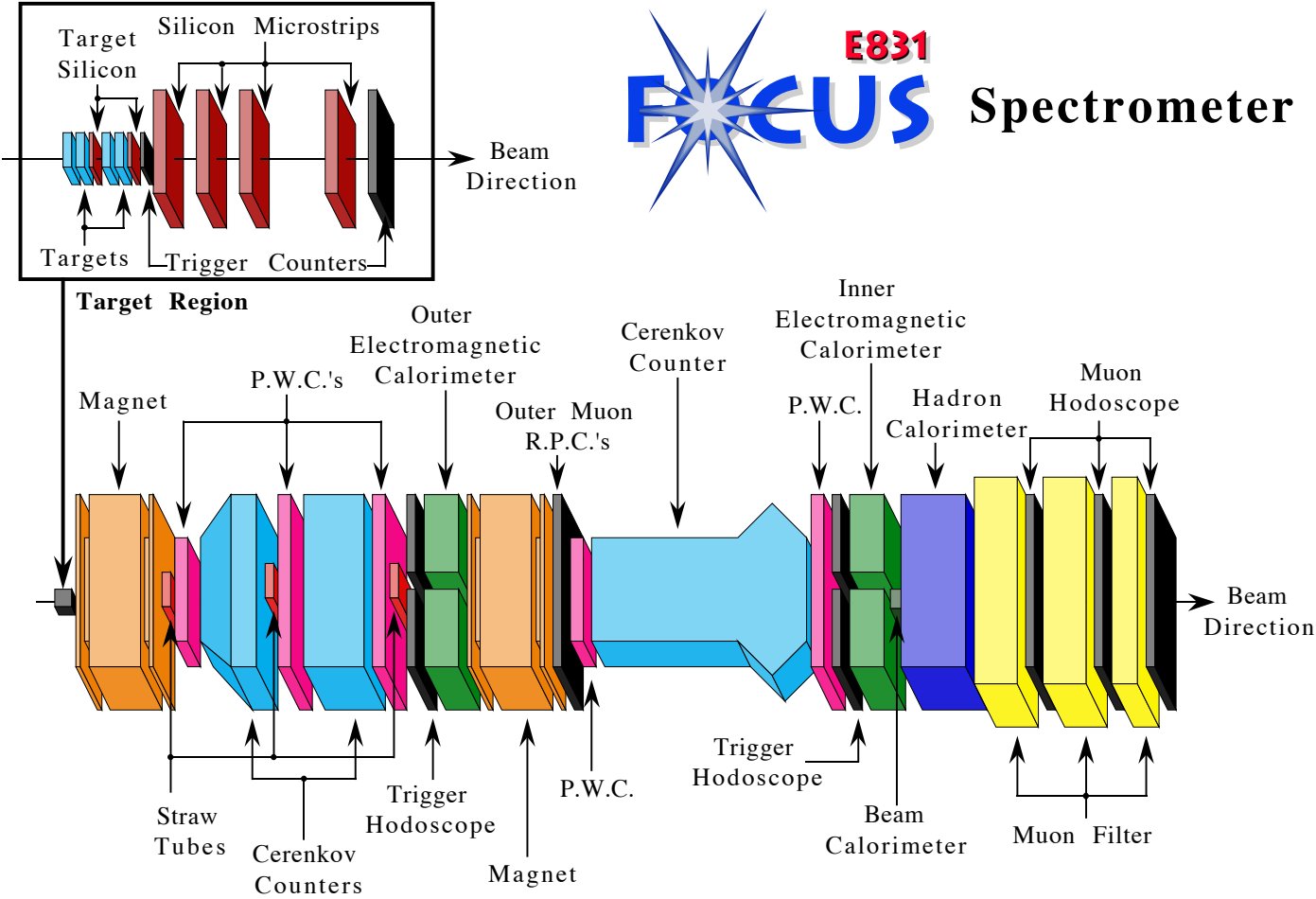}
  \end{center}
  \caption[The FOCUS spectrometer]
          {The FOCUS spectrometer.
  \label{fig_spectrometer}}
\end{figure}
}

Charged particles emerging from the target region are first tracked by two systems of silicon strip detectors.  The upstream system, consisting of four planes (two stations of two views), is interleaved with the experimental target, while the other system lies downstream of the target and consists of twelve planes of microstrips arranged in three views.  Once this initial stage of precision tracking is complete, the momentum of a charged particle is determined by measuring its deflections in two analysis magnets of opposite polarity with five stations of multiwire proportional chambers.  The measured momentum is used in conjunction with three multicell threshold $\check{\rm C}$erenkov counters to discriminate between pions, kaons, and protons.

In addition to excellent tracking and particle identification of charged particles, the FOCUS detector provides good reconstruction capabilities for neutral particles. $K^0_S$'s are reconstructed using the ``one-bend'' approximation described in Ref.~\cite{E831ks}.  Photons and $\pi^0$ are reconstructed using two electromagnetic calorimeters covering different regions of rapidity.  

Three elements of the FOCUS detector are most important for the analysis of the charm decay $D^0\rightarrow K^-\pi^+$ which we use to search for CPT violation.  First, the tracking system provides a list of charged tracks and their momenta.  Second, the particle identification system classifies the charged tracks as pions, kaons, or protons.  Third, the triggering elements require that events satisfy a certain number of requirements before they are recorded. Further information on other detector elements (e.g., the calorimeters) can be found elsewhere~\cite{EricThesis,LinkThesis}.

 \subsection{Tracking}

\begin{figure}
\begin{center}
\includegraphics[width=0.9\textwidth]{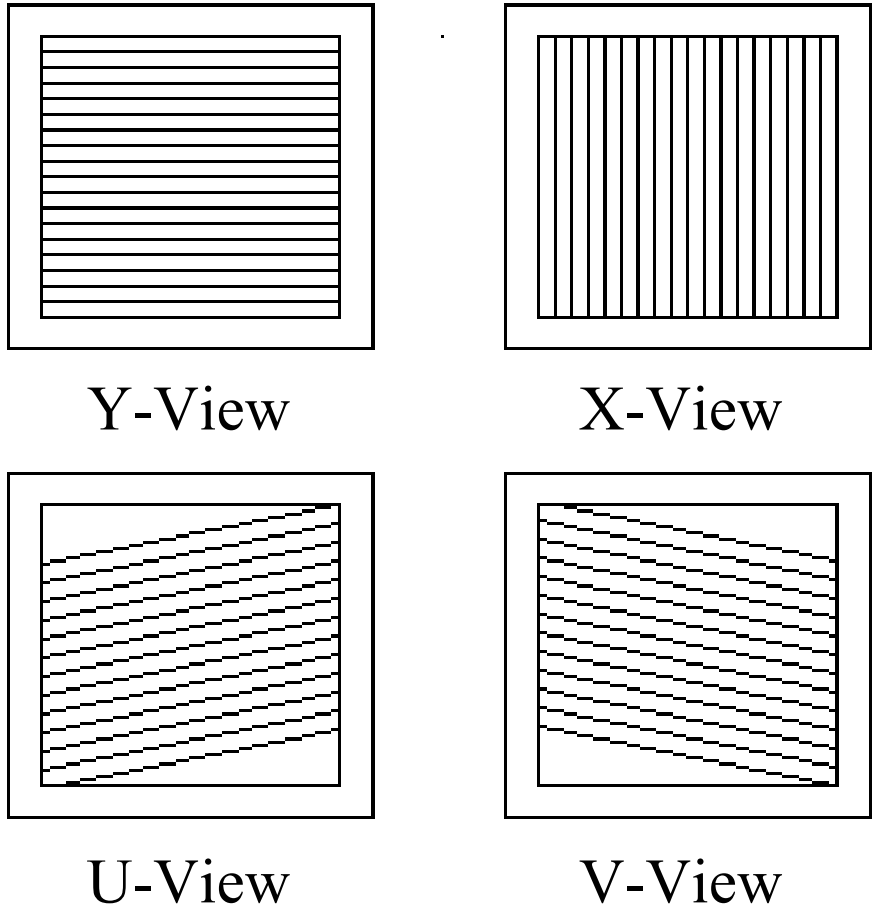}
\caption[A rough sketch of the different PWC views]{A rough sketch of the different PWC views.  Charged particles passing through these planes ionize the gas around one of the wires, the charge is collected on the nearest wire, and then recorded, giving a one-dimensional coordinate of the particle.  Combining views, the ($x,y$) position of a particle can be calculated at a given value of $z$.}
\label{fig_pwc}
\end{center}
\end{figure}

The purpose of the tracking system is both to reconstruct the paths particles have traveled through the spectrometer and to measure the momenta of these particles.  The first task is accomplished by a series of detecting planes normal to the beam direction and placed at advantageous positions throughout the spectrometer.  Each plane consists of an array of parallel silicon strips or wires, depending on the detector type, which send out a signal when a charged (ionizing) particle passes through a silicon plane or close by a wire. Knowing which wire or strip a particle has passed near or through provides a one-dimensional coordinate of the position of the particle on the detecting plane.  By grouping planes at various tilts, or views (see Fig.~\ref{fig_pwc}), into stations, an ($x,y$) coordinate can be calculated at various positions of $z$, where $z$ is the distance from the target, and $x$ and $y$ are horizontal and vertical coordinates, respectively.  Connecting the ($x,y$) coordinates from station to station ($z$ position to $z$ position) results in a track, the path a charged particle has followed through the spectrometer.


The second task, measuring a track's momentum, is accomplished by observing the deflections of the charged particle in known magnetic fields.  In FOCUS, this is accomplished by using two different large aperture dipole magnets.  The first magnet (M1) provides a vertical momentum kick of 0.5~$\gevc$, while the second magnet (M2) provides a larger vertical momentum kick of 0.85~$\gevc$ in the opposite direction.  Having different strengths for the two magnets allows sensitivity to a larger range of momentum.  A low momentum track will be measured well by M1, but may be bent out of the acceptance by M2.  A high momentum track may not be deflected enough by M1 for a good momentum measurement, but will be picked up by the stronger M2.  The momentum of a track is calculated by using
    \begin{equation} 
          p=\frac{Kick}{\Delta S}
    \end{equation} 
as the track passes through either magnet, where $Kick$ is the constant momentum kick of one of the magnets and $\Delta S$ is the change in vertical slope of the track as it passes through that magnet.

The FOCUS tracking system consists of several distinct subsystems.  The upstream system consists of silicon strip detectors placed among the target elements (referred to as the target silicon system~\cite{E831tsil}) and silicon strip detectors placed just downstream of the target region (referred to as the SSD system). 
 Charged tracks are followed through the two dipole magnets by the downstream tracking system, which consists of five stations of proportional wire chambers (PWC).  Three stations of PWC are between M1 and M2, and two are downstream of M2.



The silicon strip detectors in the upstream system are essentially reverse-biased diodes with charge collecting strips etched on the surface.  When a charged particle passes through the interior of the silicon, electron-hole pairs are created.  The internal electric field pulls the freed electrons to the surface of the silicon where they are picked up by the conducting strip, amplified, and registered in the data acquisition system.

The target silicon system, the silicon strip system placed among the target elements, is composed of two stations of two planes of silicon strip detectors with strips oriented at $\pm45^{\circ}$ from
the horizontal.  The first station is between the second and third target elements, and the second station follows immediately after the last target element (see Fig.~\ref{fig_spectrometer}).  The planes are $25\times50~\mm$ in size (the larger dimension is vertical), and the strips have a width of 25~$\micron$, giving 1024 different channels per plane.

The SSD system, the second system of silicon strip detectors, begins just downstream of the target system and extends downstream approximately $30~\cm$, still upstream of the first dipole magnet. It consists of four stations of three planes each with the silicon strips oriented vertically, and $\pm45^{\circ}$ from the horizontal.  The stations are each $6~\cm$ apart except for the last, which is separated by $12~\cm$.  The first station (i.e., the most upstream) consists of $25~\mm$ long strips.  In the central region the strips are 25~$\micron$ wide and in the outer region the strips are 50~$\micron$ wide.  The other stations consist of $50~\mm$ long strips, with widths of 50~$\micron$ in the central region and 100~$\micron$ in the outer.

Tracks in the upstream tracking system are found in three steps.  First, clusters are formed within each plane.  That is, regions where adjacent strips have fired are grouped together.  By measuring the amount of charge collected, the cluster is forced to be consistent with having been formed by a single charged track.  Second, projections are formed within each station.  In other words, clusters within planes are joined to form a very short track segment within a station.  Finally, tracks are formed by connecting the station projections.  The last step is accomplished by fitting different combinations of station projections with straight lines and taking the best fits to be the tracks.

The downstream tracking system is composed of five stations of proportional wire chambers (PWC). A PWC operates on roughly the same principle as a silicon strip detector. When a charged particle passes through a PWC, the PWC gas is ionized and the ions drift through an electric field and are collected by parallel metal wires.  The charge is collected at the end of a wire giving the one-dimensional position of a track. Arranging the PWC planes within a station at various tilts, or views, gives an ($x,y$) coordinate for a PWC station.

Five stations of four PWC planes each are interspersed throughout the FOCUS spectrometer. The planes within a station are oriented vertically, horizontally, and at $\pm11.3^{\circ}$ from the horizontal.  The first three stations (most upstream) are placed between the magnets M1 and M2, and the last two stations appear downstream of M2 on either side of the last $\check{\rm C}$erenkov counter (C3). The first and fourth stations have the dimensions of $76\times127~\cm$ and have a wire spacing of $2~\mm$.  The second, third, and fifth stations are $152 \times 229~\cm$ and have a wire spacing of $3.3~\mm$.

Tracks in the downstream system are reconstructed in three steps.  First, hits in the planes with vertical strips are connected from station to station with straight lines, referred to as $x$ view tracks.  The line segments formed from hits in this view are straight since it is the projection unaffected by the magnetic field, i.e., it is the non-bend view.  Second, the other three views (the horizontal wires, and $\pm11.3^{\circ}$ wires) are combined within each station to form short projections.  Finally, the $x$-view tracks and station projections are combined by fitting to two straight lines, one before M2 and one after, and with a bend parameter to take into account the track's bending through M2.

Once tracks have been found in the upstream and downstream tracking systems, they must be linked together. This is accomplished by refitting all the hits of the upstream and downstream tracks with three straight lines and two bend parameters corresponding to the amount of deflection resulting from M1 and M2.  With two opportunities to measure the momentum, tracks can be linked by enforcing consistency.  Doubly linked tracks, where one upstream track is linked with two downstream tracks, are allowed to accommodate the possibility of photons converting to $\epem$ pairs that do not significantly separate until after M1.

The momentum resolution for charged tracks depends on the momentum of the track and whether the track has passed through M1 and M2 or just M1.  For tracks only deflected by M1, the resolution is given by:
    \begin{equation}
          \frac{\sigma_p}{p}=0.034\cdot\frac{p}{100~\gevc}
          \sqrt{1+(\frac{17~\gevc}{p})^2} .
    \end{equation}
For tracks extending through M2 the momentum resolution is:
    \begin{equation}
          \frac{\sigma_p}{p}=0.014\cdot\frac{p}{100~\gevc}
          \sqrt{1+(\frac{23~\gevc}{p})^2} .
    \end{equation}
For low momentum tracks, the momentum resolution is limited by multiple scattering within the detector material.  The momentum resolution for high momentum tracks is limited by the spacing of the wires and strips and uncertainties in the alignment of the detector planes.

\subsection{Particle Identification}

Particle identification in FOCUS is provided by a series of three $\check{\rm C}$erenkov counters, which are based on the principle that when a particle travels through a medium with a velocity greater than $c/n$, where $c$ is the speed of light in vacuum and $n$ is the index of refraction of the medium, then the particle will radiate photons.  Being sensitive to these radiated photons, a $\check{\rm C}$erenkov counter can determine whether or not the velocity of a particle is above or below the velocity threshold, $c/n$.  This velocity threshold corresponds to different momenta thresholds for particles of different masses\footnote
{The momentum of a particle, $p$, is given by $p = \gamma mv$, where $m$ is the mass of the particle, $v$ is the velocity and $\gamma = (1-\frac{v^2}{c^2})^{-1/2}$.}, and this is what allows a $\check{\rm C}$erenkov counter to distinguish between particle types.  For example, if the velocity threshold of a $\check{\rm C}$erenkov counter were $0.9999\,c$, then the momentum threshold for a pion would be 9.87~$\gevc$, while the momentum threshold for a kaon would be 34.9~$\gevc$.  So, if a track had a momentum of 20~$\gevc$, as determined by the tracking system, then the $\check{\rm C}$erenkov counter would fire if the track were a pion, but would not fire if the track were a kaon.  In this particular example, the $\check{\rm C}$erenkov counter ideally could cleanly distinguish between pions and kaons for all tracks with momenta between 9.87 and 34.9~$\gevc$.

\begin{table}[t]
\centering
\caption[Properties of the three FOCUS $\check{\rm C}$erenkov detectors]{Properties of the three FOCUS $\check{\rm C}$erenkov detectors.
\label{tab_cerenkov}}
\begin{tabular}{|c|c|c|c|c|}
  \hline
     Counter & Material & $\pipm$ Threshold & $\kpm$ Threshold & $\ppm$ Threshold \\
             &          &     $\,\gevc$     &     $\,\gevc$    &     $\,\gevc$    \\
  \hline
     C1      & 80\% He, 20\% $\rm{N}_2$ &  8.4 & 29.8 & 56.5  \\
     C2      &      $\rm{N_2O}$         &  4.5 & 16.0 & 30.9  \\
     C3      &          He              & 17.4 & 61.8 & 117.0 \\
  \hline
\end{tabular}
\vskip 0.25in
\end{table}

By using three different $\check{\rm C}$erenkov counters filled with gases of different indices of refraction (see Table~\ref{tab_cerenkov}), FOCUS can cleanly distinguish between pions, kaons, and protons over a wide range of momentum.  Now, for example, a 20~$\gevc$ pion, a 20~$\gevc$ kaon, and a 20~$\gevc$ proton will all have different signatures.  The pion will fire all three counters C1, C2, and C3; the kaon will only fire C2; and the proton will not radiate at all.  Notice that there is ideally a clean separation between pions and kaons with momenta all the way from 4.5~$\gevc$ to 61.8~$\gevc$.  E687 used a particle identification system based only on these thresholds and logic tables.

FOCUS has improved on this system by measuring the angle with which photons are radiated by a particle traveling with a velocity above threshold.  This provides additional information about the particle's velocity, $v$, since the angle of radiation, $\theta$, is given by 
    \begin{equation}
          \cost=\frac{c}{nv} .
    \end{equation} 
Therefore the higher the velocity is above threshold, the larger the ring of the emitted photons.  The measurement of the angle has been made possible by dividing the back of the $\check{\rm C}$erenkov counters into arrays of cells, with smaller cells near the center of the counter and larger cells further out from the center.

FOCUS has implemented a system called CITADL for particle identification based on the detected rings in the counters~\cite{E831id}. The CITADL system works by assigning likelihoods to different particle hypotheses. For example, if a particle of given momentum (measured by the tracking system) were a pion, then we can calculate its velocity and the angles of radiation and thus know which cells in which counters should have fired. The likelihood for the pion hypothesis is then calculated based on the status of these cells. If a given cell should be ``on'' given the pion hypothesis, and the cell was found to be ``on'', then the total likelihood for the pion hypothesis receives a contribution of
    \begin{equation} 
          L_{\mathrm{cell}}=(1-e^{-\mu})+a-a(1-e^{-\mu}) ,
    \end{equation} 
where $\mu$ is the expected number of photoelectrons in the cell, $a$ is the accidental firing rate, and Poisson statistics has been assumed.  If the cell was found to be ``off'', then the total likelihood receives a contribution of 
    \begin{equation}
          L_{\mathrm{cell}}=1-[(1-e^{-\mu})+a-a(1-e^{-\mu})] .
    \end{equation}
The likelihoods are summed over all the cells in the ring of cells that should have fired given the pion hypothesis to give a total likelihood for the pion hypothesis:
    \begin{equation}
          L_{\pi} = \sum_{\mathrm{cells}} L_{\mathrm{cell}} .
    \end{equation}
Similarly, likelihoods are calculated for the $e^{\pm}$, $\kpm$, and $\ppm$ particle hypotheses.  

To convert the likelihoods to $\chisquare$-like measures, the CITADL system introduces the variables 
    \begin{equation}
          W_i = -2\ln(L_i) , 
    \end{equation}
where $i$ indicates the hypothesis under consideration, i.e., either $\epem$, $\pipm$, $\kpm$, or $\ppm$. The $W_i$ with the lowest value indicates the most likely particle hypothesis. Since kaons and pions dominate the hadronic final states, useful parameters for particle identification are the "pionicity", defined as 
    \begin{equation}
          \mathrm{Pionicity} \equiv W_K - W_{\pi} ,
    \end{equation}
and the ``kaonicity'', defined as 
    \begin{equation}
          \mathrm{kaonicity} \equiv W_{\pi} - W_K .
    \end{equation}
Increasing the ``kaonicity'' requirement, for example, decreases the chances a pion will be misidentified as a kaon.

  \subsection{Triggers}

Whenever an interesting event occurs in the detector, data must be read out and stored.  The trigger system is responsible for discriminating between interesting and uninteresting events.  The trigger decision takes place in several stages and there are several different triggers based on different physics questions.  The data for the $K^-\pi^+$ analyses included in the remaining chapters are obtained through the hadronic trigger.  While triggering elements are located throughout the spectrometer and serve various purposes, the hadronic trigger imposes only three simple criteria on events.

First, like all other triggers, the hadronic trigger requires a coincidence in TR1 and TR2. TR1 is just downstream of the target assembly, and TR2 is just downstream of the SSD system.  A coincidence in TR1 and TR2 guarantees that at least one charged track has passed through the SSD system.

Second, in addition to having tracks in the SSD system, the hadronic trigger requires at least two charged tracks to traverse the entire downstream tracking system. The OH and $\hxv$ detectors are located just after the last PWC station and are designed to count charged tracks.  The $\hxv$ detector covers the inner region of the acceptance and the OH detector covers the outer region.  The hadronic trigger requires either two charged tracks be detected by the $\hxv$ or one charged track register in the $\hxv$ and one in the OH.  Both the $\hxv$ and the OH include a vertical gap from top to bottom to allow $\epem$ pairs to pass.

Finally, a minimum hadronic energy of 18~$\gev$ as determined by the hadronic calorimeter is an additional requirement imposed by the hadronic trigger.  This requirement ensures the presence of hadronic tracks (as opposed to $\epem$ tracks).

\section{Data Collection}

Over the course of its running, the FOCUS experiment collected 6.5 billion events recorded on 5926 tapes, each tape holding 4.5 Gigabytes of data.  The data was collected over approximately 6500 runs, each run corresponding to roughly one hour of running time.  The data was processed in four separate stages.

(1) PassOne was where all the major reconstruction was performed, e.g., track reconstruction and particle identification.

(2) Skim1 separated the PassOne output into six large superstreams based on different physics criteria. One of the superstreems was the hadronic meson decays also called as ``SEZDEE'' (Super EaZy DEE) stream. This stream included all hadronic decays of $D$ mesons. This stream was still large (approximately 300 8mm tapes, 1.2 TB worth of data). 

(3) In Skim2, the superstreams were separated into separate substreams by requiring more specific physics criteria. One of the substreams of SEZDEE was tuned to select $D^0\rightarrow K^-\pi^+$ decays by requiring that the invariant mass of  $K^-\pi^+$ is between 1.7\gev~and 2.1\gev, have a decay length significance greater than 2.5, and a confidence level of secondary vertex greater than 1\%.  After these cuts the data was reduced to a size of $\approx 63$~GB.

(4) In the final stage, the $K^-\pi^+$ data was copied to the local disks at one of the Indiana University High Energy Physics clusters.

\chapter{Data Analysis}

In this thesis we investigate the current
experimental sensitivity for a CPT-violating signal
using data  collected by the FOCUS Collaboration during the 
1996--97 fixed-target run at Fermilab. The analysis is also described in a journal publication~\cite{ref:abaz}.
\section{Analysis Aproach}
The data analysis is as follows.
We analyze the two right-sign hadronic 
decays $D^0 \rightarrow K^-\pi^+ $ 
and $\bar{D}^0 \rightarrow K^+\pi^-$.
We use the soft pion from the decay $D^{*+}\rightarrow D^0\pi^+$
to tag the flavor of the $D$ at production, and 
the kaon charge in the decay $D^0\rightarrow K^- \pi^+$
to tag the $D$ flavor at the time of decay. (Charge conjugate modes are
assumed throughout this thesis.)
\section{Analysis Cuts}
$D^0\rightarrow K^- \pi^+$ events were selected by requiring
a minimum detachment of the secondary (decay) vertex from the 
primary (production) vertex of 5$\sigma_L$. $\sigma_L$ is the decay length error. 
The primary vertex was found using a candidate driven vertex
finder which nucleated tracks about a ``seed'' track 
constructed using the secondary vertex and the reconstructed
$D$ momentum vector. Both primary and secondary vertices
were required to have confidence level fits of greater
than 1\%.
The $D^*$-tag is accomplished by requiring the
$D^*-D^0$ mass difference to be less than 3 MeV/$c^2$ of
the nominal value~\cite{pdg2004}.

\begin{figure}[tbh]
\centerline{
\psfig{figure=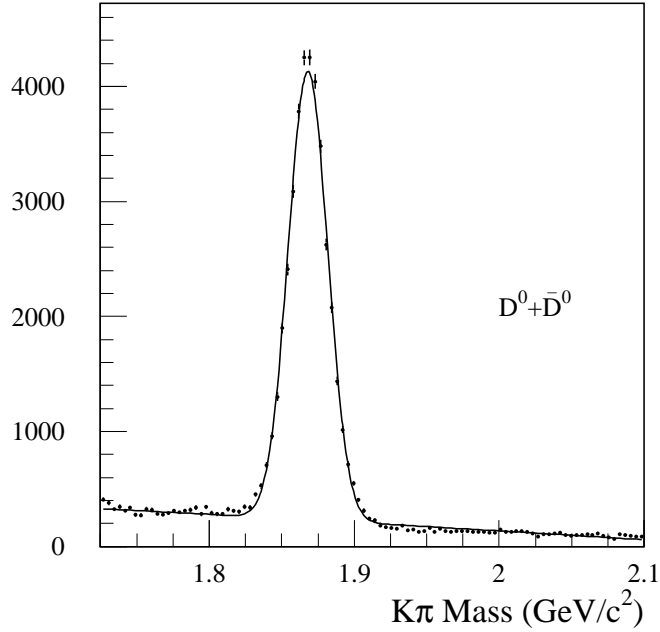,height=3.5in}
}
\caption{Invariant mass distribution for the sum of $D^0$ and 
$\overline{D}^0$ right-sign decay candidates.}  
\label{fig:signal}
\end{figure}

\begin{figure}
\includegraphics[height=2.5in]{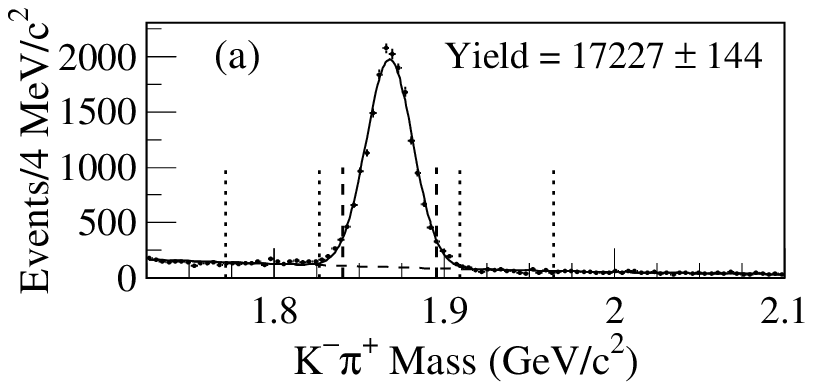}\\
\includegraphics[height=2.5in]{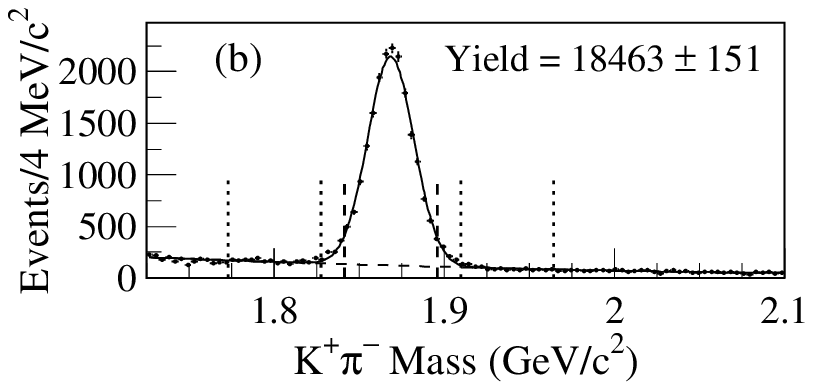}
\caption{Invariant mass of (a) $D^0\rightarrow K^-\pi^+$ and (b) 
$\d0b\rightarrow K^+\pi^-$ for data (points) fitted with a
Gaussian signal and quadratic background (solid line). 
The vertical dashed lines indicate the signal
region, and the vertical dotted lines indicate the sideband region.}  
\label{fig:dmass_split}
\end{figure}

Kaons and pions were identified using
the $\check{\rm C}$erenkov particle identification cuts.
These cuts are based on likelihood ratios between the
various stable particle hypotheses, and are computed for a 
given track from the observed firing response (``on'' or ``off'')
of all cells within the track's ($\beta$ = 1) $\check{\rm C}$erenkov 
light cone in each of three multi-cell, threshold $\check{\rm C}$erenkov
counters as described earlier.
The product of all firing probabilities
for all cells within the three $\check{\rm C}$erenkov cones
produces a $\chi^2$-like variable called
$W_i \equiv $ -- 2$\times$log(likelihood) where $i$ ranges
over electron, pion, kaon and proton hypotheses.
For the $K$ and the $\pi$ candidates, we require $W_i$ to be no more
than 4 greater than the smallest of the other three hypotheses
($W_i - W_{min} < 4$) which eliminated candidates that are highly to have been
 misidentified. 
In addition, $D^0$ daughters must satisfy the slightly stronger
$K\pi$ separation criteria $W_\pi - W_K > 0.5$ for the $K$
and $W_K - W_\pi > -2 $ for the $\pi$.
Doubly misidentified $D^0 \rightarrow K^-\pi^+$ candidates are removed
by imposing a hard $\check{\rm C}$erenkov cut on the sum of 
the two separations 
$((W_\pi - W_K)_K + (W_K - W_\pi)_\pi > 8)$.
Primary vertices that lie in the TR1 region are poorly reconstructed
so we exclude events in TR1, by
imposing  the $z$ coordinate of the primary vertex~$<2$~cm.
Fig.~\ref{fig:signal} shows the invariant mass distribution
for the two $D^*$-tagged, right-sign decays
$D^0\rightarrow K^-\pi^+$ and $\bar{D}^0\rightarrow K^+\pi^-$.
Fig.~\ref{fig:dmass_split} shows the invariant mass distributions
for right-sign decays split up into particle and antiparticle.
A fit to the mass distribution is carried out using a Gaussian
function to describe the signal and a second-order polynomial for the
background. The fit yields $17\,227 \pm 144$ $D^0$ and
 $18\,463 \pm 151$ $\d0b$ signal events. 

\begin{figure}[tbh]
{\hbox
{\psfig{figure=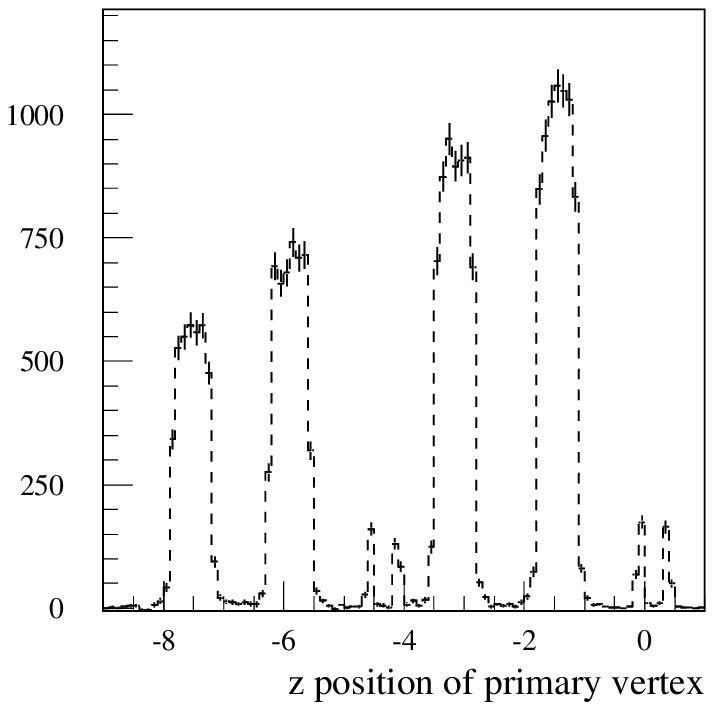,height=2.5in}
\hskip 0.25in
\psfig{figure=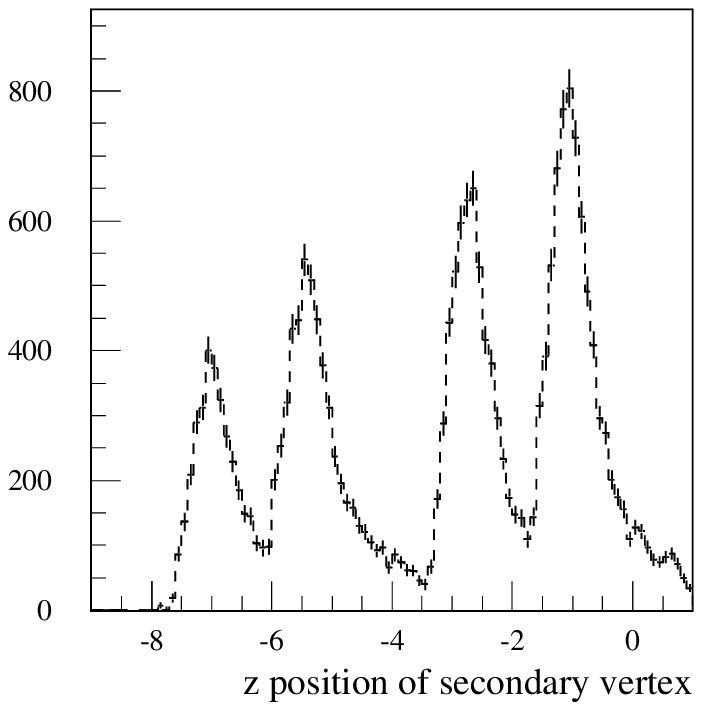,height=2.5in}}
}
\caption{(Left) $z$ position of the primary vertex of $D$'s for 
$run number > 9750$ and (right) $z$ position of $D$'s secondary vertices for 
$run number > 9750$. The dashed line is to guide the eye.}  
\label{fig:z_vertexes}
\end{figure}

Fig. \ref{fig:z_vertexes} shows the primary and secondary vertices for
$D^0$'s for the run period 6 which has 4 target segments interwoven with
target silicons. Most of the primary vertices lie within the target
segments and some in the target silicons. The contours of the
target segments and target silicons can be seen. About 60\% of $D^0$ decays occur outside of target segments.  

\begin{figure}[tbh]
{\hbox
{\psfig{figure=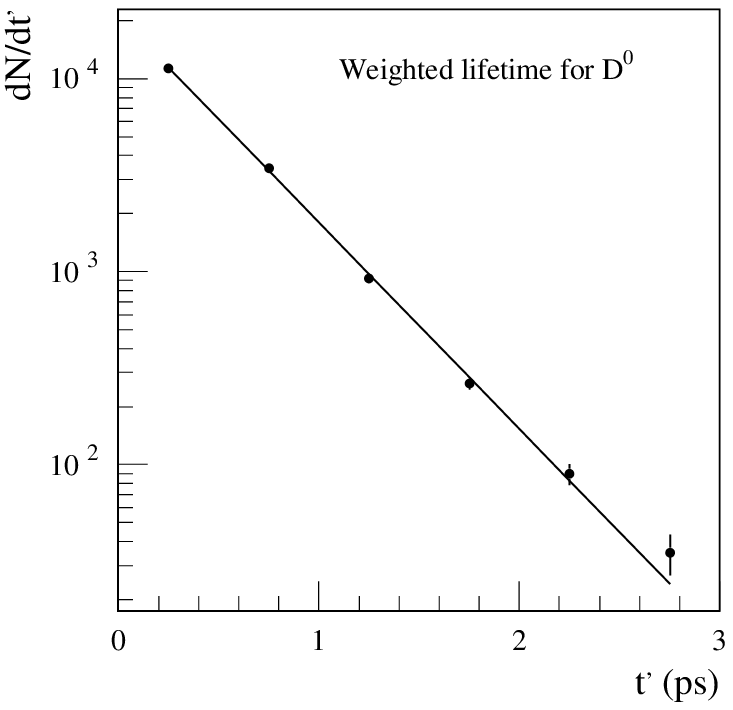,height=3.0in} 
\hskip 0.25in 
\psfig{figure=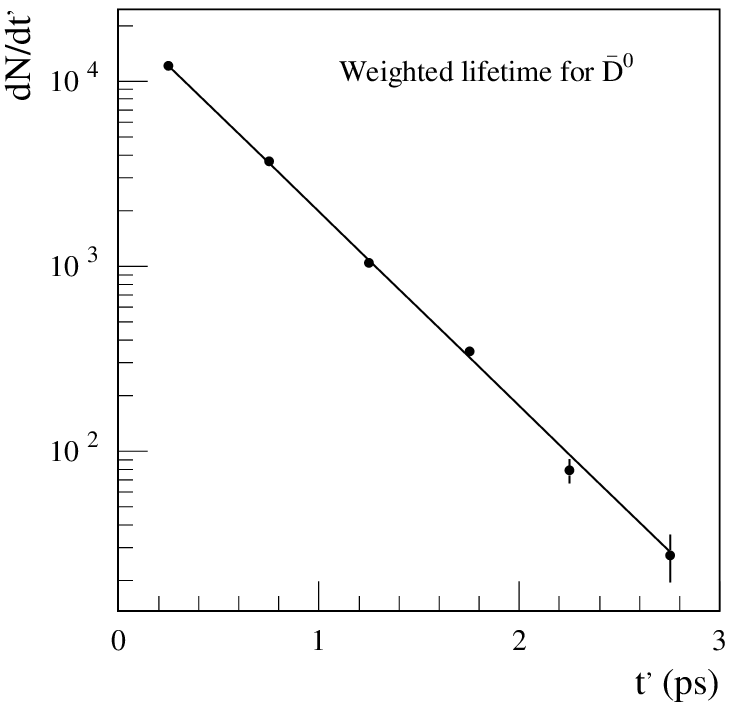,height=3.0in}}
}
\caption{Background
subtracted reduced proper time distributions for 
$D^0$ and $\d0b$. }  
\label{fig:ptime}
\end{figure}

The reduced proper time is a traditional lifetime variable used
in fixed-target experiments that uses the detachment between
the primary and secondary vertex as the principal tool in
reducing non-charm background. The reduced proper time is
defined by $t'=(\ell - N\sigma_\ell)/(\beta\gamma c)$ where
$\ell$ is the distance between the primary and secondary
vertex, $\sigma_\ell$ is the resolution on $\ell$, and $N$
is the minimum detachment cut required to tag the charmed
particle through its lifetime.
Fig.~\ref{fig:ptime} shows reduced proper time distributions
for the two right-sign decays:
$D^0\rightarrow K^-\pi^+$ and $\bar{D}^0\rightarrow K^+\pi^-$.

Table~\ref{tab_yields} shows a summary of the fits for
$D^0\rightarrow K^-\pi^+$ and $\bar{D}^0\rightarrow K^+\pi^-$. It
gives us an overall picture of yields, signal to background ratios,
masses and lifetimes.\footnote{From a simple exponential fit.} The above cuts have been chosen to maximize the signal to background ratio.
\\
\\
\begin{table}[t]
\centering
\caption[Summary of Yields]{Summary of the fits for 
$D^0\rightarrow K^-\pi^+$ and $\bar{D}^0\rightarrow K^+\pi^-$.
\label{tab_yields}}
\begin{tabular}{|c|c|c|c|}
\hline
Parameter&$\overline{D^0}$&$D^0$&$D^0+\overline{D^0}$\\
\hline
\hline
Yield&$18287 \pm 235$&$17085 \pm 224$&$35342 \pm 322$\\
\hline
$S/B$&$10.24$&$11.11$&$10.12$\\
\hline
${\rm Mass}~({\rm MeV}/c^2)$&$1868.50 \pm 0.11$&$1867.80 \pm 0.11$&$1868.10 \pm 0.08$\\
\hline
$\sigma({\rm MeV}/c^2)$&$13.51 \pm 0.10$&$13.63 \pm 0.10$&$13.57 \pm 0.07$\\
\hline
$\tau(fs)$&$412.8 \pm 3.8$&$405.7 \pm 3.8$&$409.6 \pm 2.7$\\
\hline
\end{tabular}
\vskip 0.25in
\end{table}

\begin{figure}[tbh]
{\hbox
{\psfig{figure=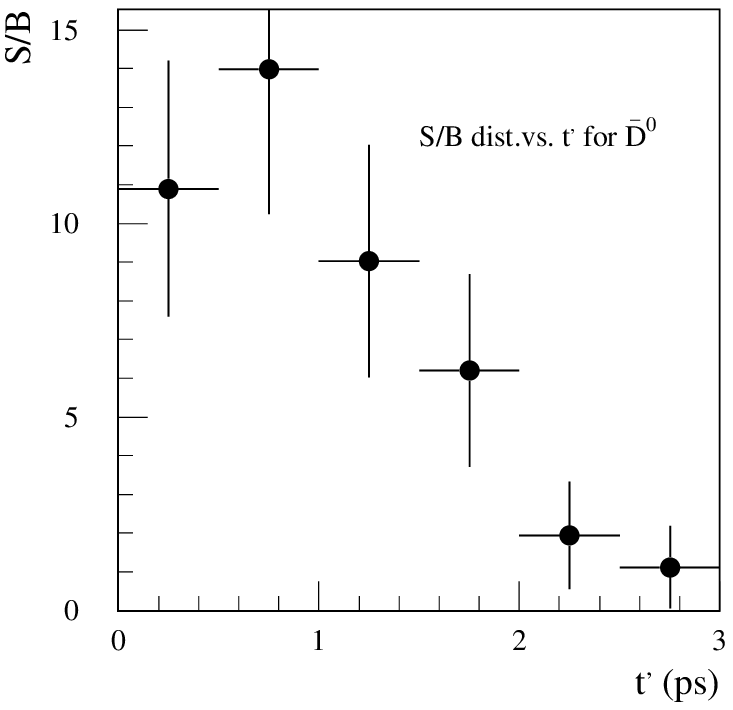,height=2.5in}
\hskip 0.25in
\psfig{figure=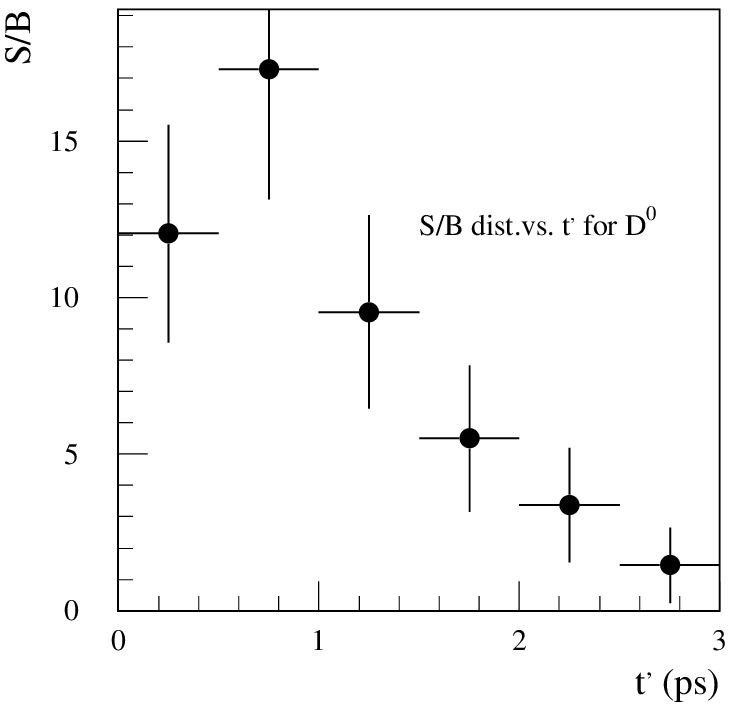,height=2.5in}}
}
\caption{ ${S}\over{B}$ in bins of reduced proper time for $\d0b$
(left) and $D^{0}$ (right).}  
\label{fig:sb_dist}
\end{figure}

It is useful to know how how the signal to background ratio is
distributed in bins of reduced proper time. 
We denote $S_{i}$ as the amount of signal in bin $i$ and $B_{i}$ the
amount of background in the same bin. When we apply sideband
subtraction, each event carries a weight and thus errors of each bin
will depend on signal to background ratio. The smaller this ratio, the
larger the errors. When there is only signal then the error is equal to the
square root of the bin content. Let's see quantitatively what happens.
When the sideband lines are chosen as in Fig.~\ref{fig:dmass_split},
 we get a formula that connects error with signal to background ratio
$err_{i}=\sqrt{S_{i}(1+1.5\times B_{i}/S_{i})}$.
Based on this formula, we extract signal to background ratio per each
bin when we know $err_{i}$ and $S_{i}$.
Fig.~\ref{fig:sb_dist} shows the distribution of signal to background
ratio in bins of reduced proper time for $\d0b$ and  $D^{0}$. Both
show that signal to background ratio decreases in large $t'$. This is
due to the fact that contamination from other charm mesons is more
likely at larger $t'$ values than for smaller values.

\section{Results for the Asymmetry}

\begin{figure}
\includegraphics[height=2.5in]{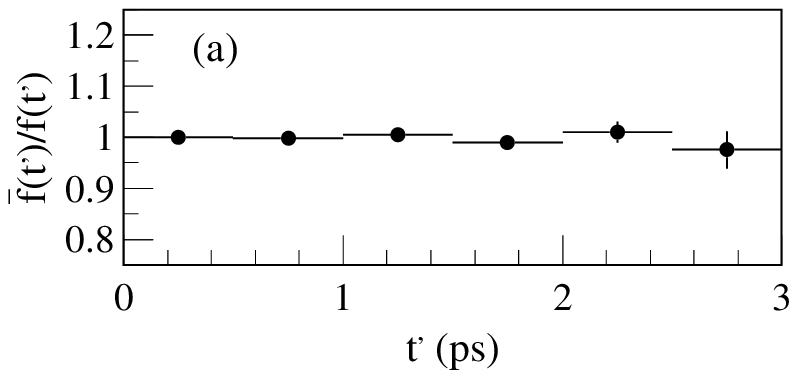}\\
\includegraphics[height=2.5in]{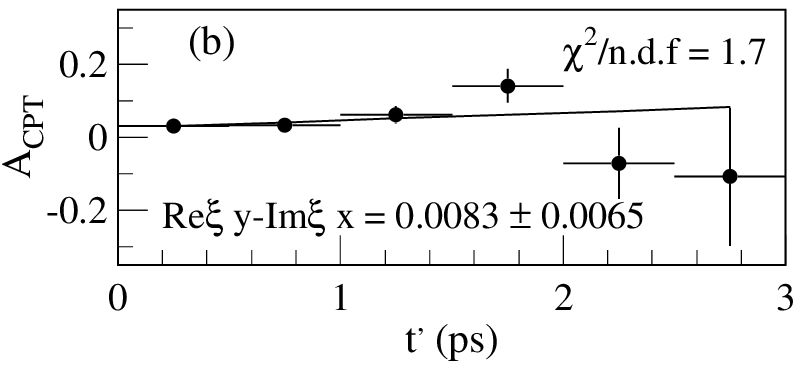}\\
\includegraphics[height=2.5in]{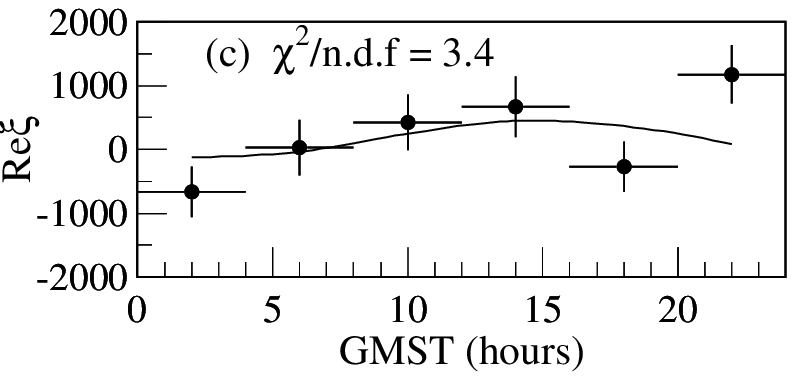}
\caption{(a) The ratio of the corrections, and (b) $A_{\rm CPT}$ as a function of reduced proper time. The data
points represent the $A_{\rm CPT}$ as given in Eq.~\ref{eq:asym_exp} 
and the solid line represent the fit given in functional form by 
Eq.~\ref{eqn:acpt}: (c) ${\rm Re}\,\xi$ as a function of Greenwich Mean
Sidereal Time~(GMST).}  
\label{fig:main_results}
\end{figure}

We plot the difference in right-sign events between $\d0b$ and
$D^0$ in bins of reduced proper time $t'$. The background subtracted
yields of right-sign $D^0$ and $\d0b$ were extracted by properly
weighting the signal region~($-2\sigma, +2\sigma$), the low mass
sideband~($-7\sigma, -3\sigma$) and high mass sideband~($+3\sigma,
+7\sigma$), where $\sigma$ is the width of the fitted signal Gaussian. 
For each data point, these yields were used in forming the ratio:
\begin{equation}
A_{\rm CPT}(t') = {{\overline{Y}(t')-Y(t'){{\overline{f}(t')}\over{f(t')}}}\over{\overline{Y}(t')+Y(t'){{\overline{f}(t')}\over{f(t')}}}},
\label{eq:asym_exp}
\end{equation}
where $\overline{Y}(t')$ and $Y(t')$ are the yields for
 $\d0b$ and $D^{0}$ and $\overline{f}(t')$, and $f(t')$ are
their respective correction functions. 
In the absence of detector acceptance corrections, this is 
equivalent to $A_{CPT}$ as defined in Eqn.~\ref{eq:asym}.  
The functions $\overline{f}(t')$ and $f(t')$
account for geometrical acceptance, detector and reconstruction
efficiencies, and the absorption of parent and daughter
particles in the nuclear matter of the target.
The correction functions are determined using a detailed Monte
Carlo~(MC) simulation using \textsc{PYTHIA}~\cite{pythia61}.
The fragmentation is done using the Bowler modified Lund string model. 
\textsc{PYTHIA} was tuned  using many production 
parameters to match various data production variables such as charm 
momentum and primary multiplicity.
The shapes of the $f(t^{\prime})$ and $\overline{f}(t^{\prime})$
functions are obtained
by dividing the reconstructed MC $t^{\prime}$ distribution by a pure
exponential with the MC generated lifetime. 
Fig.~\ref{fig:fot} shows these corrections. Detector resolution effects cause less than 8\% change in the $t'$ distribution as measured by deviations from a pure exponential decay.
The ratio of the correction functions, shown in Fig.~\ref{fig:main_results}(a),
enters explicitly in Eq.~\ref{eq:asym_exp} and its effects on the
asymmetry are less than 1.3\% compared to when no corrections are applied. 
Due to the QCD production mechanism for photoproduced charm mesons,  
more $\d0b$ than $D^0$ are produced in the FOCUS data sample. This has
been previously investigated in photoproduction by
E687, in which the production asymmetries were
studied in the context of a string fragmentation model~\cite{e687}. The effect
on the $A_{CPT}$ distribution is to add a constant, production-related
offset, which is accounted for in the fit. 

The $A_{\rm CPT}$ data in Fig.~\ref{fig:main_results}(b) are fit to a 
line using the form of Eq.~\ref{eqn:acpt_sim} plus a constant offset.
 The allowed fit parameters are a constant
production asymmetry parameter $\alpha$ and 
 ${\rm Re}\,\xi~y-{\rm Im}\,\xi~x$.  
The value of $\Gamma$ used in the fit is taken as $\Gamma=1.6\times10^{-12}$~GeV~\cite{pdg2004}.
The result of the fit is:
\begin{equation}
 {\rm Re}\,\xi~y-{\rm Im}\,\xi~x=0.0083 \pm 0.0065.
\end{equation}
 We also report $\alpha$ for completeness:
\begin{equation}
 \alpha=0.026 \pm 0.009.
\end{equation}
 If one assumes mixing parameter $x, y$ values of 5\%(current 95\% C.L. upper limits) and Im~$\xi=0$, one obtains for Re~$\xi$, ${\rm Re}\xi=0.17 \pm 0.13$. We infer one
standard deviation errors on Re~$\xi$ of
$0.13$, and 95\% confidence level upper bounds of $0.26$.

\section{Results for Coefficients of Lorentz Violation}

Any CPT and Lorentz violation within the Standard Model can be
described by the SME 
proposed by Kosteleck\'y {\em et al.}~\cite{colladay-kostelecky}.
In quantum field theory,
the CPT-violating parameter $\xi$ must generically depend on lab momentum,
spatial orientation, 
and sidereal time~\cite{kostelecky-prl,kostelecky}.
The SME can be used to show that Lorentz violation in the $D$ system
is controlled by the four vector $\Delta a_{\mu}$. 
The precession of the experiment with the earth relative to
the spatial vector $\vec{\Delta a}$ would
modulate the signal for CPT violation, thus
making it possible to separate the components of $\Delta a_{\mu}$. 
The coefficients for Lorentz violation depend on the flavor of the 
valence quark states and are model independent. 
In the case of FOCUS, where $D^0$ mesons in the lab frame
are highly collimated in the forward direction and under the
assumption that $D^0$ mesons are uncorrelated, the $\xi$ parameter assumes the 
following form~\cite{kostelecky} outlined earlier:
\begin{eqnarray}
\xi(\hat{t}, p)  & = & {\gamma(p)\over{\Delta\lambda}}[ \Delta a_0 + \beta \Delta a_Z {\rm cos} \chi 
+ \beta {\rm sin} \chi (\Delta a_Y {\rm sin} \Omega \hat{t} + \Delta a_X {\rm cos} \Omega \hat{t}) ].
\label{eq:xii}
\end{eqnarray}
$\Omega$ and $\hat{t}$ are the sidereal frequency and time respectively, 
$X, Y, Z$ are non-rotating coordinates with $Z$ aligned along the
Earth's rotation axis, $\Delta\lambda=\Gamma(x-iy)$, and 
$\gamma(p) = \sqrt{1 + p^2_{D^0}/m_{D^0}^2}$. Binning in sidereal time 
$\hat{t}$ is very useful because it 
provides sensitivity to components $\Delta a_X$ and $\Delta a_Y$.    
Since Eq.~15 of Ref.~\cite{kostelecky} translates into ${\rm Re}\,\xi\,y - {\rm Im}\,\xi\,x = 0$,
setting limits on the coefficients of Lorentz violation requires expanding
the asymmetry in Eq.~\ref{eq:asym} to higher (non-vanishing) terms. 
In addition,
the interference term of right-sign decays with DCS decays must also be included since it gives a comparable
contribution. One can follow the procedure given by equations
[16] to [20] of Ref.~\cite{kostelecky} where the basic transition
amplitudes $\langle f|T|\overline{P^0}\rangle $ and $\langle \overline{f}|T|P^0\rangle $ are not
zero but are DCS amplitudes. After Taylor expansion the asymmetry can be
written as:
\begin{eqnarray}
A_{\rm CPT} &=& \frac{{\rm Re}\,\xi (x^2 + y^2) (t/\tau)^2}{2x}
\left[ \frac{xy}{3} (t/\tau) + \sqrt{R_{\rm DCS}}\left(
x\,\cos{\delta}+ y\,\sin{\delta} \right) \right],
\label{eq:new_asym}
\end{eqnarray}
where $R_{\rm DCS}$ is the branching ratio of DCS relative to
right-sign decays and $\delta$ is the strong phase between the DCS
and right-sign amplitudes.
We searched for a sidereal time dependence by dividing our data sample into
four-hour bins in Greenwich Mean Sidereal 
Time (GMST)~\cite{JeanMeeus}, where for each bin we repeated
our fit in $t^\prime$ using the asymmetry given by
Eq.~\ref{eq:new_asym} and extracted ${\rm Re}\,\xi$.
The resulting distribution, shown in Fig.~\ref{fig:main_results}(c),
was fit using
Eq.~\ref{eq:xii} and the results for the expressions involving 
coefficients of Lorentz violation in the SME were:
\begin{equation}
C_{0Z}\equiv N(x,y,\delta)(\Delta a_0 + 0.6\,\Delta a_Z)
=(1.0 \pm 1.1)\times 10^{-16}~{\rm GeV},
\end{equation}
\begin{equation}
 C_{X}\equiv N(x,y,\delta)\Delta a_X=(-1.6 \pm 2.0)\times 10^{-16}~{\rm GeV},
\end{equation}
 and 
\begin{equation}
C_{Y}\equiv N(x,y,\delta)\Delta a_Y=(-1.6 \pm 2.0)\times 10^{-16}~{\rm GeV},
\end{equation}
 where $N(x,y,\delta)=[{xy}/3 + 0.06\,(x\,\cos{\delta}+ y\,\sin{\delta})]$ 
is the normalization factor. 
The angle between the FOCUS spectrometer axis and the Earth's
rotation axis is approximately $\chi = 53^\circ~({\rm cos}{\chi} =
0.6)$. We average over all $D^0$ momentum so $\langle \gamma(p) \rangle
\approx \gamma(\langle p \rangle) = 39$ and $\beta \approx 1$. We also
compare with the previous measurements
for the kaon $r_K$ and $B$ meson $r_B$ by constructing a similar
quantity $r_D$~\cite{kostelecky99},
$r_D=|\Delta\Lambda|/m_{D^0}=\beta^\mu\Delta a_{\mu}/m_{D^0}=|\overline{\xi}||\Delta\lambda|=\gamma(p)|\Delta
a_0 + 0.6\,\Delta a_Z|/m_{D^0}$. The result for $N(x,y,\delta)\,r_D$ is:
\begin{equation}
N(x,y,\delta)\,r_D=(2.3 \pm 2.3)\times 10^{-16}~{\rm GeV}.
\end{equation}
 Although it may
seem natural to report $r_D$, the parameter $r_D$ (and $r_K$, $r_B$) has
a serious defect: in quantum field theory, its value changes 
with the experiment. This is because it is a combination of the 
parameters $\Delta a_{\mu}$ with coefficients controlled by the $D^0$ meson 
energy and direction of motion. 
The sensitivity would have been best if $\chi = 90^\circ$.

\section{Monte Carlo}
\begin{figure}[tbh]
\centerline{
\psfig{figure=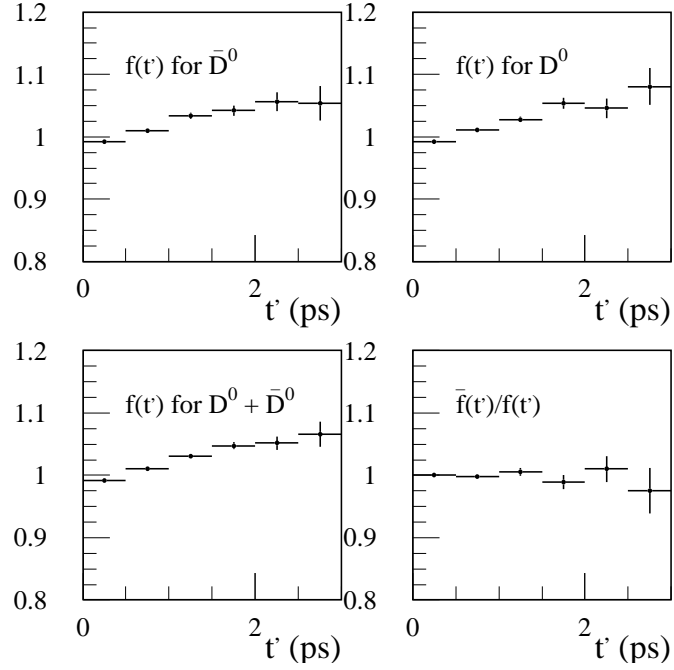,height=3.5in}
}
\caption{$f(t')$ corrections and their ratio for  
$D^0$ and $\d0b$. }  
\label{fig:fot}
\end{figure}

To understand the corrections, we analyzed simulated Monte Carlo
events. Our Monte Carlo simulation includes the \textsc{PYTHIA} Model for
photon gluon fusion and incorporates a complete simulation of all
detectors and trigger systems, with known multiple scattering and
absorption effects. The default Monte Carlo flag which is responsible
for scattering and absorption effects include a simulation of $D^0$
and $\bar{D^0}$ cross sections set at half the cross section for a pion.
The Monte Carlo was prepared such that after trigger
requirement and analysis cuts as in the data, we reconstruct
50 times the data statistics. 
Fig.~\ref{fig:fot} shows the corrections $f(t')$ for $D^0$,
$\bar{D^0}$, $D^0$+$\bar{D^0}$ and the ratio of ${\bar{D^0}}\over{D^0}$. The
deviations are less than 8\% for individual $f(t')$. Furthermore
they cancel out when we take the ratio (of the order of 1.3\%)
${\bar{f(t')}}\over{f(t')}$. The ratio is the only combination used when we form the asymmetry, so our detector corrections on the
asymmetry are very small.    

\begin{figure}[tbh]
\centerline{
\psfig{figure=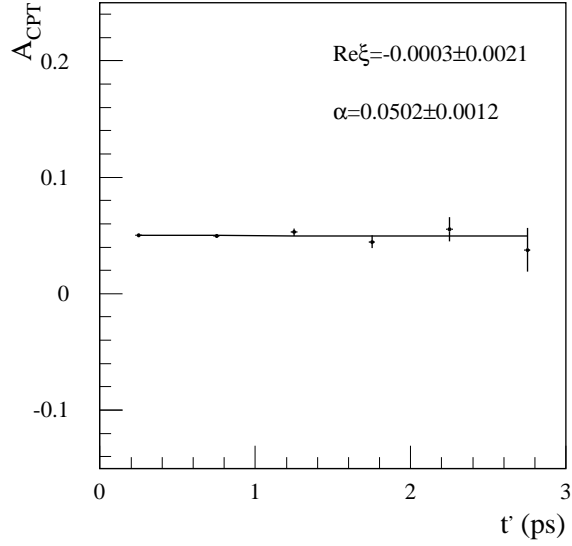,height=3.0in}
}
\caption{Asymmetry in Monte Carlo fitted with Eq.~\ref{eqn:acpt}}  
\label{fig:asym_monte}
\end{figure}

Fig.~\ref{fig:asym_monte} shows the asymmetry in Monte Carlo by
fitting it with the function in Eq.~\ref{eqn:acpt}. There is enough
data statistics in the Monte Carlo sample to demonstrate that there is no slope in the
asymmetry, i.e., only a small value of $Re\xi =-0.0003 \pm 0.0210$, consistent with zero, could result from these corrections. Thus a significant slope in observed real data should be attributed
to CPT. The $D^0$, $\overline{D^0}$ production asymmetry in Monte Carlo is $\alpha=0.052 \pm 0.001$.

\section{Systematic Uncertainties}

Previous analyses have shown that MC absorption corrections are very small~\cite{focusycp}.
The interactions of pions and
kaons with matter have been measured, but no equivalent data exists for
charm particles. To check for any systematic effects associated
with the fact that the charm particle cross section is unmeasured, 
we examined several variations of $D^0$ and $\d0b$ cross sections.
The standard deviation of these
variations returns systematic uncertainties of $ \pm 0.0017$, 
$\pm 0.3\times 10^{-16}$~GeV, $\pm 0.0\times 10^{-16}$~GeV, and
$\pm 0.1\times 10^{-16}$~GeV to our measurements of 
${\rm Re}\,\xi~y-{\rm Im}\,\xi~x$, $C_{0Z}$, $C_X$, and $C_Y$ respectively. 

We also investigated parent ($D^0$,$\d0b$) and daughter $(K,\pi)$
absorption separately. The study showed that the flat corrections in
MC are small, not only because absorption effects are small, but also
because of a cancellation due to two competing effects. The 
$D^0$ has a slightly higher absorption rate than the $\d0b$, and
the net absorption rate of a ($K^-,\pi^+$) from a $D^0$ is slightly
lower than the net absorption rate of a ($K^+,\pi^-$) from the $\d0b$. 

In a manner similar to the S-factor method used by the Particle Data
group PDG~\cite{pdg2004}, we made eight statistically independent
samples of our data to look for systematic effects. We split
the data in four momentum ranges and two years. The split in year was
done to look for effects associated with target
geometry and reconstruction due to the addition of four silicon planes
near the targets in January, 1997~\cite{E831tsil}. 
We found no contribution to our measurements of ${\rm Re}\,\xi~y-{\rm
Im}\,\xi~x$ and $C_{0Z}$. The contributions to
$C_X$ and $C_Y$ were $\pm 1.3\times 10^{-16}$~GeV and $\pm 1.6\times 10^{-16}$~GeV respectively.   
We also varied the bin widths and the position of the sidebands to
assess the validity of the background subtraction method and the
stability of the fits. The standard deviation of these
variations returns systematic uncertainties of $\pm 0.0012$, $\pm 0.3\times
10^{-16}$~GeV, $\pm 0.9\times 10^{-16}$~GeV, and $\pm 0.5\times
10^{-16}$~GeV to our measurements of 
${\rm Re}\,\xi~y-{\rm Im}\,\xi~x$, $C_{0Z}$, $C_X$, and $C_Y$ respectively.
Finally, to uncover any unexpected systematic uncertainty, we varied
our $\ell/\sigma_\ell$ and $W_\pi - W_K$ requirements and
the standard deviation of these variations returns systematic uncertainties of
$\pm 0.0036$, $\pm 1.5\times 10^{-16}$~GeV, $\pm 1.0\times 10^{-16}$~GeV,
and $\pm 1.1\times 10^{-16}$~GeV to our measurements of 
${\rm Re}\,\xi~y-{\rm Im}\,\xi~x$, $C_{0Z}$, $C_X$, and $C_Y$ respectively.
Contributions to the systematic uncertainty are summarized in 
Table~\ref{tb_syst1}. 
Taking contributions to be uncorrelated, 
we obtain a total systematic uncertainty
of $\pm 0.0041$ for ${\rm Re}\,\xi\,y - {\rm Im}\,\xi\,x$, 
$\pm 1.6\times10^{-16}$~GeV for $C_{0Z}$, 
$\pm 1.9\times10^{-16}$~GeV for $C_X$, and
$\pm 2.0\times10^{-16}$~GeV for $C_Y$.

\begin{table}
\caption{\label{tb_syst1}Contributions to the systematic uncertainty.}
\begin{center}
\begin{tabular}{|c|c|c|c|c|}
\hline
Contribut. & ${\rm Re}\,\xi\,y - {\rm Im}\,\xi\,x$ & $C_X$~(GeV) &$C_{0Z}$~(GeV) &  $C_Y$~(GeV)\\
\hline \hline
Absorption  & $ 0.0017$ & $0.0\times 10^{-16}$ & $0.3\times 10^{-16}$ & $0.1\times 10^{-16}$\\
Split sample & $  0.0000$ & $ 1.3\times 10^{-16}$ & $  0.0\times 10^{-16}$ & $ 1.6\times 10^{-16}$\\
Fit variant  & $  0.0012$ & $ 0.9\times 10^{-16}$ & $  0.3\times 10^{-16}$ & $ 0.5\times 10^{-16}$\\
Cut variant  & $  0.0036$ & $ 1.0\times 10^{-16}$ & $  1.5\times 10^{-16}$ & $ 1.1\times 10^{-16}$\\
\hline
{\bf Total}       & $  {\bf 0.0041}$ &  $ {\bf 1.9\times 10^{-16}}$ & $  {\bf 1.6\times 10^{-16}}$ & $  {\bf 2.0\times 10^{-16}}$\\
\hline
\end{tabular}
\end{center}
\end{table}

\vspace{5mm}

To see further details on the assessment of systematic errors, see Appendix~\ref{app:systematics}.

\chapter{Conclusions}
We have performed the first search for CPT and Lorentz violation in
neutral charm meson oscillations. We have measured:
\begin{equation}
{\rm Re}\,\xi\,y - {\rm Im}\,\xi\,x = 0.0083 \pm 0.0065~({\rm stat}) \pm 0.0041~({\rm syst}),
\end{equation}
which leads to a 95\% confidence level limit of:
\begin{equation}
-0.0068 < ({\rm Re}\,\xi\,y - {\rm Im}\,\xi\,x )< 0.0234.
\end{equation}
As a specific example, assuming $x=1\%$ and ${\rm Im}\,\xi = 0$ and $y = 1\%$~(current central values for mixing), one finds: 
\begin{equation}
{\rm Re}\,\xi = 0.83 \pm 0.65~({\rm stat}) \pm 0.41~({\rm syst}),
\end{equation}
 with a 95\%
confidence level limit of
\begin{equation}
 -0.68 < {\rm Re}\,\xi < 2.34.
\end{equation} Within the SME, we set three independent first limits on the
expressions involving coefficients of Lorentz violation of:
\begin{equation}
 (-2.8<N(x,y,\delta)(\Delta a_0 + 0.6\,\Delta
a_Z)<4.8)\times 10^{-16}~{\rm GeV},
\end{equation}
\begin{equation}
 (-7.0<N(x,y,\delta)\Delta
a_X<3.8)\times 10^{-16}~{\rm GeV},
\end{equation}
 and
\begin{equation}
 (-7.0<N(x,y,\delta)\Delta a_Y<3.8)\times
10^{-16}~{\rm GeV}.
\end{equation}
 As a specific example, assuming $x=1\%$, $y = 1\%$~(current central values for mixing) and
$\delta=15^\circ$~(current theoretical prediction)
one finds the 95\% C.L. limits on the coefficients of Lorentz violation of:
\begin{equation}
 (-3.7<\Delta a_0 + 0.6\,\Delta
a_Z<6.5)\times 10^{-13}~{\rm GeV},
\end{equation}
\begin{equation}
 (-9.4<\Delta a_X<5.0)\times 10^{-13}~{\rm GeV},
\end{equation}
and 
\begin{equation}
(-9.3<\Delta a_Y<5.1)\times 10^{-13}~{\rm GeV}.
\end{equation}
The measured values are consistent with no significant CPT or Lorentz invariance violation.

\null\vfill
\begin{center}
\Huge Part II
\end{center}
\null\vfill
 
\chapter{Introduction to \lb}  
\label{ch:introlb}

The UA1 experiment at CERN announced the discovery of \lb\ baryon in 1991~\cite{UA1lb}. They measured the production fraction times branching ratio to be:

$F(\Lambda_b)\times B(\Lambda_b\rightarrow J/\psi \Lambda) = (1.8 \pm 1.0) \times 10^{-3}$.

Later on, in 1996 both ALEPH~\cite{alephlb} and DELPHI~\cite{delphilb} measured the \lb\ mass in the decay $\Lambda_{c}\pi$. Each experiment found only 4 candidates.  \lb\ was unambiguously observed by CDF 110 pb$^{-1}$ Run I data~\cite{cdflb} with a mass of $5621\pm 4~({\rm stat}) \pm 3~({\rm syst})$~MeV, and a production fraction times branching ratio of
 \begin{eqnarray}
F(\Lambda_b)\times B(\Lambda_b J/\psi \Lambda) = (3.7 \pm 1.7~({\rm stat}) \pm 0.7~({\rm syst})) \times 10^{-4}.
\label{eq:cdf}
\end{eqnarray}

Since the signal consisted of only 20 events, only a mass measurement was made; lifetime measurement in this mode required more data.
In the second part of this thesis, we report a preliminary measurement of the lifetime of \lb\ in the fully reconstructed decay mode \lbdec.
Measuring the lifetime in this decay mode is particularly interesting, since no other measurement has been published in a fully reconstructed decay mode.

\begin{figure}[h!tb]
\begin{center}
\includegraphics[height=12.0cm]{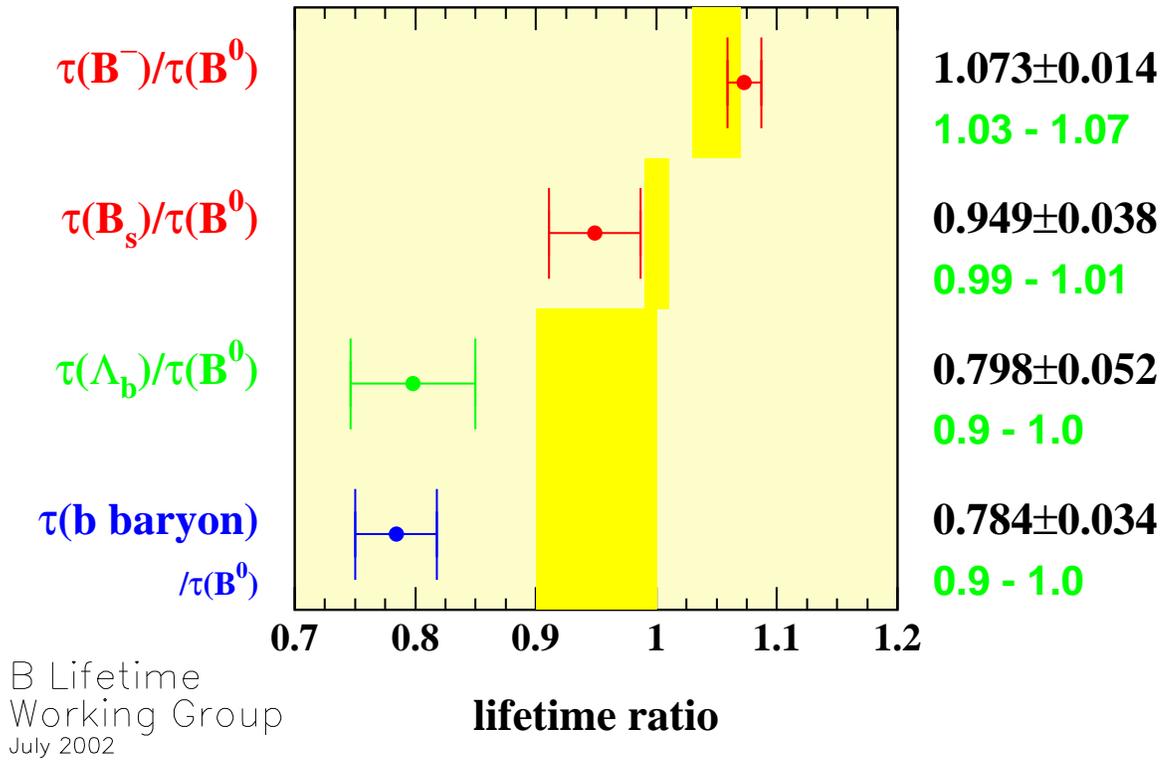}
\caption{World averages of $b$ hadron lifetimes compared to older
theoretical predictions (yellow or dark grey bands).}
\label{fig:theory}
\end{center}
\end{figure}

\begin{figure}[h!tb]
\begin{center}
\includegraphics[height=8.0cm]{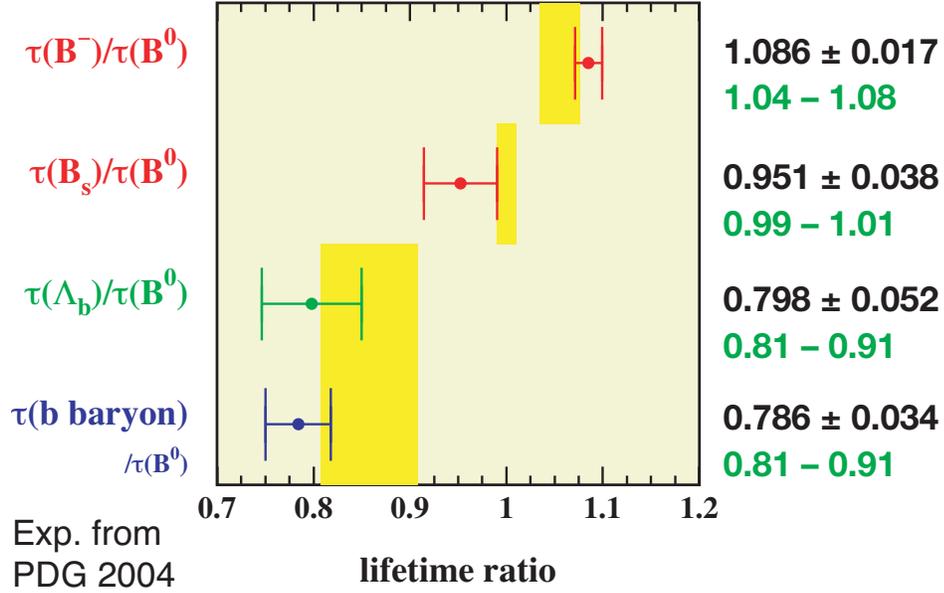}
\caption{World averages of $b$ hadron lifetimes compared to more recent
theoretical predictions (yellow or dark grey bands).}
\label{fig:theory2004}
\end{center}
\end{figure}

More importantly, there was a long standing discrepancy in the measured value in $\Lambda^0_b$ and $b$-baryon lifetimes compared to
theoretical predictions~\cite{pdg2004}. There was concern about this disagreement. Figure~\ref{fig:theory} shows the lifetime ratios where the yellow bands are theoretical predictions.
The experimental world average for $\tau(\Lambda_b)/\tau(B^0)$ is $0.797 \pm 0.052$, while theory predicted the value to be between 0.9 and 1. 
However the most recent calculations, only in the past year, show less of a discrepancy~\cite{pdg2004}. Figure~\ref{fig:theory2004} shows the recent results. 
 To understand any discrepancy, we need to consider clean decays of
$\Lambda^0_b$ baryon such as the exclusive \lbdec\ decay.
 

The \lb\ heavy baryon is an outstanding system to understand quark dynamics and
 Heavy Quark Effective Theory~(HQET) and the operator product expansion (OPE)~\cite{1mbcorrection}. The quark content of \lb\ is ({\it udb})
where the {\it b} quark is separated from the {\it (ud)} quarks that form their
own spin-0 system.
The \lbdec\ decay receives only small nonspectator contributions and
hence its theoretical calculation is relatively straight forward.

The decay mode \lbdec\ is said to be fully reconstructed, because all of the final state particles leave tracks in the detector. Thus the \lb\ full momentum and invariant mass may be determined. This is in contrast to semileptonic decay modes such as $\Lambda_b\rightarrow \Lambda_c \ell \nu$ that contain neutrinos. Neutrinos are neutral and interact very weakly; they leave no signal in the D\O~detector. The semileptonic decays of \lb\ have larger branching ratio and therefore are more abundant in our detector; however the sample is not as pure as in the case of fully reconstructed decay like \lbdec. 

Since the \lb\ is more massive than the $B^+$ and $B^0$ mesons, it is currently produced only at the Tevatron. D\O\ and CDF are the only currently operating detectors that can study it (in comparison to the $B$ factories, Belle and BaBar, operating at the $\Upsilon(1S)$). 

\begin{figure}
\begin{center}
\epsfxsize=4.5 in
\epsfbox{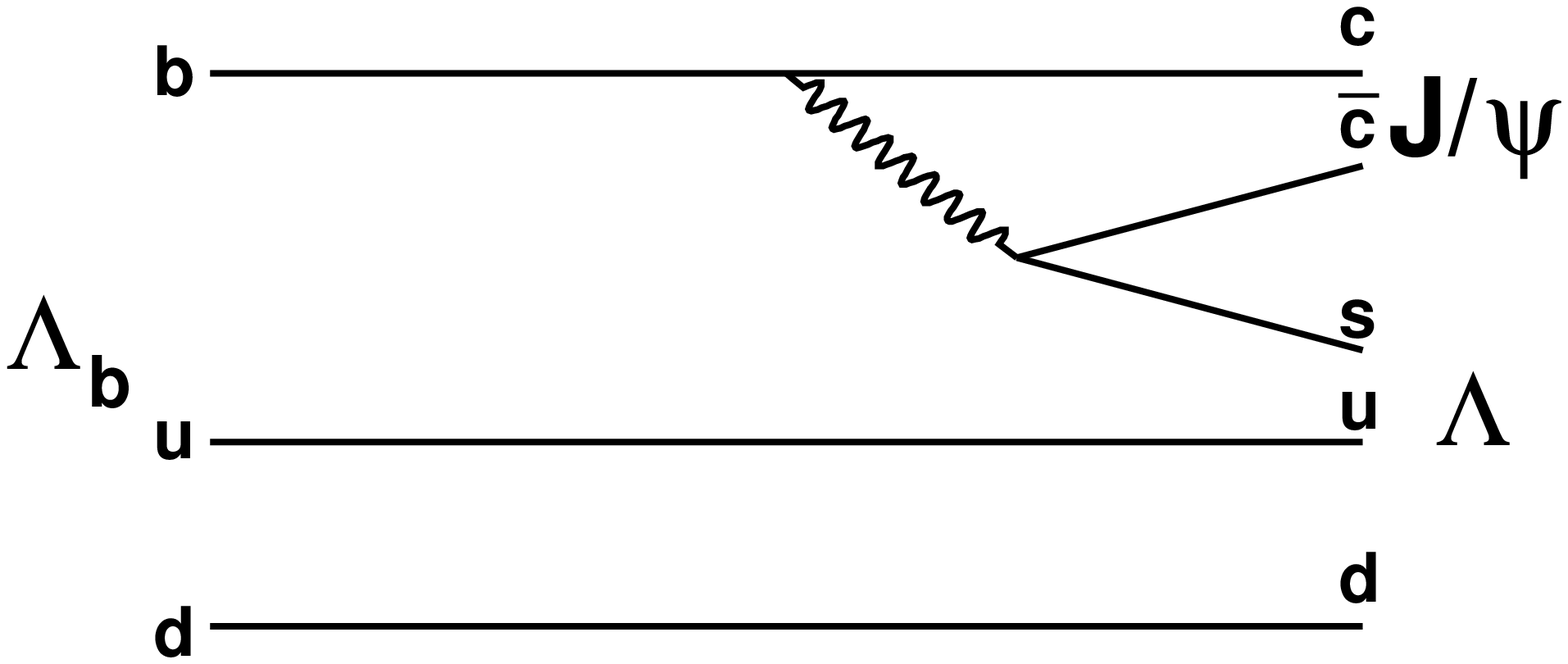}
\end{center}
\caption{Feynman diagram for the decay \lbdec.}
\label{fig:feynman_lb}
\end{figure}

The decay \lbdec\ is a color-suppressed decay that proceeds through an internal $W$ decay. The Feynman diagram is shown in Figure~\ref{fig:feynman_lb}. The decay is color-suppressed because the colors of the quarks from the virtual $W$ must match the colors of the $c$ quark and the remaining diquark system.

A brief description of the analysis follows. We reconstruct \lb\ candidates as described in Chapter~\ref{ch:recolb} and first establish the \lb\ mass signal. In addition to reconstructing \lbdec\, we also reconstruct \bddecks.~This serves as a control sample, since it has similar topology to that of \lb\ but the $B^0$ is more abundant and has a well-known lifetime.
For each candidate, we obtain the value of proper decay length. We then perform  an unbinned maximum likelihood fit to the distribution of proper decay lengths in the data, to extract the value of $c\tau(B^0)$ and $c\tau(\Lambda_b)$. 
A first result was presented at Lepton-Photon conference in Summer 2003. The signal and lifetime presented in this thesis are found with additional data until January 2004. Even more data, in an analysis continued by members of CINVESTAV group, resulted in a submitted result~\cite{eduardo}.

\chapter{Theory Predictions}
\label{ch:theorylb}

\section{Spectator Model}

In the Spectator Model of hadrons, a heavy quark inside a hadron is bound to the lighter ``spectator'' quarks. When the interactions of the heavy quark with the lighter quarks are small, we can estimate the weak decay of the heavy quark separately. In this approximation, all the hadrons containing a given heavy quark have the same lifetime. This approximation is more valid when the quark at hand is heavier. In the simple spectator model, the decay width of a $b$ hadron is:
\begin{eqnarray}
\Gamma(b)=\frac{9|V_{cb}|^{2}G^{2}_{F}m^{5}_{b}}{{192\pi^3}}.
\label{eq:naivedecay}
\end{eqnarray}
This is the formula for the muon decay width, with the addition of CKM matrix element $V_{cb}$ for quark coupling ($b$ to $c$). We assume $b$ decays mostly to $c$. Plugging in $m_b\approx4.2$~GeV~\cite{pdg2004} and $|V_{cb}|\approx0.045$~\cite{Vcd} this gives $\tau=1.2$~ps. Despite the simplicity of it, the experiments show that this model is not sufficient. Experiments show the hierarchy of $b$-quark lifetimes to be 
$\tau(\Lambda_b)<\tau(B^0)\approx\tau(B_s)<\tau(B^{+})$

\section{Present Findings}
In the previous section we stated that the ``naive'' spectator model is insufficient to explain the hierarchy of lifetimes of heavy hadrons. In a hadron one cannot neglect the strong interactions between the heavy quark and the lighter quarks, therefore a theory that includes these interactions is needed to explain the observed hierarchy. In the Heavy Quark Expansion~(HQE) one uses the dimensional operators that contain $1/m_b$ terms~\cite{fabrizio2}.
 It has been shown that this theory explains well the hierarchy in the $B$-meson lifetimes and also it predicts the ratios of the lifetimes of $B$-mesons. Predictions agree with experimental measurements. The agreement between theory and experiment gives us some confidence that quark-hadron duality, which states that smeared partonic amplitudes can be replaced by the hadronic ones, is expected to hold in inclusive decays of heavy flavors.
Figure~\ref{fig:theory2004} shows a summary of the current world average of $b$ hadron lifetimes compared to theory. According to Fig.~\ref{fig:theory2004} for the $B$ mesons, we have these experimental values and theoretical predictions: 
\begin{eqnarray}
\frac{\tau(B_u)}{\tau(B_d)}|_{ex}=1.086 \pm 0.017, ~\frac{\tau(B_u)}{\tau(B_d)}|_{th}=1.06 \pm 0.02, 
\label{eq:bratio1}
\end{eqnarray}
\begin{eqnarray}
\frac{\tau(B_s)}{\tau(B_d)}|_{ex}=0.951 \pm 0.038, ~\frac{\tau(B_s)}{\tau(B_d)}|_{th}=1.00 \pm 0.01, 
\label{eq:bratio2}
\end{eqnarray}
which show agreement of theoretical predictions and experimental measurements.

For a long time, the low measured value of the ratio $\tau(\Lambda_b)/\tau(B_d)$ has been a challenge for the theory. 
 According to Fig.~\ref{fig:theory2004} for the ratio $\tau(\Lambda_b)/\tau(B_d)$ we have:
\begin{eqnarray}
\frac{\tau(\Lambda_b)}{\tau(B_d)}|_{ex}=0.798 \pm 0.052, 
\label{eq:lbratio2}
\end{eqnarray}
\begin{eqnarray}
\frac{\tau(\Lambda_b)}{\tau(B_d)}|_{th}=0.86 \pm 0.05. 
\label{eq:lbratio2theo}
\end{eqnarray}
 This ratio is inconsistent with the initial theoretical predictions, as shown in Fig.~\ref{fig:theory}. However, recent next-to-leading order~(NLO) calculations of perturbative QCD~\cite{qcdprediction} and $1/m_b$ corrections~\cite{1mbcorrection, fabrizio2} to the spectator effects reduced the discrepancy, yielding result as shown in Fig.~\ref{fig:theory2004} which is in better agreement with experimental measurement.

\section{Formalism}

Inclusive decay rates can be calculated in the HQE. We use the optical theorem to relate the decay width to the imaginary part of the matrix element of the forward scattering amplitude:

\begin{eqnarray}
\Gamma(H_b\rightarrow X) = 1/{2m_b}\langle H_b|{\it T}|H_b\rangle , 
  {\it T}={\it Im}\,i\int{d^{4}xT\{H_{eff}(x)H_{eff}(0)\}}.
\label{eq:decaywidthgeneral}
\end{eqnarray}
Here $H_{eff}$ represent an effective $\Delta B=1$ Hamiltonian at the scale $\mu=m_b$, 
\begin{eqnarray}
H_{eff}=\frac{4G_F}{\sqrt(2)}V_{cb}\sum [c_1 Q^{u'd'}_1 + c_2 Q^{u'd'}_2] + h.c.,
\label{eq:hamiltonian}
\end{eqnarray}
where $c_i$ are the Wilson coefficients, $d'$ and $u'$ are quark flavor eigenstates, and $Q_1$ and $Q_2$ are the four-quark operators. 
The energy release is large in the heavy quark limit, therefore an OPE can be constructed for Eq.~\ref{eq:decaywidthgeneral}, which results in series of local operators of increasing dimension suppressed by powers of $1/m_b$ as shown~\cite{bigi}:

\begin{eqnarray}
\Gamma(H_b\rightarrow f) &=& \frac{G^2_F m^5_b}{192\pi^3}|KM|^2\bigg[c_3(f)\frac{\langle H_b|\bar{b}b|H_b\rangle }{2M_{H_b}}  + \frac{c_5(f)}{m^2_b} \frac{\langle H_b|\bar{b}i\sigma_{\mu\nu}G_{\mu\nu} b|H_b\rangle }{2M_{H_b}} +  \nonumber \\      
&&+\sum_{i} \frac{c^{(i)}_6(f)}{m^3_b} \frac{\langle H_b|(\bar{b}\Gamma_iq)(\bar{q}\Gamma_ib)|H_b\rangle }{2M_{H_b}} + O(1/m^4_b)\bigg], 
\label{eq:operatorexpansion}
\end{eqnarray}
where the dimensionless coefficients $c_i(f)$ depend on the parton level characteristics of $f$~(such as the ratios of the final state masses to $m_b$). KM denotes the appropriate combination of weak mixing angles. $G_{\mu\nu}$ is the gluonic field strength tensor. The summation in the last term is over the four-fermion operators with different light flavors $q$.

As we showed in the last section, at leading order in the heavy quark expansion, all heavy hadrons have the same lifetime. The situation changes at higher orders. At order $1/m^2_b$ the difference between meson and baryon lifetimes has to do with their structure. The ratio of lifetimes of \lb\ and $B_d$ is
\begin{eqnarray}
\frac{\tau(\Lambda_b)}{\tau(B_d)} =1 + \frac{1}{2m^2_b}[\mu^2_\pi(\Lambda_b)-\mu^2_\pi(B_d)] + \frac{C_G}{m^2_b}[\mu^2_G(\Lambda_b)-\mu^2_G(B_d)] + O(1/m^3_b),
\label{eq:ordersquare}
\end{eqnarray}
where $C_G\approx 1.2$~\cite{bigi,neubert}. $\mu^2_\pi$ and $\mu^2_G$ represent kinetic energy and chromomagnetic interaction corrections~\cite{bigi}. At this order in HQE, the difference is mainly driven by the fact that light quarks in \lb\ appear in a $J^P = 0^+$ quantum state, reducing any correlations of spins between the heavy-quark and the light quark-gluon cloud. This results in 
$\mu^2_G(\Lambda_b)=0$. Matrix elements of kinetic energy operators cancel each other to a large degree, that results in a difference of at most 1--2\%, which is not sufficient to explain the observed pattern of lifetimes. 

Dimension six operators, that enter at the $1/m^3_b$ level, are the main contributors. An important subgroup of these operators involves four-quark operators, whose contribution is also enhanced due to the phase-space factor $16\pi^2$. These effects are usually called Weak Scattering~(WS), Weak Annihilation~(WA), and Pauli Interference~(PI). They introduce major differences in the lifetimes of all heavy mesons and baryons~\cite{bigi,neubert,shifman,guberina}. Their contribution to the lifetime ratios are directed by the matrix elements of $\Delta B =0$ four-fermion operators~\cite{fabrizio2}: 
\begin{eqnarray}
{\it T}_{spec} ={\it T}^u_{spec} + {\it T}^{d'}_{spec} + {\it T}^{s'}_{spec},
\label{eq:dbzero}
\end{eqnarray}
where ${\it T}_i$ terms contributing to Eq.~\ref{eq:decaywidthgeneral} are expressed in terms of the four-quark operators $O^q_i$. They are defined as,
$O^q_1 =\overline{b_i}\gamma^\mu(1-\gamma_5)b_i\overline{q_j}\gamma_\mu(1-\gamma_5)q_j$, $O^q_2 =\overline{b_i}\gamma^\mu\gamma_5 b_i\overline{q_j}\gamma_\mu(1-\gamma_5)q_j$. The recent progress has been in understanding lifetimes by concentrating on computing the next-to-leading order~(NLO) QCD corrections to Wilson coefficients of these operators. Also a great deal of progress has been made in calculating matrix elements of these operators in quark models and on the lattice. At NLO one can parametrize the meson-baryon lifetime ratio as:

\begin{eqnarray}
\frac{\tau(\Lambda_b)}{\tau(B_d)} \approx 0.98 -(d_1 + d_2 \bar{B})r -(d_3\epsilon_1 + d_4\epsilon_2) - (d_5B_1 + d_6 B_2) + \delta_{1/m},
\label{eq:to3rd}
\end{eqnarray}
where the scale dependent parameters $d_i$ are defined in Ref.~\cite{neubert}, and $r=|\psi^{\Lambda_b}_{bq}(0)|^2/|\psi^{B_q}_{bq}(0)|^2$ is the ratio of the wave functions at the origin of the \lb\ and $B_q$ mesons.
The $\delta_{1/m}$ term represents contributions of order $1/m^4_b$ and higher.
In Ref.~\cite{1mbcorrection,fabrizio2}, higher-order corrections have been calculated and they shift the ratio of the lifetimes by $-4.5$\% in addition to the $\approx 10\%$ given by up to order $1/m^3_b$. In particular, when one goes to the calculation of the subleading $1/m_b$~($1/m^5_b$) as shown in Ref.~\cite{fabrizio2}, the inclusion of these corrections brings into agreement the theoretical predictions and experimental measurements of the ratio of lifetimes of \lb\ baryon and $B_d$ meson. Even though these calculations bring into better agreement theory with experimental measurements, one cannot conclude that the agreement is for certain because of the size of theoretical and experimental uncertainties. The discrepancy of $\tau(\Lambda_b)/\tau(B_d)$ ratio could show up again if future measurements at the Tevatron and later at LHC would find the mean value to stay the same, with shrinking errors.
In this thesis, we add another clean contribution to the world-average measurement of the lifetime ratio.

\chapter{The \D0 ~Detector for Run~II}

\begin{figure}
\begin{center}
\epsfxsize=6 in
\epsfbox{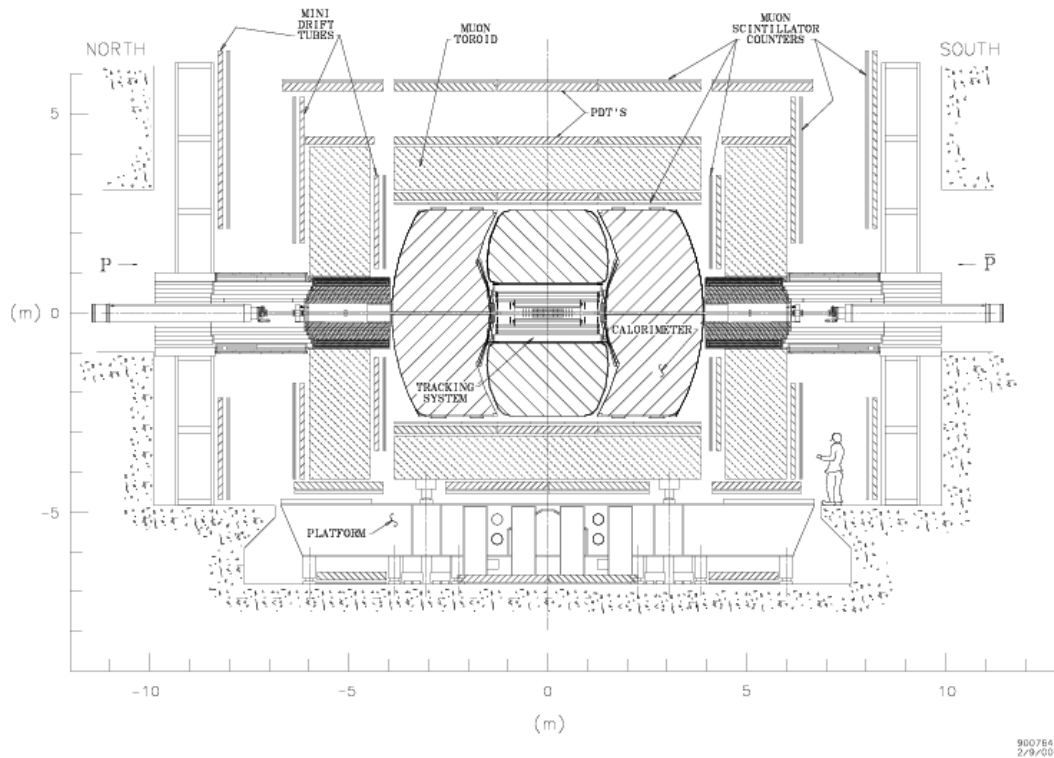}
\end{center}
\caption{Elevation view of the upgraded \D0~detector.}
\label{fig-d0det}
\end{figure}

This chapter describes the \D0 detector and the Tevatron upgrade at Run~II.
It is based on Diehl's review~\cite{physics:cdfd0}.
Fig.~\ref{fig-d0det} shows an elevation view of the \D0 detector.

First in section~\ref{detector-coordinate} we define the coordinate system 
used in \D0 detector.  Section~\ref{detector-Tevatron} describes the upgrades 
to the world's highest energy accelerator, the Tevatron, and its current status.
Since the tracking system and muon spectrometer are crucial to $B$ physics, emphasis will be placed on the muon detector and central tracking system in this
chapter.
Section~\ref{detector-silicon} describes the new silicon vertex system.
Section~\ref{detector-cft}  describes the upgrades to the central fiber tracking system.
Section~\ref{detector-cals}
describes upgrades to the calorimeter systems. Section~\ref{detector-muons}
describes the upgrades to the muon detectors. Section~\ref{detector-trig}
describes the trigger systems.  Section~\ref{detector-daq} describes the data acquisition system.

\boldmath
\section{The \D0 ~Coordinate System}
\label{detector-coordinate}
\unboldmath

At \D0, the primary coordinate system has the $z$-axis 
along the proton beam direction, and the positive $y$-axis pointing up,
so that $(x,y,z)$ make a right-handed Cartesian system.

Sometimes cylindrical ($r,\phi, z$) coordinates are used, 
as are spherical ($r,\phi, \theta$) coordinates, 
$r$ and $\theta$, giving respectively
the perpendicular distance and the angle from the $z$-axis.
The azimuthal coordinate, $\phi$, gives the angle from the $x$-axis
in the $x-y$ projection.

The angular variables are defined so that
$\theta=0$ is along the positive $z$-axis direction, and $\phi=\pi/2$ is parallel to 
the positive $y$-axis. 

The rapidity $y$ is defined as
\begin{equation}
y = \frac{1}{2} \ln {\frac{E + p_z}{E - p_z}},
\label{eq:rapidity}
\end{equation}

where $E$ is the energy, and $p_z$ is the particle momentum on the $z$-direction.

The variable of pseudorapidity is often more convenient, which is defined as
\begin{equation}
\eta = - \ln [\tan{\frac{\theta}{2}}].
\label{eq:pseudorapidity}
\end{equation}

In the limit that $m \ll E$ (where $m$ is the invariant mass),
the pseudorapidity approximates the true rapidity. 

``Transverse" momentum ($p_T$) is also commonly used.
$p_T$ is the momentum vector projected onto a plane perpendicular to the 
beam axis:
\begin{equation}
p_T = p \sin{\theta}.
\label{eq:pt}
\end{equation}

This is particularly useful due to the fact that 
in a high energy collision, many of the products of the collision escape down the
beam pipe, so the momenta along the beam of the colliding partons are unknown.
However, their transverse momenta are very small compared to their momenta
along the beam, so momentum can be considered to be conserved in the
transverse plane.


\boldmath
\section{The Run~II Accelerator Upgrade}
\label{detector-Tevatron}
\unboldmath

In Run~II, beginning in March 2001, the Tevatron 
collides protons with antiprotons at a center-of-mass
energy of 1.96 TeV, which is a slightly higher energy than the 1.8 TeV available
in Run~I.  In the first 3 years, four times the Run~I integrated luminosity has
been collected (by the end of August 2004).
We show in Fig.~\ref{fig_tevatron} a schematic of the Fermilab Tevatron Collider.

In Run~II, a large advance has been made with the
construction of the Main Injector, a 150 GeV synchrotron, built in a
separate tunnel from the Tevatron. The Main Injector has replaced the function
of the Main Ring for antiproton production.  It produces $2\times 10^{11}$
antiprotons per hour~\cite{holmes}, four times the rate of the old
Main Ring.  A new permanent magnet ``Recycler Ring"~\cite{recycler} allows
recovery and reuse of uncollided antiprotons when the Main Injector has
produced enough to merit injecting a new store into the Tevatron.
Instantaneous luminosities in the range $5-20 \times 10^{31} {\rm cm}
^{-2} {\rm s}^{-1}$
were available early in Run~II.  Because the number of protons per
bunch is near the limit of the Tevatron, the number of bunches is
increased from Run~I's 6 bunches of protons and 6 of antiprotons to 36
 of each species. With an instantaneous luminosity
of $20 \times 10^{31} {\rm cm}^{-2} {\rm s}^{-1}$, an average of 5.2 $p\bar{p}$ collisions
occurs each bunch crossing. Increasing the number of bunches 
decreases the average number of collisions per bunch crossing.
By decreasing the bunch width as the store is depleted by collisions, 
the luminosity can be maintained at an optimal level.

Such changes have profound implications for the detectors.  Where in the
past, there were 3.5 microseconds between each beam crossing, in Run~II this
is 396 ns for 36 bunches.
This required all the front-end electronics from the detector 
to be replaced with electronics capable of
faster response.  Aside from the issues of shorter times between collisions,
the detector improves its ability to identify leptons,
energetic photons and charged particles, and particles emerging from
a secondary vertex (daughters of long-lived parents).  Increasing the
collision rate by an order of magnitude also requires substantial
upgrades of the trigger and data acquisition systems to handle the
increased data flow.


\boldmath
\section{Silicon Microvertex Tracker (SMT)}
\label{detector-silicon}
\unboldmath

The purposes of the silicon detectors are to identify tracks of charged
particles that decay before they reach the outer tracker, to reconstruct
the decay vertices of $b$ hadrons, and to extend or improve the track-finding
and momentum resolution. Because the proton and antiproton beams are very
thin $(\sigma_{\rm transverse}<40~\mu {\rm m})$ and because their transverse
positions can be
maintained very close to the center of the beampipe, a typical
$b$-hadron decay vertex can be identified by the impact parameters of the
tracks of $\sim$ $300~\mu {\rm m}$. This requires fine-pitched detectors oriented
perpendicularly to the trajectories of the charged particles and mounted
as close as practical to the decay vertices. The fine pitch and small radius
necessary to achieve sufficient impact parameter resolution also
improves the momentum resolution of the tracker systems.

The accelerator parameters contribute to the design decisions.
The interactions are spread out over a length  $\sigma_z=25$ cm,
setting the length scale and motivating the detectors to have
extended geometries.
The expected integrated luminosity implies a radiation dose of $\leq 1$ Mrad
during the life of the silicon detector. This, in turn, forces the sensors and
readout chips to be radiation-hard and operated at cold temperatures, which
mitigates some of the adverse effects of radiation.

The basic detector design is the ``barrel geometry'' comprised of layers of 
silicon detectors arranged in plates perpendicular  to the beampipe.  
In addition to measuring separated vertices and improving the momentum resolution, 
D\O \ chose to use silicon detectors
to extend the tracking coverage to high pseudorapidity with disk-shaped
assemblies comprised of wedge-shaped silicon wafers arranged in plates
oriented perpendicular to the beamline. 

\begin{figure}
\begin{center}
\epsfxsize=4.5 in
\epsfbox{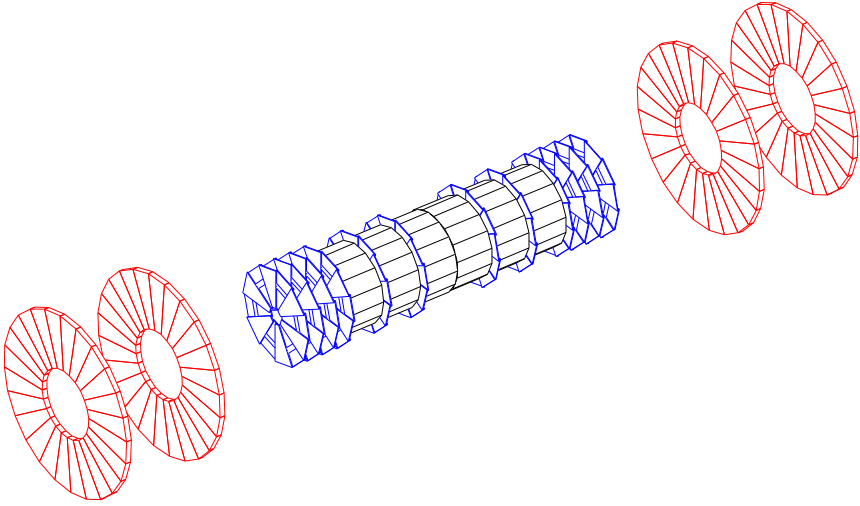}
\end{center}
\caption{D\O \ silicon detector. The figure shows the configuration of the 6 barrel
modules, the 12 ``F-disks", and the 4 ``H-disks".}
\label{fig-d0-silicon}
\end{figure}

The D\O \ silicon microvertex detector~\cite{lipton}
consists of six barrel modules, twelve small
disks (``F-disks"), and four large disks (``H-disks").
The mixed barrel/disk geometry provides silicon sensors arrayed at normal
incidence, as is optimal for good tracking resolution, to charged particles
with $|\eta |< 3$.
Figure~\ref{fig-d0-silicon} shows the D\O \ silicon barrel modules with
barrel detectors parallel to the beamline and disks perpendicular to the
beamline.

The six barrel modules are constructed with 4 layers of ladder assemblies
with considerable overlap. Each barrel module is 12.4 cm in length and the
total length of the barrel is 76.2 cm.  The 4-layer coverage corresponds to
the region $\eta=1.5$ for interactions at $z=0$.  Barrel layers one and three
are constructed from double-sided silicon sensors with axial $(r-z)$ and
$90^\circ$ stereo layers, except for the modules on each end of the barrel,
modules one and six,
which have single-sided axial strips.  These are single-sided because
stereo tracking is dominated by the information from the F-disks.
Barrel layers two and four are constructed from double-sided detectors
with axial and $2^\circ$ stereo layers in all six barrel modules.
All of the detectors are AC-coupled to the readout electronics.

The 12 ``F-disks" are comprised of 12 trapezoidal wedges arranged into a plate
with a hole for the beampipe. The active area inner radius is at 2.5 cm
from the center of the beampipe and the outer is at 9.8 cm.
The detectors are AC-coupled, double-sided, with strips angled
at $\pm 15^\circ$ with respect to the vertical. The pitch of the p-side and
n-side detectors is $50~\mu {\rm m}$ and  $62.5~\mu {\rm m}$, respectively,
and the silicon wafers are 300~$\mu {\rm m}$ thick.  Naturally, the strips are
of different length depending on their locations
on the wedge. The six innermost F-disks attached to the outer sides of the
six barrel modules. The three additional F-disks are attached to each
outer side of the barrel assembly, with the effect of extending the acceptance
of the silicon system to higher pseudorapidity, especially for interactions
which occur at larger $z$.

The four large ``H-disks" are located at $z=\pm 94$ cm and $z=\pm 126$ cm.
The inner radius of the active area is at 9.6 cm from the center of the
beampipe and the outer is at 23.6 cm. The detectors are AC-coupled,
single-sided, with 40 $\mu {\rm m}$ pitch strips
(pairwise readout makes the effective pitch 80 $\mu {\rm m}$)
angled at $\pm 7.5^\circ$
from the vertical. Each plane has wedges glued together back-to-back to
provide a $15^\circ$ stereo angle. These forward disks are necessary to
provide track stubs for forward particles which would otherwise exit the
region of full solenoidal magnetic field without hitting the outer tracker.
They cover a pseudorapidity range of approximately $2\le |\eta| \le 3$.

All of the three detector types are able to withstand a radiation dose
greater than 1 Mrad.  The dose depends on the detector location and integrated
luminosity. The innermost layer is expected to receive 0.5 Mrad/fb$^{-1}$.
The effect of the radiation damage is to increase the leakage current,
increase the bias voltage necessary for full depletion, and decrease the
signal-to-noise ratio.  The effects are temperature dependent and can be
reduced by operated the detectors at low temperatures. The detectors
are operated at temperatures between $5-10^\circ$ C using a cooling
mixture of deionized ethylene glycol and water.

The D\O \ silicon detectors are read out using the SVX~II
chip~\cite{FNALTM1892}. SVX~II is a 128 channel, radiation hard CMOS chip
mounted directly on the ``High Density Interconnects" (HDI's), kapton-based
flexible circuit, wire-bonded to the sensors at the ends of the ladders
and wedges. Each of the 128 channels in an SVX~II chip features a
preamplifier, a 32 stage analog pipeline with 4 $\mu {\rm s}$ delay,
an 8-bit Wilkinson type ADC, and a latch-based sparsified readout.
The chip is programmable for any interaction time from 132 to 396 ns.
The pipeline depth, ADC ramp rate, preamplifier bandwidth, and thresholds
are downloadable to each SVX~II chip.

Connections to the outside world continue from the HDI's to 8-ft
long ``low-mass" cables that join the ends of the HDI's to unpowered
``transition cards" mounted on the ends of the central calorimeter
cryostat. Ultimately, these signals are gathered by a sequencer board
connected by optical link to VME readout buffer electronics in the movable
counting house. The D\O \ silicon detector has 792,576 channels.

\boldmath
\section{Outer Tracker: Central Scintillating Fiber Tracker (CFT)}
\label{detector-cft}
\unboldmath

The outer tracker~\cite{physics:cdfd0} is the charged particle tracker at largest radius
within the calorimeter.  Outer trackers perform two functions.  The
first is measuring the momentum and charge of particles
produced in the collision, and the second is to provide
pattern recognition assistance for the silicon detectors.

A particle with non-zero charge $q$ and momentum $p$ in a
solenoidal magnetic field along the $z$-direction
of strength $B$ will travel in a helix with radius $r$
given by

\begin{equation}
 r = \frac{p_T}{qB},
\end{equation}

where $p_T = \sqrt{p_x^2 + p_y^2}$.  Therefore, by measuring
the track's curvature in the $r-\phi$ plane, we
effectively measure $p_T$.  By measuring the
track's direction in the $r-z$ plane, we measure
$p_T/p_z$, which completes our measurement of the
3D momentum vector of the particle.

Tracks in these detectors typically have several
dozen hits, which allows for highly efficient
and pure identification of these tracks.  Silicon
trackers in these detectors have of order a
half-dozen hits, and while they can be used
standalone, a much better way to use them is to
find the track in the outer tracker and project
this track back into the silicon.  Once the
approximate trajectory in the silicon detector
has been established, the tracking algorithm
can search the silicon detector for hits and
use these hits to improve the track measurement.
This technique uses each component to its best
advantage: the silicon tracker measures the
track's point of origin and initial direction,
and the outer tracker measures the track's
momentum.

\subsubsection{The D\O \ Solenoid  and Scintillating Fiber Tracker}

One substantial improvement to the D\O \ detector for Run~II was
the addition of a superconducting solenoidal magnet providing a solenoidal field of 2T parallel to the $z$-axis.  
Having a magnetic field enhances the D\O \ detector. 
It has the ability to measure the momentum and charge of leptons and hadrons.
The muon momentum can be measured in the central magnetic field, 
before the muon has scattered in the steel of the muon toroids,
and this allows for the comparison of the muon's momentum as measured in the solenoid to the
momentum as measured in the toroids.  We can see that the muon
momentum resolution is much improved after matching a
muon local track with central track.  

\begin{figure}
\begin{center}
\epsfxsize=4.5 in
\epsfbox{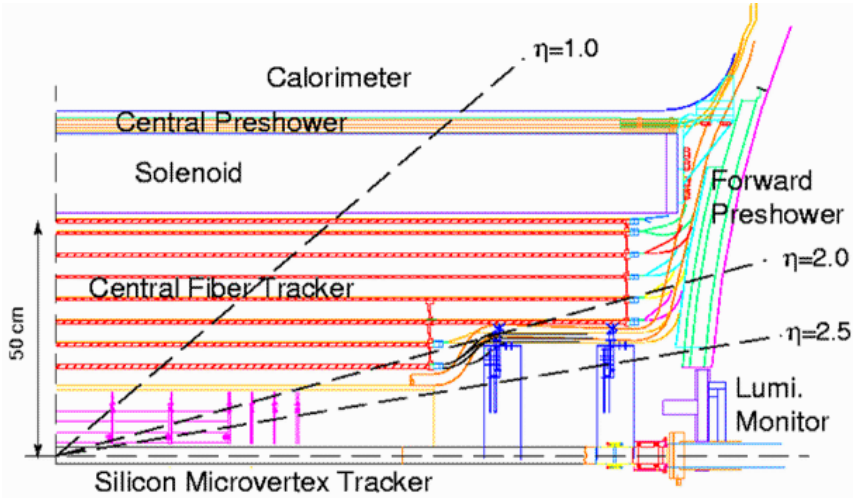}
\end{center}
\caption{D\O 's scintillating fiber tracker.}
\label{fig-d0-tracker}
\end{figure}

Mechanically, the D\O \ outer tracker is a
simple device.  Layers of scintillating fiber
are placed on carbon-fiber composite cylinders.
When a particle travels through one of these
fibers, the scintillator emits light, which is
totally internally reflected down the fiber,
coupled to a clear fiber, and transported
to a solid-state light detector.  Fig.~\ref{fig-d0-tracker}
shows a side view of the tracker.

In detail, the D\O \ central fiber tracker (CFT)~\cite{wayne} consists of
scintillating fibers mounted on eight concentric cylinders made from
a composite of layers of high modulus of elasticity carbon fiber
sandwiching structural foam. The cylinders range in radius from 20 to 50 cm
and  are about 2.5 m in length, except for the inner two,
which are about 1.7 m in length. The scintillating fibers are
double clad, 835~$\mu$m in diameter, and
are constructed in ribbons each 128 fibers wide composed of
a ``doublet" layer of fibers with the centers of one of the single layers
in the space between the fibers of the other single
layer. There are eight doublet axial (aligned along the beam
axis) layers of scintillating fiber, as well as eight doublet stereo
layers that make a $\pm 3^{\circ}$ angle with the beam axis.
The outer (8th) layer is at the largest possible radius.
The 7th layer is as close to the outer layer as is possible. The
inner layer is at the least possible radius.
The detector is divided into 80 sectors in $\phi$. Each pie-shaped
slice has 960 fibers and the entire detector therefore
has 76,800 channels.

The scintillation light from the fibers is totally internally
reflected down the length of the fiber. A connection is made at the
edge of the tracker to  a clear fiber that transports the light to a
solid-state light detector called a Visible Light Photon Counter (VLPC).
The number of photons available at the VLPC's,
for a charged particle which traverses through the center of a fiber,
varies from 10 to 40, depending on the charged particle's pseudorapidity
and on attenuation due to the  distance from the clear fiber connector.
The VLPC's are small silicon devices which
have an array of eight photo sensitive areas, each 1 mm in diameter,
on their surface. They operate at temperatures from 6 to 15 K,
have a quantum efficiency of over 80\%, and have a gain of
20,000 to 50,000.  The high quantum efficiency is important
because of the low number of photons.

The momentum resolution is currently $\sigma(p_T)/p_T=0.13\% \cdot p_T$
when scintillating fiber tracker measurements are combined with the silicon tracker.
The $p_T$-dependent impact parameter resolution is currently 30~$\mu {\rm m}$ in the $x$ and $y$ direction, and 42~$\mu {\rm m}$ in the $z$ direction
for high momentum tracks, if SMT and CFT information are combined.


\boldmath
\section{Calorimeters and Preshower Detectors}
\label{detector-cals}
\unboldmath

   Because of the dependence of the Tevatron's physics program 
on lepton identification and jet energy measurement,
calorimetry is a critical aspect of Run II experiments.
The calorimeters are used for identification and measurement of the
electron, neutrino and jets from the decays of the top
quarks in these events.

   The calorimetry is divided into two parts, electromagnetic
and hadronic.  Nearest the vertex is the electromagnetic
calorimeter that measures the energy of electrons and photons
(including those from $\pi^0$ and $\eta$ decay) and has
improved position measurement at the point of maximum
shower development.  Farther in radius is the hadronic
calorimeter which measures the energy of hadrons as
they interact with the material of the calorimeter.
Muons deposit a small amount of energy (due to ionization)
in both sections, and the lack of a large energy deposit
can be used to identify a particle as a muon.  Neutrinos
deposit no energy at all in these calorimeters, but
the absence of energy deposition appears as a momentum
imbalance in the transverse plane, also called ``missing
$E_T$".

D\O's hermetic, radiation-hard uranium and liquid-argon
calorimeter~\cite{d0runI} consists of three separate cryostats: the Central
Calorimeter (CC), and the two Endcap Calorimeters (EC's). Each is segmented
into an electromagnetic section, a hadronic section, and a coarse hadronic
section (inside to outside), with many layers of sampling. Each is
divided into pseudoprojective towers covering $\eta \times \phi = 0.1
\times 0.1$ rad.  The readout of the electromagnetic section has four layers of
longitudinal segmentation.  The third electromagnetic layer,
at EM shower maximum, has segmentation $\eta \times \phi = 0.05 \times 0.05$ rad.
The readout of the hadronic sections have 4 (5) longitudinal layers in the
CC (EC's). There are no projective cracks.
The calorimeter  provides hermetic coverage to $|\eta|<4$.
The energy resolution is $\sigma_E/E=15\%/\sqrt{E (GeV)} \oplus 0.4\%$
for electrons and photons.  For charged pions and
jets the resolutions are approximately $50\%/\sqrt{E (GeV)}$ and
$80\%/\sqrt{E (GeV)}$, respectively.  

The Inter Cryostat Detectors (ICD's) augment the \D0  liquid argon
calorimeters by providing a measurement of the energy in between the
central and endcap cryostats. This improves the energy measurement for
jets that straddle the intercryostat region and improves the resolution
of the missing transverse energy measurement. 
Sixteen new ICD detector segments
form an annular ring of 1/2" thick scintillator covering
$1.1\le |\eta| \le 1.4$ on the hadronic section of the inner end of each
EC cryostat. The sixteen segments are further segmented into sections of size
$\Delta \eta \times \Delta \phi =0.1 \times 0.1$.
Each section has an embedded wavelength shifting fiber
to collect light. These are in turn connected to long clear fibers which carry
the light to photomultiplier tubes located underneath the cryostats, in a
region with reduced magnetic field from the solenoid.

\begin{figure}[htbp]
\begin{center}
\epsfxsize=5.5 in
\epsfbox{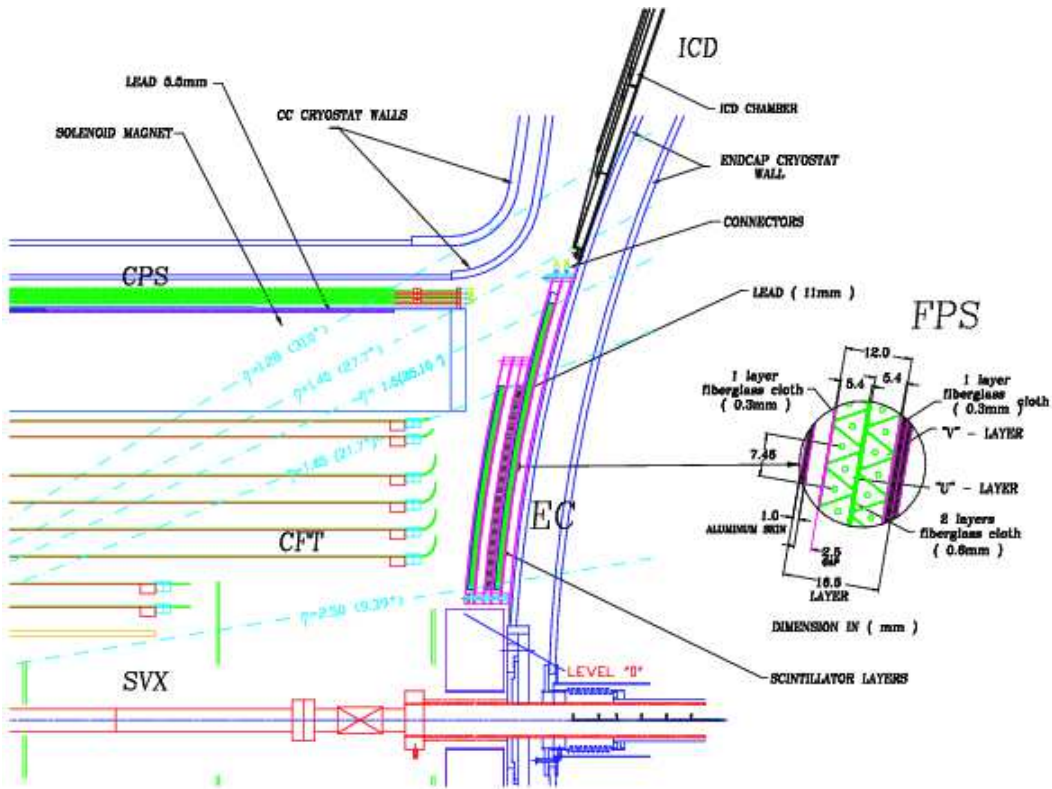}
\end{center}
\caption{One quarter $r-z$ view of the end of the \D0 \ trackers
and the start of the EC, indicating the Central Preshower, the Level-0
detector, the solenoid magnet and calorimeter cryostats, and the Forward
Preshower detector.  The Forward Preshower detector is shown in detail
in the inset. }
\label{fig-d0-fps3}
\end{figure}

The primary purpose of the Central and Forward Preshower Detectors (CPS and FPS) 
is to exploit the difference
between energy loss mechanisms of electrons and photons with that of the
backgrounds, principally hadronic jets with leading $\pi^0$'s, to improve
the trigger and offline purity.  Secondly, they provide a precision
measurement of the starting point of the electromagnetic showers.
The CPS is cylindrically shaped, mounted on the outside of the solenoid
magnet, and covers the region $|\eta|\le 1.3$. The FPS are shaped like
annular rings, mounted on the inside of the EC's, and cover the region
$1.5\le|\eta|\le 2.5$. The FPS and CPS are shown in Fig.~\ref{fig-d0-fps3}.

The D\O \ Run~II luminosity monitor (Level-0) consists of two
arrays of plastic scintillation counters located on the inside faces
of the EC's and arranged symmetrically around the beampipe.
The pseudorapidity coverage is $2.7 \le |\eta| \le 4.4$. Because the
solenoid field is $\sim 1$~Tesla in that region, short magnetic field
resistant photomultiplier tubes are used to read the light. A coincidence
of hits in the counters on both sides of the interaction point
provides the simplest indication that an inelastic collision occurred.

A new near beam detector called the ``Forward Proton Detector (FPD)" is available
for Run~II. It is a series of small, retractable scintillating fiber
detectors placed a few millimeters from the beamline in the region 20--60
meters from both sides of the interaction point. They are triggered by
small scintillation counters and read out by multi-anode PMT's. Their
purpose is to identify scattered protons and anti-protons in diffractive
events.

\boldmath
\section{Muon Systems}
\label{detector-muons}
\unboldmath

\begin{figure}[htbp]
\begin{center}
\epsfxsize=6.0 in
\epsfbox{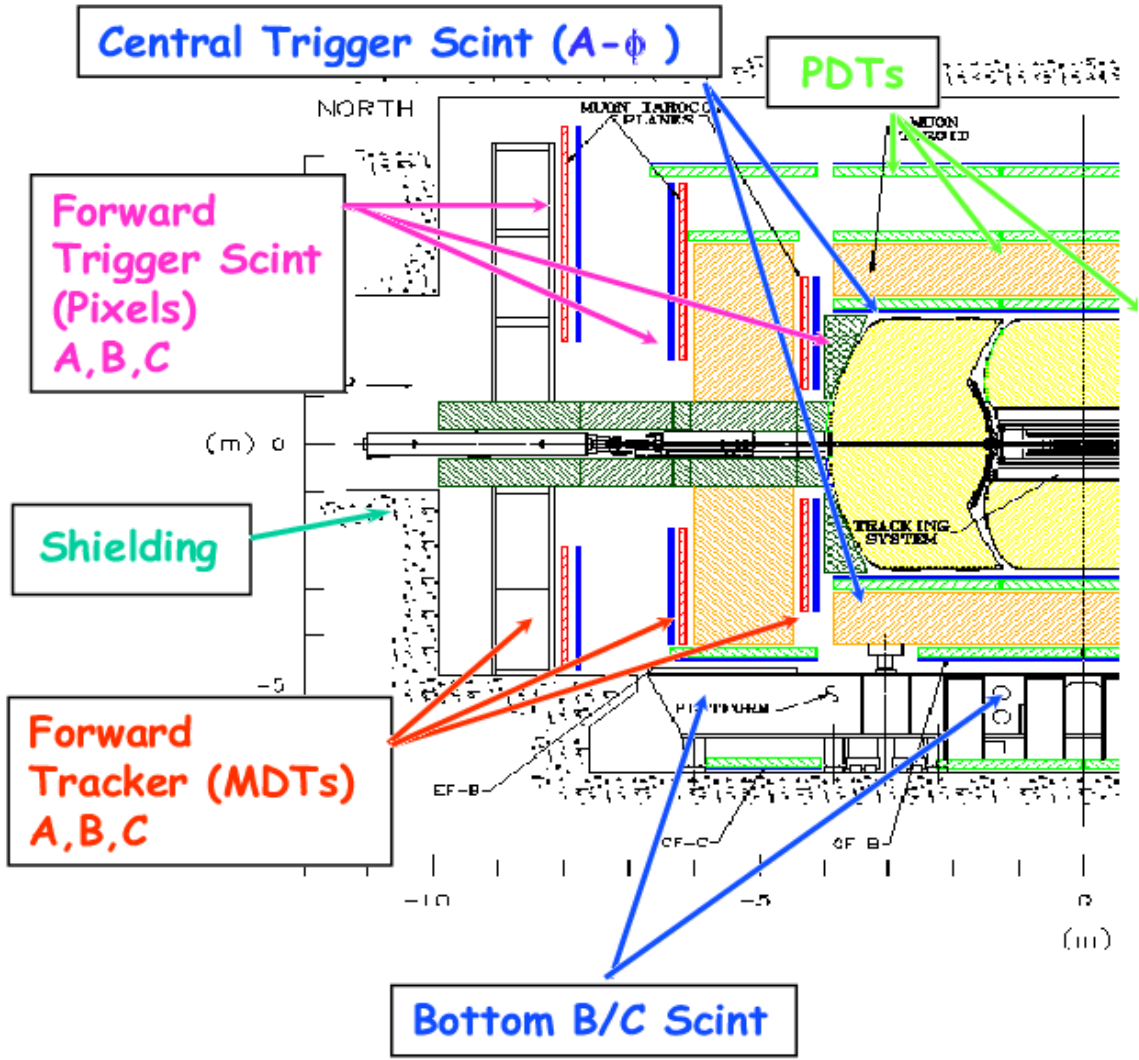}
\end{center}
\caption{Half $r-z$ view of the \D0 \ muon subdetector.}
\label{fig-d0-muon-1}
\end{figure}

\begin{figure}[htbp]
\begin{center}
\epsfxsize=5.0 in
\epsfbox{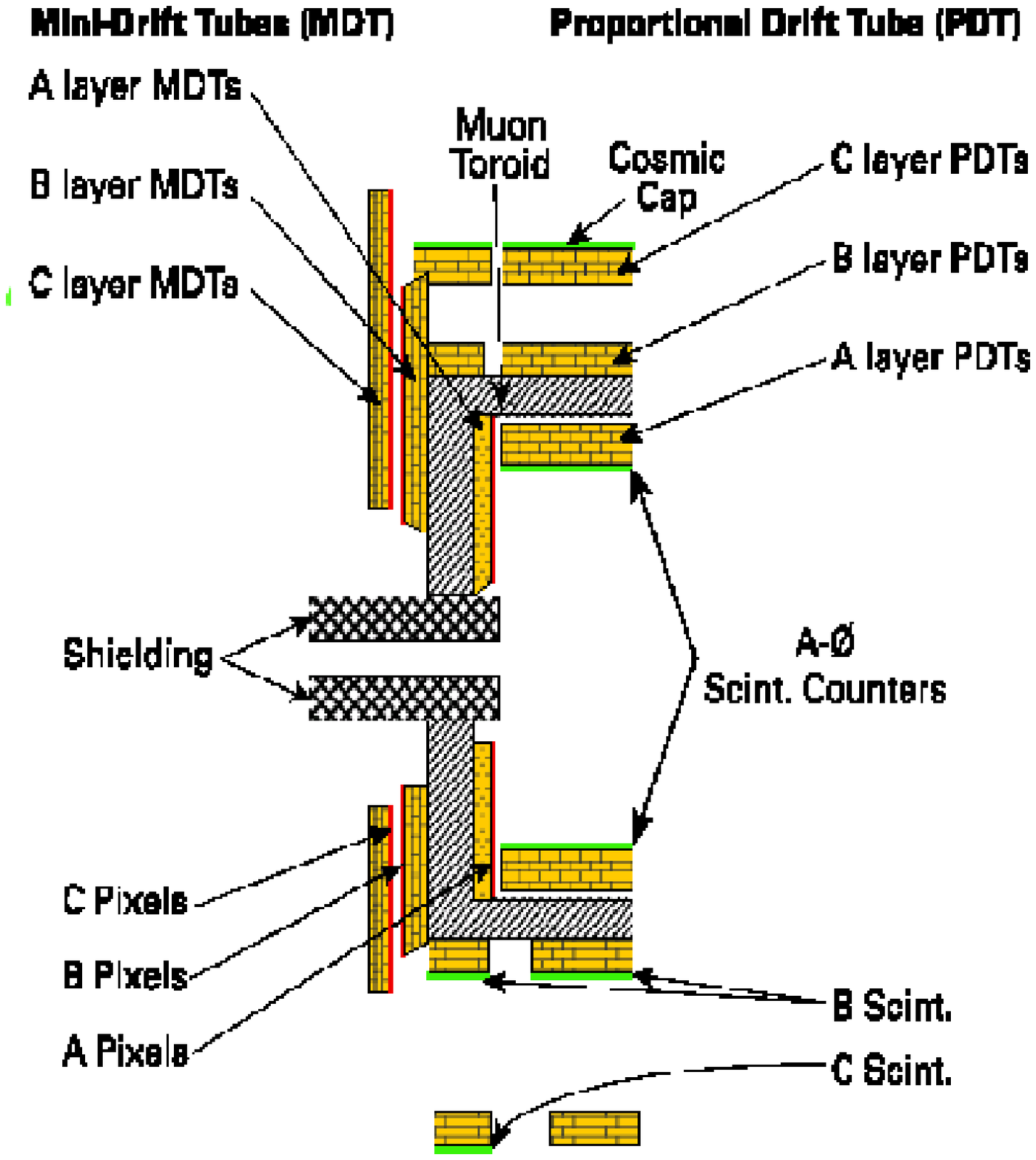}
\end{center}
\caption{Illustration of the \D0 \ muon subdetector components.} 
\label{fig-d0-muon-2}
\end{figure}

  The muon detection strategy at \D0 relies on the
penetration power of muons.  Several meters of absorber (including
the calorimeters) absorbs the vast majority of hadrons, and any
charged particle that penetrates this material is inferred to be
a muon.  Because they are at large radius, muon detectors are large,
and to keep costs reasonable, they have very coarse granularity:
typically they are single wire proportional chambers with drift
times in excess of a microsecond.
Fig.~\ref{fig-d0-muon-1}, and ~\ref{fig-d0-muon-2} show an $r-z$ view of the \D0 detector,
and all components are illustrated in the graph.

\begin{figure}
\begin{center}
\includegraphics[height=15.0cm,width=160mm]{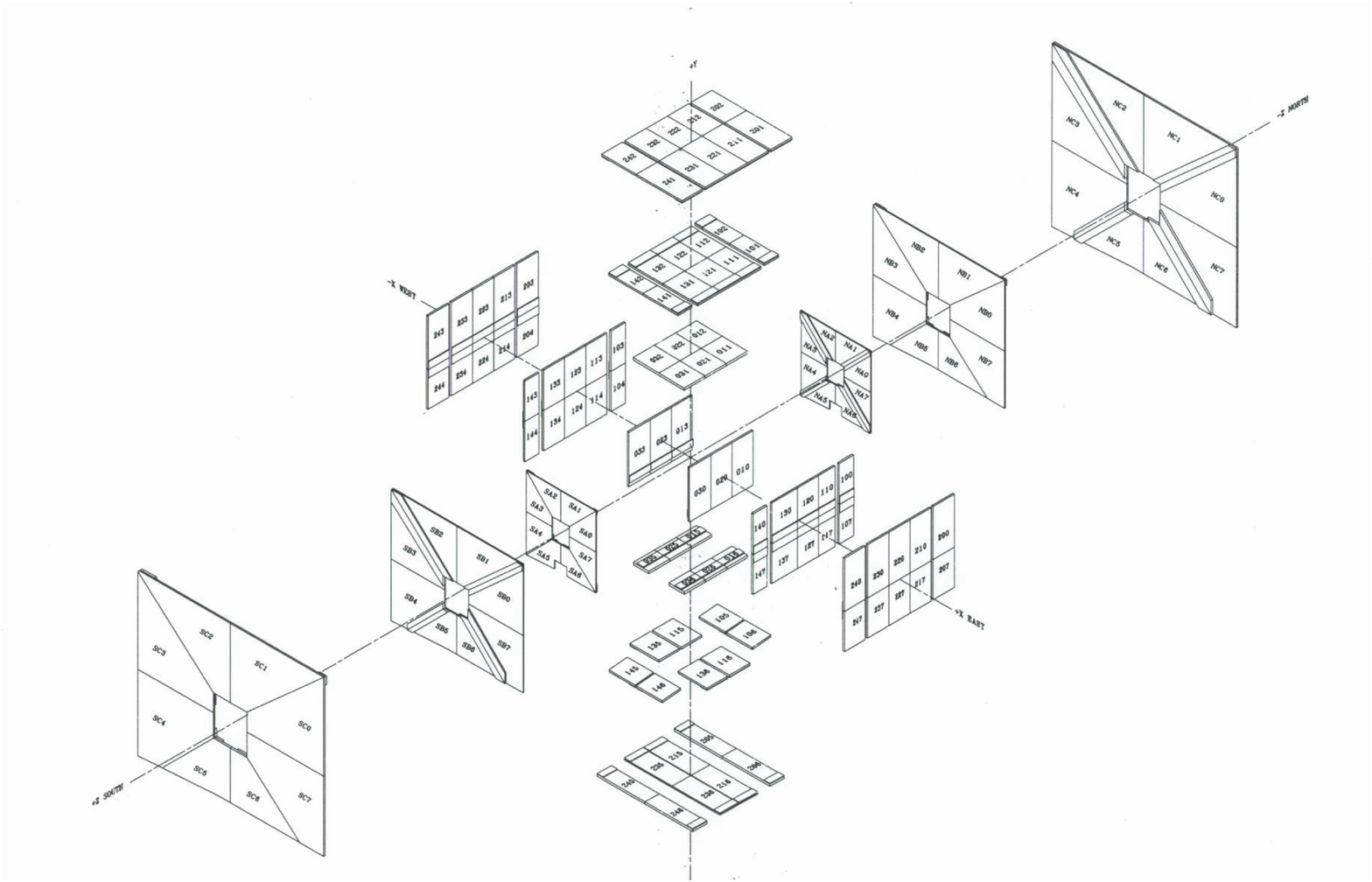}
\caption{Layout of PDTs and MDTs.}
\label{fig-MuDrift}
\end{center}
\end{figure}

The muon detector consist of scintillator and drift tubes, with effectively 
complete coverage out to $|\eta| < 2$. As seen in the layout, 
the detector is split at $|\eta|$ of 1 into a central and forward system. 
Each has 3 layers (usually called A,B,C with A between the calorimeter 
and iron and the other two outside the iron) of drift tubes. 
In the central region are proportional drift chambers (called PDTs). 
In the forward region are minidrift tubes (called MDTs).
Figure~\ref{fig-MuDrift} shows the layout of PDTs and MDTs.

  In Run~II, scintillator counters are added adjacent to the chambers.  
Scintillator has a response time measured in nanoseconds, so that the
coincidence between the counter and the chamber unambiguously
determines the bunch crossing.  Because the timing resolution
is substantially better than the minimum required to resolve
bunch crossings, we gain the ability to use timing to reject
certain backgrounds: particularly out of time particles
produced upstream of the interaction point and cosmic rays.

There is 2 or 3 layers of scintillator coverage with the forward 
scintillators sometimes called pixels, the central A-layer counters 
called A-phi, and the BC counters called the cosmic cap. Scintillator time 
is read out with both a 15--20 ns ``trigger'' gate and a 80--100 ns 
``readout'' gate. 
The D\O \ detector relies on layers of scintillation counters and drift
chambers to identify muons and measure their position and momentum. 

The muon system has three large toroid magnets, one central and
two forward, which act as absorber in addition to the calorimeter,
provide a structure on which to mount the muon detectors, and provide
a bend to the muons so the detectors can measure the momentum.
The calorimeters contain between 7 and 10 interaction lengths of material,
depending on the pseudorapidity. The thickness of the calorimeters
plus the toroid magnets varies between 13 to 14 interaction lengths for
$|\eta|\le 0.9$, 10 to 15 interaction lengths for $0.9 \le |\eta| \le 1.2$,
and 18 to 20 interaction lengths for $1.2 \le |\eta| \le 2.0$.  Muons
with momentum greater than $\sim 1.4$ GeV/$c$ ($\sim 3.5$ GeV/$c$) penetrate
the calorimeter (toroid magnet) at $\eta=0$.
In addition, new massive shielding structures isolate the muon detectors 
from backgrounds generated near the beampipe and accelerator elements.

\subsection{Central Muon System Drift Chambers}
The central muon drift chambers were retained from Run~I, but their electronics
have been replaced. The drift chambers are made from extruded aluminum cells
of 4-inch width and lengths up to 228 inches. The wires in the cells
are parallel to the field in the toroid magnets so that the bending of the
track in the toroids takes place in the drift ordinate. 
Refer to Fig.~\ref{fig-d0-muon-1},~\ref{fig-d0-muon-2}.

\begin{figure}[htbp]
\begin{center}
\epsfxsize=4.5 in
\epsfbox{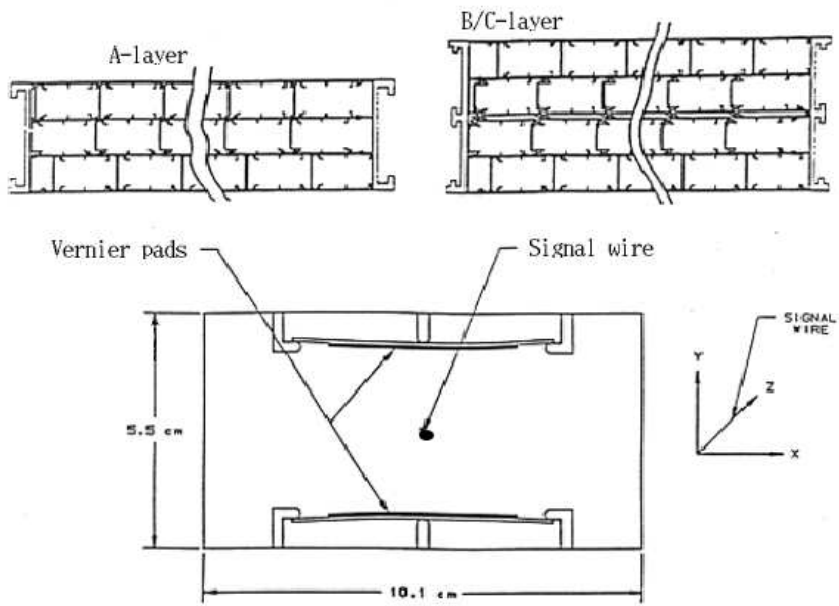}
\end{center}
\caption{Geometry of a PDT cell.}
\label{fig-pdt-cell}
\end{figure}

Individual drift chambers (PDT's) in the C- and B-layers consist of three
staggered decks of up to 24 cells each.  Drift chambers in the A-layer consist
of four staggered decks of 24 cells each, except for the ones on the bottom,
which are three deck PDT's. The top and bottom of each cell has a copper-clad
cathode pad. The copper has a milled cut separating it into an inner and
outer pad such that the width of the inner pad alternately increases and
decreases along its length. The wavelength of the vernier is 24 inches.
Pairs of wires are connected through a delay chip at the end away from the
front-end electronics. Fig.~\ref{fig-pdt-cell} shows the geometry of a PDT cell.

On passing through a cell, a muon will cause a
hit in the cell and a hit in the neighbor cell which is some time later
depending on the muon's proximity to the far end.  Charge is accumulated on
the inner and outer pads of the cell through which the muon passed. The drift
time is derived from the sum of the two cells times.  The distance along
the wire is derived from the difference. The normalized difference of the
integrated pad charge provides the distance along a pad wavelength.

The Run~I A-layer PDT's were rebuilt so as to increase their effective
lifetimes.  The ends of the PDT's were removed.  The cathode pads, which
outgassed a dielectric that coated the anode wires in Run~I, were replaced
with new G-10 pads that do not outgas.  The lifetime of the B- and C-layer PDT's
is long enough so that aging won't pose a problem for these detectors.

The PDT's use a drift gas composed of 80\% argon, 10\% CF$_4$, and
10\% CH$_4$. The maximum drift time is approximately 500 ns, longer
than the bunch spacing. This poses no problem for the electronics, which
records each hit in as many crossings in that it could have occurred.
The time measurements are made in 1.8 ns bins.  The drift ordinate resolution
is $\sim 500~\mu {\rm m}$ per hit, limited by the fluctuations in the drift
time due to the gas. Normally the drift velocity is about 0.1 mm/ns.

The muon momentum is calculated from the bend in the toroid magnet as
determined from the difference in slopes between the line formed from the
interaction point and the A-layer hits and the line through the B- and
C-layer hits. The momentum resolution is expected to be
$\sigma(1/p)=0.18/p \otimes 0.005$ with $p$ in GeV/$c$. This momentum resolution
is worse than that expected from the fiber tracker measurement, especially
at low $p_T$.

\subsection{Central Muon System Scintillation Counters}
An important part of the upgrade is two new layers of scintillation counters.
These detectors not only tag the bunch crossing from which the muons originate
for the slow drift chambers, but also reject background particles which leave
hits at times other than expected from a muon originating at the interaction
point.  Refer to Fig.~\ref{fig-d0-muon-1},~\ref{fig-d0-muon-2}. 

The A-layer contains 630 ``A-phi" counters, each approximately 32" long
and spanning $4.5^\circ$ in azimuthal angle. There are 9 counters in a row
spanning $-1.0 \le \eta \le 1.0$.  Each row has a slight overlap with a
neighboring row so as to minimize the cracks between counters. The counters
are made from 1/2 inch thick Bicron scintillator plates with many Bicron
BCF'92 wavelength shifting fibers
embedded in deep grooves. The fibers collect and transmit the light to a
single photomultiplier tube (PMT).
The counters have a time resolution of $\sigma = 4$ ns and are expected to
discriminate between muons produced in the collisions and the background,
that is composed of particles backscattered from the calorimeter exit, and 
arrives 14 ns later than a muon.
The ``A-phi" counters span 93\% of the azimuthal angle. There is a gap
in the A-layer coverage where the calorimeter is supported by the detector
platform.

A layer of scintillation counters has been added on the outside of the
muon toroid magnet. The 240 ``Cosmic Cap" C-layer scintillation counters
were deployed late in the second half of
Run~I and previously have been described in
detail~\cite{d0scintnim}.  These counters are between 81.5 and 113 inches
long and 25 inches wide. Eight of them are mounted on the outside of each
C-layer PDT on the top and sides of the central muon detector.
Underneath the toroid magnet, the three layer coverage is broken up because
of the support structure for the central platform and toroid magnets.
120 new ``Cosmic Bottom" counters are arrayed on bottom C-layer and
B-layer PDT's.
The Cosmic Cap and Cosmic Bottom counters are made from 1/2 inch thick
Bicron scintillator with Bicron BCF'91A wavelength shifter fibers embedded
in grooves. Each counter is read out with two photomultiplier tubes.
One advantage of using two PMT's is that coincidental tube noise is
improbable. Another is the immediate redundancy available in case one
of the PMT's fails.

The scintillation counters have an LED pulser calibration system capable
of providing a clocked, timed, amplitude-controlled photon pulse. Each PMT
is connected by a light-shielded optical fiber to a light-tight box which
houses bundles of LED's glued into a clear acrylic block. A single box
may provide the photon pulse for up to 100 PMT's. The stability of the
photon pulse is monitored with a light-sensitive diode housed in the
clear block. This system allows the timing and amplitude to be
monitored and controlled.

The muon system has a wide range of options available for triggering.
Triggers may be composed of coincidences of in-time hits in scintillation
counters, hits in layers of the PDT's, and tracks found in the central
tracker.  Low-$p_T$ muons would rely on hits only in the A-layer detectors
and fiber tracker. High-$p_T$ muons would also use scintillation counters
and PDT's in the B- and C-layers.

\subsection{Forward Muon System Drift Tubes}

\begin{figure}[htbp]
\begin{center}
\epsfxsize=4.0 in
\epsfbox{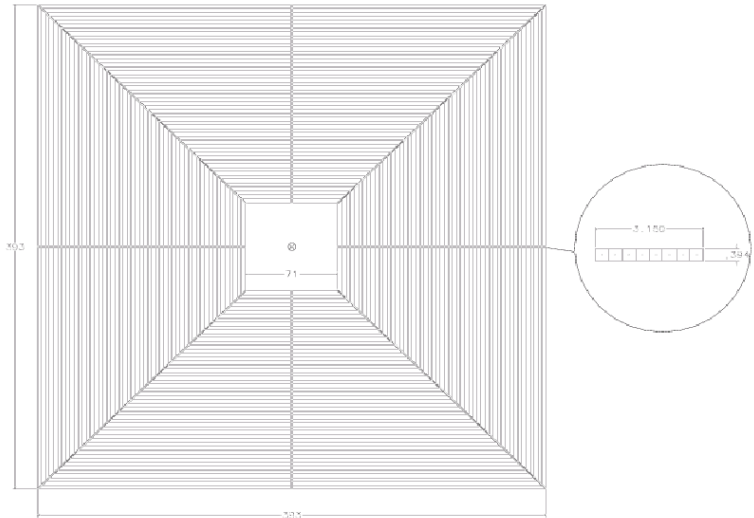}
\end{center}
\caption{\D0~forward muon mini-drift tube (MDT) plane. The octant
boundaries are shown.}
\label{fig-d0-mdt}
\end{figure}

The Mini-Drift Tube (MDT) system~\cite{abramov}
is comprised of three planes of drift-tubes, with one plane in front of, and
two planes behind the forward toroid magnet.
 The layers are divided into octants with
tubes of different length depending on position in the octant.
As in the central region, the MDT A-layer has four decks of drift tubes and
the B and C-layers have three decks each. The drift tubes, made from long
aluminum extrusions of eight 1 cm square cells, are contained in  plastic
sleeves.  Wires in the cells are oriented parallel to the magnetic field
of the forward toroid magnet.  The sleeves of tubes are mounted on an
aluminum support structure which also provides mechanical support for the
infrastructure. A plane of MDT counters is shown in Fig.~\ref{fig-d0-mdt}.

\begin{figure}[ht]
\begin{center}
\epsfxsize=4.5 in
\epsfbox{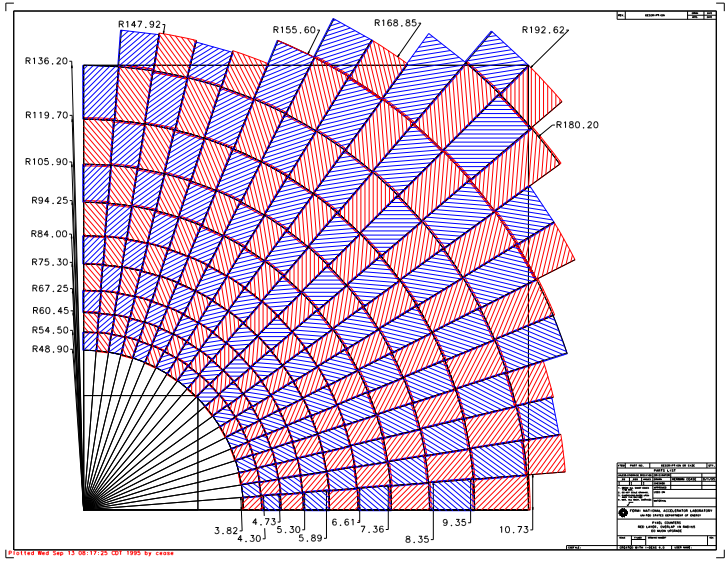}
\end{center}
\caption{Two D\O \ forward muon pixel octants.}
\label{fig-d0-mupixels}
\end{figure}

The MDT's use a non-flammable gas mixture composed of 90\% CF$_4$ and
10\% CH$_4$, with a maximum drift time of about 60 ns.  The momentum
resolution is limited by multiple Coulomb scattering in the iron toroid
and the hit resolution of the detector. The MDT electronics uses a coarse
digitization of the drift time (18.8 ns time bin).  The momentum resolution
is roughly $\sigma(p_T)/p_T = 0.2 $.  Importantly, it is on par with the
resolution of D\O 's central  tracker in the forward region where the full
coverage of the fiber tracker's layers has ended.

\subsection{Forward Muon System Scintillation Counters}
Three layers of ``Pixel" scintillation counters~\cite{abramov}
are added to the forward region (approximately $1.0 \le |\eta| \le 2.0$).
Their primary role is muon triggering.  The $\sim 4800$ pixel counters
have segmentation $0.1 \times 4.5^\circ$ in $\Delta \eta \times \Delta \phi$.
Most of the trapezoidal shaped pixel counters are made from 0.5" thick Bicron
scintillator with wavelength shifting bars. They are each read out by a
single PMT. The counters are held in protective aluminum containers with a
steel fastener, on one corner, which mates to the magnetic shield of the PMT.
A few of the counters have special space constraints and are made with
wavelength shifting fiber so as to allow more flexibility in the orientation
of the PMT and magnetic shield.  The same kind of LED pulser calibration
system used in the central muon scintillators is used for the Pixel counters.

The counters are assembled into octants of $\sim 100$ counters each.
The octants provide mechanical support for the counters and their
infrastructure.  These are mounted directly onto the A- and
B-layer sides of the forward toroid magnets and onto the inside face of the
C-layer support frame.  Two octants are shown in Fig.~\ref{fig-d0-mupixels}.

Forward muon triggers are formed from coincidences in the three layers of
scintillation counters consistent with a muon of a given momentum. The
MDT's provide a $p_T$ measurement at trigger level, especially important
in the fiducial region where the fiber tracker has reduced coverage.

\subsection{Shielding}
The purpose of shielding is to shield the muon detectors from backgrounds
generated at high $|\eta|$ from the interaction
of the beam jets with forward elements of the detector and accelerator
hardware, such as the beampipe and low-beta quadrupole magnet. 
The shielding is built in several large, movable sections extending from the
endcap calorimeters, through the forward toroid magnets, to the Tevatron
tunnels. The shields themselves totally contain the accelerator elements
within the collision hall, including the low-beta quadrupole magnet, inside
a case of 20 inches of iron, six inches of polyethylene, and two inches of
lead.  The shielding is shown in Figs.~\ref{fig-d0-muon-1},~\ref{fig-d0-muon-2}.


\boldmath
\section{Trigger}
\label{detector-trig}
\unboldmath

One of the defining features of hadron collider physics is the
necessity to select the small fraction of all bunch crossings containing
interesting collisions.
A fast selection process, called the ``trigger", sorts events into
categories of various levels of interest.
An event that passes the trigger is written to magnetic tape for later
analysis.  An event that fails the trigger is lost forever.
An event can fail to pass the trigger because the collision was a well
understood process, because the event was mistaken for a well understood
process (trigger inefficiency), or because the trigger or data acquisition
were busy processing previous collisions (dead time).

There are three trigger levels. After Level-1 trigger, the event rate
is reduced from 7 MHz to 10 KHz. Level-2 trigger reduces the event rate
10 times more. We can further reduce the event rate 20 times by the
Level-3 software trigger. After Level-3, 50 events are written to tape
per second.

\subsection{Trigger Level-1}
The first element of the trigger is the formation of
trigger primitives.  These are collections of a few
bits of data that represent the status of various detector
elements.  For example, one set of trigger primitives are
the calorimeter cells with energy above a particular threshold.
Another would be which muon chambers have detected a particle,
and whether this particle is consistent with a high $p_T$
muon, a low $p_T$ muon, or neither.  Another would be
the number and position of tracks found in the outer tracker (silicon
is not in the trigger at this stage).

A limited amount of processing is then applied to these trigger
primitives.  Typically, this is performed in Field Programmable Gate
Arrays (FPGA's) with inputs from the front-end detector electronics.
In D\O, for instance, Level-1 muon primitives are formed from combinations
of in-time hits in scintillation counters, coincidences of hit cells
in the drift chambers, and tracks formed from hit patterns in the
axial scintillating fibers of the central tracker.

\begin{figure}[htbp]
\begin{center}
\epsfxsize=5.0 in
\epsfbox{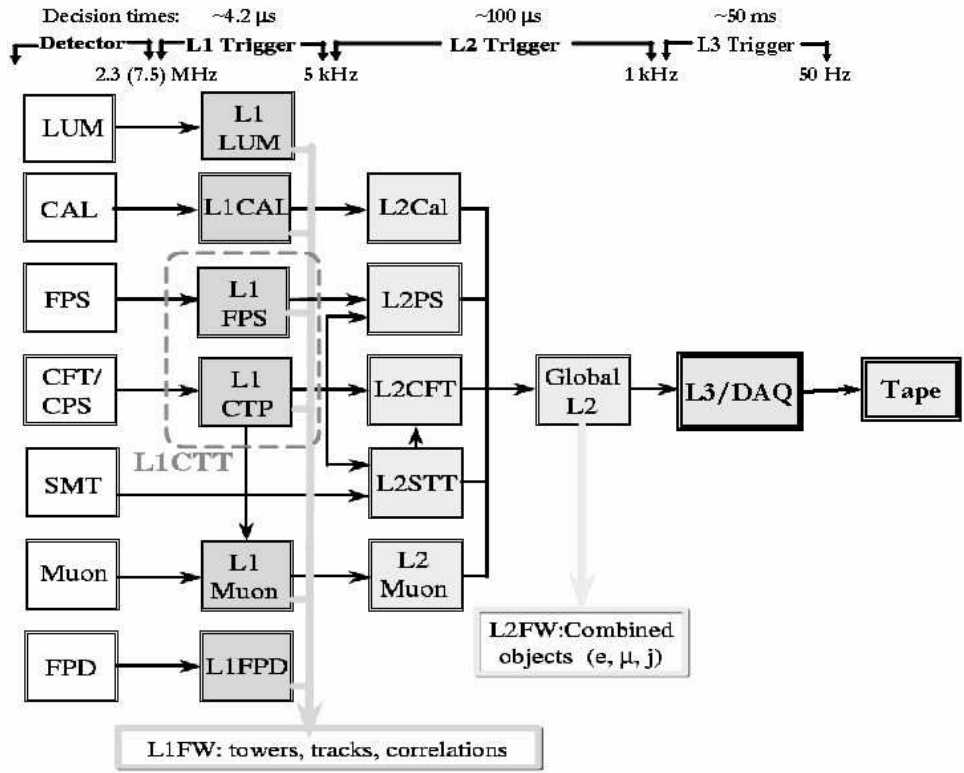}
\end{center}
\caption{Design of the D\O \ trigger system.}
\label{fig-d0trigger}
\end{figure}

The primitives are sent to the global Level-1 trigger. Combinations
of trigger primitives are compared against a runtime programmable
list containing the definitions of triggers to be used in the run.
The D\O \ Level-1 trigger system can support up to 128 different unique
triggers.
If the trigger primitives satisfy at least one of the triggers, the
event is passed to the next trigger level.
A block diagram of D\O's trigger is shown in Fig.~\ref{fig-d0trigger}. 

It takes time to form the primitives and to make a Level-1 trigger
decision. Meanwhile collisions continue to occur.  Since the trigger
needs to be ready for the next crossing while it is processing 
an event, the data from the detectors is placed in a pipeline (a
microprocessor's ``assembly line'' for executing program instructions;
a pipelined function unit in a processor separates the execution 
of an instruction into multiple stages) 
which increments every crossing.  

The D\O \ Level-1 trigger has a 
deadtimeless output rate of 10~kHz. If any single one of the 128 combinations is
positive, and the DAQ system is ready for acquisition, then the Level-1 framework
issues an accept, and the event data is digitized and moved into a series of
16 event buffers to await a Level-2 trigger decision.

The central Level-1 trigger logic is performed locally in the detector
octants. 
A muon low ${\it p_T}$ trigger is defined using only centroids found in the A-layer,
while a high ${\it p_T}$ trigger is defined by
using correlations between centroids found in the A-layer and B- or C-layer.
Four thresholds (2, 4, 7 and 11 GeV/$c$) are defined using the CFT information.
The information for each octant in each region is combined in the muon trigger
manager, which produces global muon trigger information. The muon trigger
manager makes a trigger decision based on the ${\it p_T}$ threshold (2, 4, 7 and 11 GeV/$c$),
pseudorapidity region ($ |\eta|<1.0, |\eta|<1.5$ and $|\eta|<2.0$), quality (Loose,
Medium and Tight) and multiplicity information. This trigger decision is sent
to the Level-1 Trigger Framework where it is included in the global physics trigger
decision.

\subsection{Trigger Level-2}

At Level-2, the Level-1 decision is confirmed, or not, using
the additional time to provide more and better information.
This is done in two stages.  The first stage, called the
preprocessor stage, uses FPGA's to refine the trigger primitives found
by Level-1 and to prepare the data for the second stage.
The second stage, called the global processor stage, combines and correlates
information from the trigger primitives.

The preprocessor FPGA's have about 50 $\mu$s to perform refinements to the
trigger primitives.  For example, the first stage photon primitive,
formed from a single tower at Level-1, is required to have a shape
that looks like an isolated electromagnetic shower at Level-2. The
additional time is used to compare the energy of the tower of interest
with  nearby towers.  Additional related information can also be used,
such as energy in the shower maximum. Because of the increased accuracy
of the measurement, thresholds can be set tighter in Level-2,
producing a great deal of the rejection in the global processor stage.

The global processor stage for \D0 is a VME-based CPU card
using a Digital Alpha processor.
Using a general purpose CPU at this level of the trigger
provides a great deal of flexibility -- essentially
all of the Level-1 primitives are available at Level-2,
where they can be combined in ways not possible
at Level-1.  For instance, Level-1 can produce
photon triggers and it can produce jet triggers.
However, Level-2 provides the ability to correlate
the two in ways not possible by the dedicated hardware
of Level-1. For example, requiring that the jet
be opposite the photon in $\phi$ and having similar
$E_T$.  This can be used to select events of
a given topology with particularly interesting
kinematics, for example.  Perhaps more importantly,
though, is that this design provides added
flexibility for implementing new triggers during the
Run~II even if they have not been anticipated before the
run begins. This might be in response to better
than promised luminosity from the accelerator,
or it might be in response to early physics results.

Level-2 is where the information from the silicon vertex 
trigger (SVT) becomes available.  Layers of silicon are read out into this
trigger, which uses an associative memory and roads
provided from the outer tracker to identify silicon
tracks.  This provides improved momentum resolution,
but more importantly, it also provides impact
parameter resolution for each track sufficient to
identify particles from displaced vertices.  Since
the vast majority of two track triggers are not
from heavy flavor decay, the SVT provides three
orders of magnitude rejection.  
Silicon information doesn't just improve the impact parameter resolution, it also improves
the momentum resolution, because the two tend to be highly correlated.  
The maximum deadtimeless output rate of Level-2 is about 1000~Hz
at D\O.

By shifting a 3-tube wide window over all the cells in an octant, 
and looking for wire triplets with a matching scintillator hit, the muon
preprocessor first finds track stubs separately in the A-layer and the 
BC-layer. The track stubs found are reported to an ALPHA preprocessor 
board that matches track stubs in the A layer with that in the BC-layer,
and creates Level-2 objects from matched or unmatched stubs.
These Level-2 objects hold the $\phi, \eta$ and $p_T$ of the muon, and are
reported to the Level-2 global processor. Upon a Level-2 Accept, the Level-2 objects
are sent to Level-3 for more precise muon track reconstruction.

\subsection{Trigger Level-3}

The third level trigger is often described as an event filter.
It is a software-based system characterized by parallel data
paths that transfer data from the detector front-end crates to a
farm of processors. It reduces the input rate of 1 kHz to an output rate
of 50 Hz. 

D\O's farm of 500 parallel commodity CPU's builds the event into the offline
format, runs a modified (for faster execution time) version of
the offline reconstruction on the event, and makes a decision to accept
or reject the event.  If the event is accepted, it is already in or close
to the proper format for offline analysis.  Additionally,
this trigger level can be used to characterize
the event and decide whether an event should receive
priority in reconstruction.  Reconstructing a small
fraction (say 1\%) of events in an ``express stream"
can be used to provide rapid feedback on the detector's
performance and .  The overall output rate of
Level-3 is about 50~Hz, with some variation depending
on luminosity and dataset selection requirements.

Running what is effectively the offline reconstruction
online also provides an excellent monitor of the health
of the experiment.  The full offline reconstruction lags
a day or two behind the data taking in order to use the
final calibration constants, but the Level-3 reconstruction
lags only a fraction of a second.  Serious problems can
therefore be detected before a large amount of data is
collected.  As an additional benefit, because Level-3
looks at the output of Level-2 and offline looks at the
output of Level-3, monitoring at Level-3 examines many
times more events than offline monitoring, also improving
the probability to spot trouble sooner.

Using commodity processors has a number of advantages.
First, the nature of an event filter naturally lends itself
to parallel processing: each CPU processes an entire event,
with a supervisor process assigning incoming events to
CPUs that finish their events and become ready for new ones.
Second, these computers are inexpensive, and getting
more so with time, and finally, the system is highly expandable:
additional CPU's can be added at a later date, until the
bandwidth into Level-3 becomes the limiting factor.

The muon Level-3 trigger utilize some aspects of the offline muon
reconstruction. Level-3 muons have more complete information on the
vertex and inner tracking components that will improve
momentum resolution, and has the ability to require that
multiple muons come from the same vertex. 
Similar fits are done in Level-1 as in the final offline reconstruction.
Requirements on matching the muon track to the inner tracking
can reduce remnant combinatorics plus punchthroughs.
Level-3 also uses the calorimeter information to reduce combinatorics,
and separate muons into isolated and non-isolated. 
Level-3 improves on Level-2's ability to separate muon sources into
prompt, slow, or out-of-time by fitting the available scintillator
hits along a track to the particle's velocity. Level-3 can remove a 
cosmic ray muons both by their being out-of-time and by looking for
evidence of a penetrating track on the opposite side of the detector.
Level-3 can also clean up single muon events that Level-1 and 
Level-2 identified as dimuons, such as those which pass through 
the FAMUS--WAMUS overlap region.






\boldmath
\subsection{$B$ Triggers with \D0}
\label{detector-btriggers}
\unboldmath

$B$-hadron observability depends strongly on the detector capabilities
to trigger on soft lepton(s) present in semi-leptonic channel or on
$J/\psi$'s produced in $B$ decays. Hadronic $B$ triggers are not considered in
the following.

The Level-1 muon hardware trigger is based on the combination of low $p_T$
track candidates measured in the CFT, spatially matched with hits in the
scintillator planes and/or drift chambers. Single muon events with
$p_T^{\mu}>4$~GeV/$c$ run prescaled currently, and di-muons with 
$p_T^{\mu}>2$~GeV/$c$ are expected to run unprescaled.

The electron trigger is aimed at soft electron pair detection. 
Level-1 candidates are selected separately in EM calorimeter trigger towers
($\Delta\eta\,=\,\Delta\varphi\,=\,$0.2) with a transverse energy deposit
$E_T>2.0$~GeV, and in the tracking system with a low $p_T$
track coincident with pre-shower cluster. Electron candidates of both systems
are then required to match within a quadrant in $\varphi$ 
and to have opposite signs. These electron triggers are not used in the following analysis. 

Level-2 triggers include setting of the invariant mass window 
and angular cuts in di-lepton channels to select $J/\psi$ decays 
and improve background rejection.

\boldmath
\section{Data Acquisition}
\label{detector-daq}
\unboldmath

\begin{figure}[htbp]
\begin{center}
\epsfxsize=3.5 in
\epsfbox{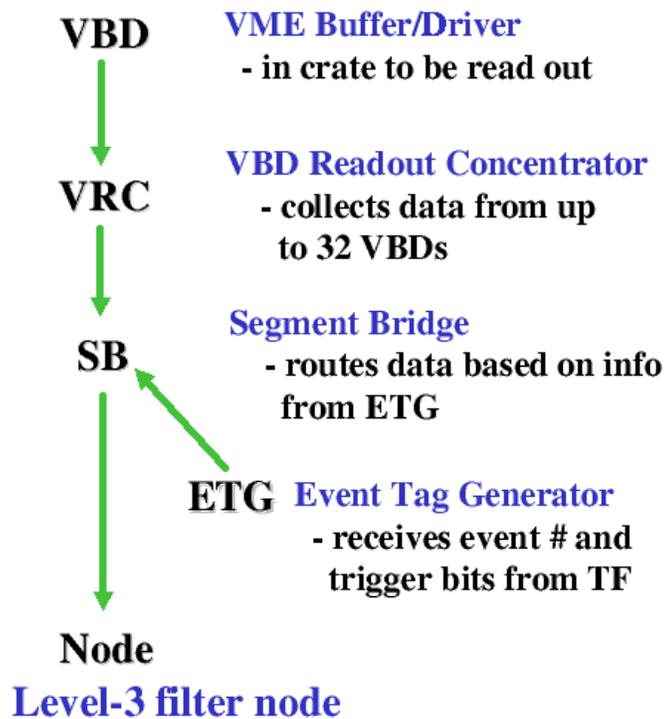}
\end{center}
\caption{Simplified data flow in Level-3/DAQ for D\O ~Run~II.}
\label{fig-d0daq}
\end{figure}

The data flow in trigger Level-3/data acquisition is illustrated in
Fig.~\ref{fig-d0daq}.
Events are reconstructed on the FNAL processor farm system, with 
that portion dedicated to \D0 capable of 
matching the 50 Hz data acquisition rate. Following reconstruction, 
data are stored on a tightly coupled disk and robotic tape system, 
and made available for analysis on a centralized analysis processor. 

\chapter{Establishing the \lb\ Signal}
\label{ch:recolb}

\section{Data and Monte Carlo Event Samples}

The data used corresponds to an integrated luminosity of 
approximately 225~pb$^{-1}$ collected between April 2002 and 
January 2004. This analysis is described in Ref.~\cite{leptonp_lambdab} and repeated in Ref.~\cite{leptonp_lambdab_update}.

\subsection{Dimuon data sample}

The data used in this analysis was part of a skim that was 
called the ``D'' skim, which included 
either $D$ mesons for semileptonic studies or $J/\psi$'s 
for exclusive channels. The dimuon requirements on this skim are:

\begin{itemize}
\item two muons with opposite charge;
\item $p_{T}>1$ GeV for each muon;
\item number of CFT hits $>$~1 for each particle;
\item at least 1 muon with $nseg=3$; $nseg$ is a muon quality cut that describes the hits in the muon detector layers.
\item for $nseg>0$, $p_{T}(\mu)>1.5$~GeV; 
for $nseg=0$,  $p_{T}>1$ GeV;
\item 2.5~GeV~$<$~Mass$(\mu^+\mu^-) < 3.6$~GeV;
\item for a muon with $nseg=0$:
     \begin{itemize}
     \item $p_{T}(\mu^+\mu^-)>4.0$ GeV;
     \item $p(\mu) < 7$~GeV;
     \item $p_{T}$ of second muon $>2.5$~GeV; and
     \item $\chi^2$ of global muon track fit~$<$~25 for both muons.
     \end{itemize}
\end{itemize}


\subsection{Primary vertex reconstruction}

To determine the primary vertex, the AATrack package~\cite{AATrack} is used: 
\begin{verbatim}
AATrack v0-10-06-01
beam spot file version 2.09
\end{verbatim} 
that follows the procedure as outlined in Ref.~\cite{primary}. A primary vertex is determined for each event by minimizing a $\chi^2$ function that depends on all the tracks in the event and a term that represents the beam spot constraint.
The beam spot is the run-by-run average beam position, where a run typically lasts several hours. The beam spot is stable during the periods of time when the proton and the antiproton beams are kept colliding continously, and can be used as a constraint for the primary vertex fit. The initial primary vertex candidate and its $\chi^2$ are obtained using all tracks. Next, each track used in the $\chi^2$ calculation is removed temporarily and the $\chi^2$ is calculated again; if the $\chi^2$ decreases by 9 or more, this track is discarded from the PV fit. This procedure is repeated until no more tracks can be discarded. The PV resolution achieved by using this procedure is 25~$\mu {\rm m}$.

\subsection{Secondary vertex reconstruction}
To reconstruct secondary vertices
(i.e.,  $J/\psi$, $\Lambda$ and \lb\ vertices), 
the BANA package~\cite{bana} is used.

\subsection{\lb\ Candidate Selection}
\begin{figure}[h!tb]
\begin{center}
\includegraphics[height=12.0cm,width=100mm]{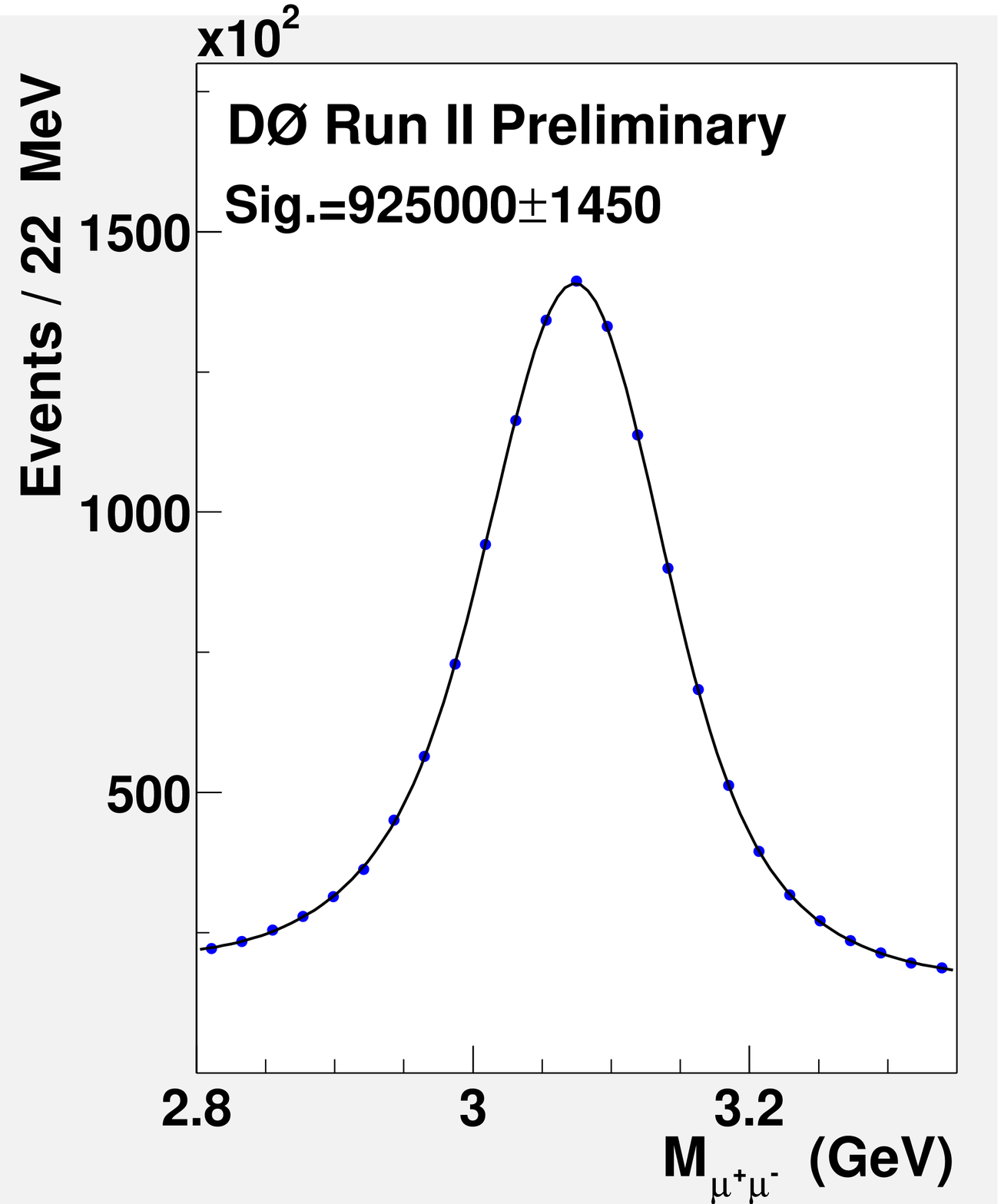}
\caption{Invariant mass distribution of the ($\mu^+\mu^-$) system for
all the events. The signal is described by a 
double Gaussian function and the background by a second order polynomial. 
The signal mean is $3070.0 \pm 0.1$~MeV and the fitted 
widths are $\sigma_{1}=55 \pm 1$~MeV and $\sigma_{2}=100 \pm 1$~MeV.}
\label{fig:jpsi_incl}
\end{center}
\end{figure}

\begin{figure}[h!tb]
\begin{center}
\includegraphics[height=12.0cm,width=100mm]{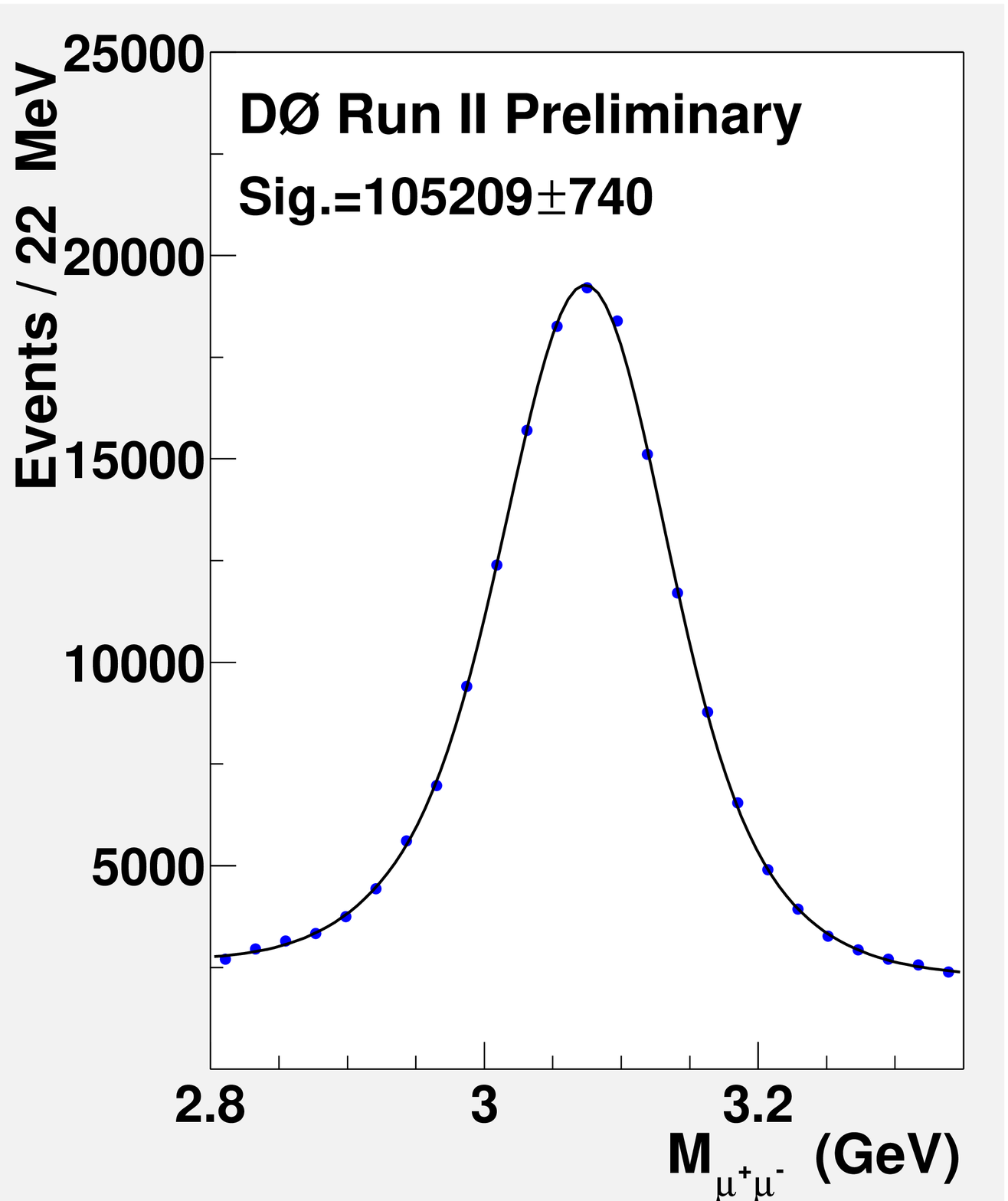}
\caption{Invariant mass distribution of the ($\mu^+\mu^-$) system for
all the events containing a $\Lambda$ candidate. The signal is described by a 
double Gaussian function and the background by a second order polynomial. 
The signal mean is $3069 \pm 1$~MeV and the fitted 
widths from double Gaussian function are $\sigma_{1}=52 \pm 1$~MeV and $\sigma_{2}=91 \pm 5$~MeV.}
\label{fig:jpsi_excl}
\end{center}
\end{figure}

Further cuts are used to select $J/\psi$ mesons:
\begin{itemize}
\item Require that a primary vertex reconstructed.
\item For dimuons forming a vertex, require:
      \begin{itemize}
      \item distance from primary vertex to $J/\psi$ vertex $<10$~cm.
      \item Limited number of track measurements downstream ($<1$) 
of the vertex and missed hits upstream ($<5$) of the vertex.
      \end{itemize}
\item 2.8~GeV~$<$~Mass$(\mu^+\mu^-) < 3.35$~GeV; and
\item number of SMT hits for each muon $>=1$.
\end{itemize}
Figure~\ref{fig:jpsi_incl} and Fig.~\ref{fig:jpsi_excl} show the 
$(\mu^+\mu^-)$ invariant mass for all inclusive events 
and for events  containing a $\Lambda$ candidate, respectively. 

To select $\Lambda \rightarrow p \pi^-$ candidates, we require:
\begin{itemize}
\item two oppositely charged tracks forming a vertex;
\item the higher momentum track is assumed to be the proton;
\item limited number~($<1$) of track measurements downstream of the vertex;
\item number of CFT hits $\ge 1$ for each track;
\item 1.105~GeV~$<$~Mass$(p \pi) < 1.125$~GeV;
\item $p_{T}(\Lambda)>0.4$~GeV; and
\item $\Lambda$ candidates falling within a $K^0_{S}$ mass window
of 0.465~GeV~$<$~Mass$(\pi^+\pi^-) < 0.52$~GeV are removed.
\end{itemize}

\begin{figure}[h!tb]
\begin{center}
\includegraphics[height=12.0cm,width=100mm]{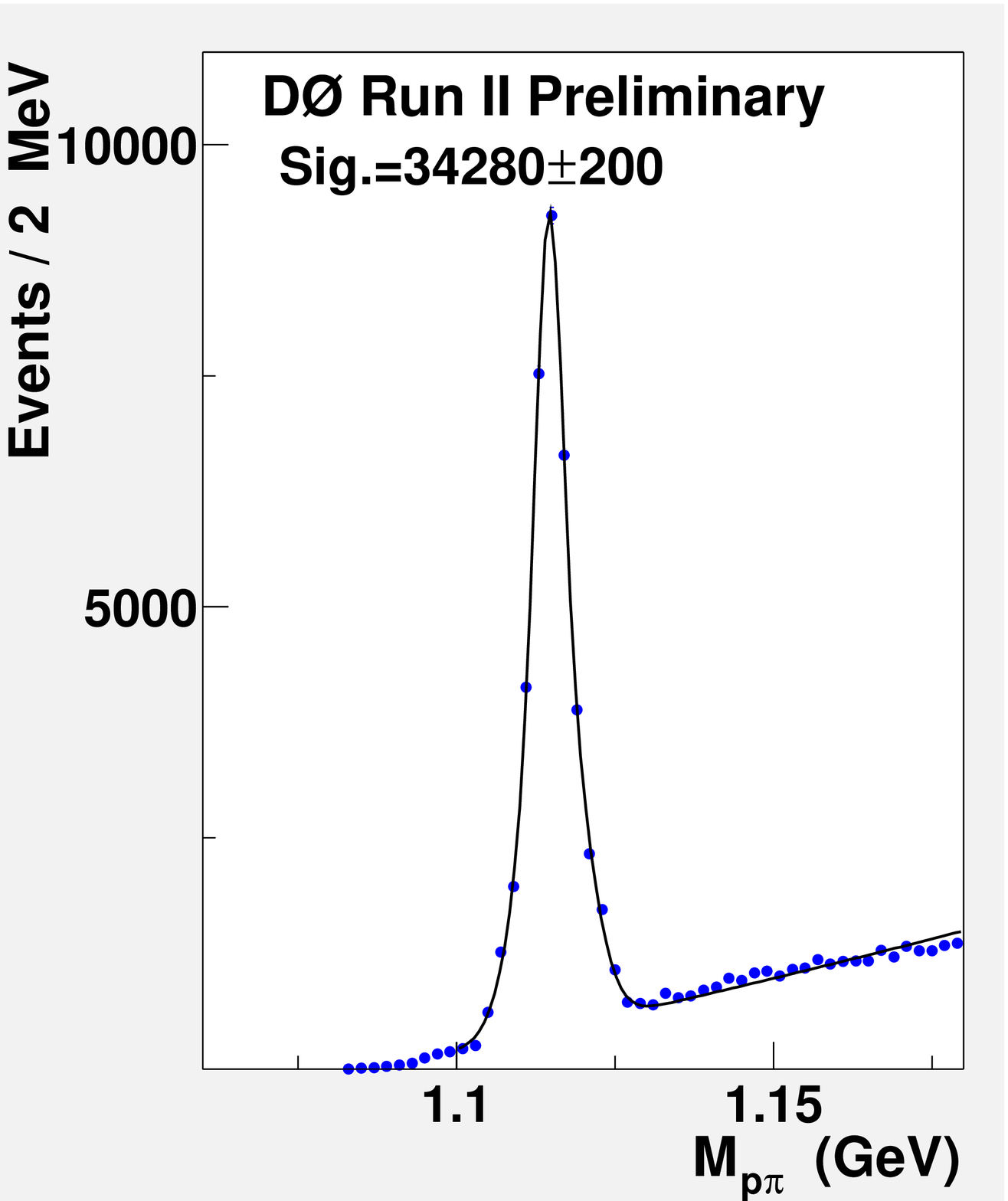}
\caption{Invariant mass distribution of the ($p,\pi$) system for
all the events containing a $J/\psi$ candidate. The signal is described by a double
Gaussian function and the background by a second order polynomial. 
The signal mean is $1115.3 \pm 0.1$~MeV and the fitted widths 
are $\sigma_{1}=4.6 \pm 0.1$~MeV and $\sigma_{2}=2.0 \pm 0.1$~MeV for a double Gaussian function.}
\label{fig:lam_excl}
\end{center}
\end{figure}

Figure~\ref{fig:lam_excl} shows the invariant mass of $(p \pi)$  
for all events containing a $J/\psi$ candidate. 

Once good $J/\psi$ and $\Lambda$ candidates have
been found in the same event,  the two muons 
from $J/\psi$ are combined with the reconstructed neutral $\Lambda$ track
 to form a \lb\ vertex. Further requirements are made on the vertex:

\begin{itemize}
\item vertex $\chi^2(\Lambda_{b})<10$;
\item $p(\Lambda_b^0) > 5$~GeV;
\item $xy$ distance from $J/\psi$ vertex to 
$\Lambda$ vertex $>$~0.3~cm;
\item assuming that the higher momentum track is the
proton, $p(p) > 0.8$~GeV; and
\item collinearity of \lb\ $>$~0.99, where collinearity is 
defined as the cosine of the angle between the momentum of \lb\ in 
the $xy$ plane and the direction from primary to secondary in
the  $xy$ plane.

\end{itemize}

\begin{figure}[h!tb]
\begin{center}
\includegraphics[height=12.0cm,width=100mm]{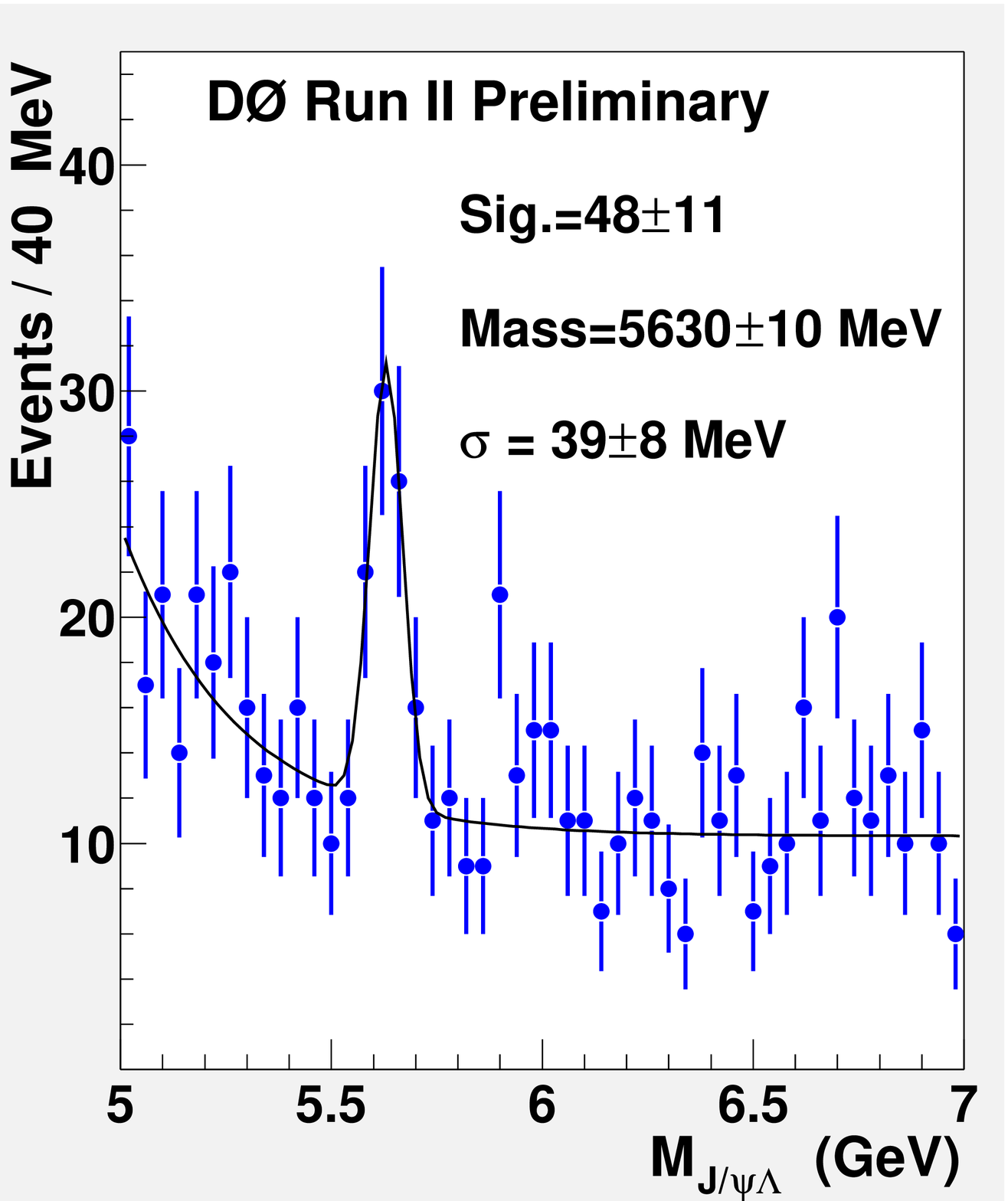}
\caption{Invariant mass distribution of the ($J/\psi,\Lambda$) system for
all $\Lambda^0_b$ candidates. The signal is described by a Gaussian function and the background by a constant plus an exponential function.}
\label{fig:mass_jpsilam}
\end{center}
\end{figure}

Figure~\ref{fig:mass_jpsilam} shows the invariant mass 
of $(\mu^+\mu^-\Lambda)$
subject to the above cuts. 
The signal is modeled with a Gaussian function, and 
the background by a constant plus an exponential function. The exponential function is taken for the fact that there is a large shoulder on the left of the signal due to partially reconstruced $B$ mesons. 
The number of signal events 
extracted from the fit is $ 48 \pm  11$ with a mean
of $5630 \pm 10$~MeV and a width of $\sigma = 39 \pm 8$~MeV.

\begin{figure}[h!tb]
\begin{center}
\includegraphics[height=6.0cm,width=60mm]{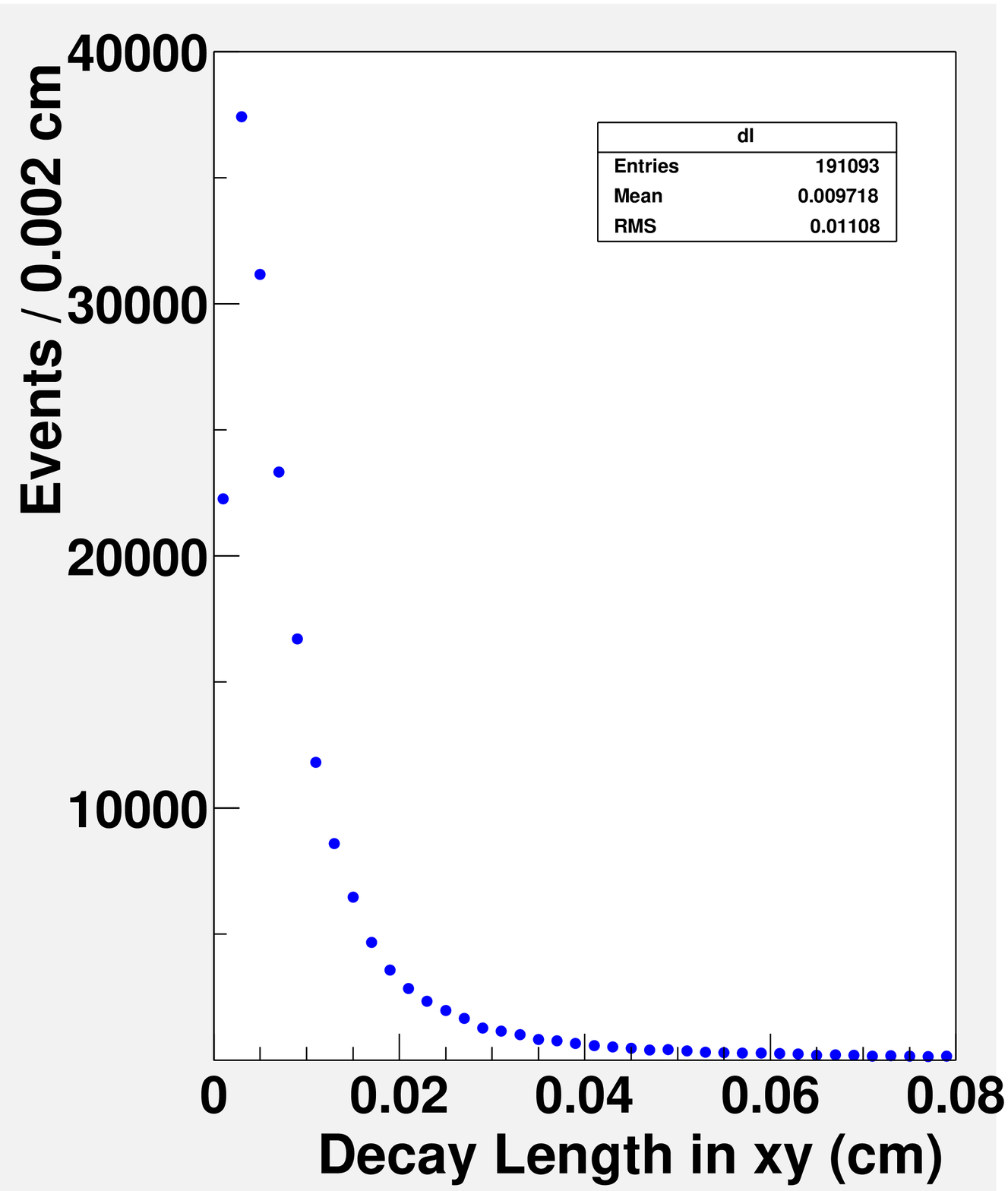}
\includegraphics[height=6.0cm,width=60mm]{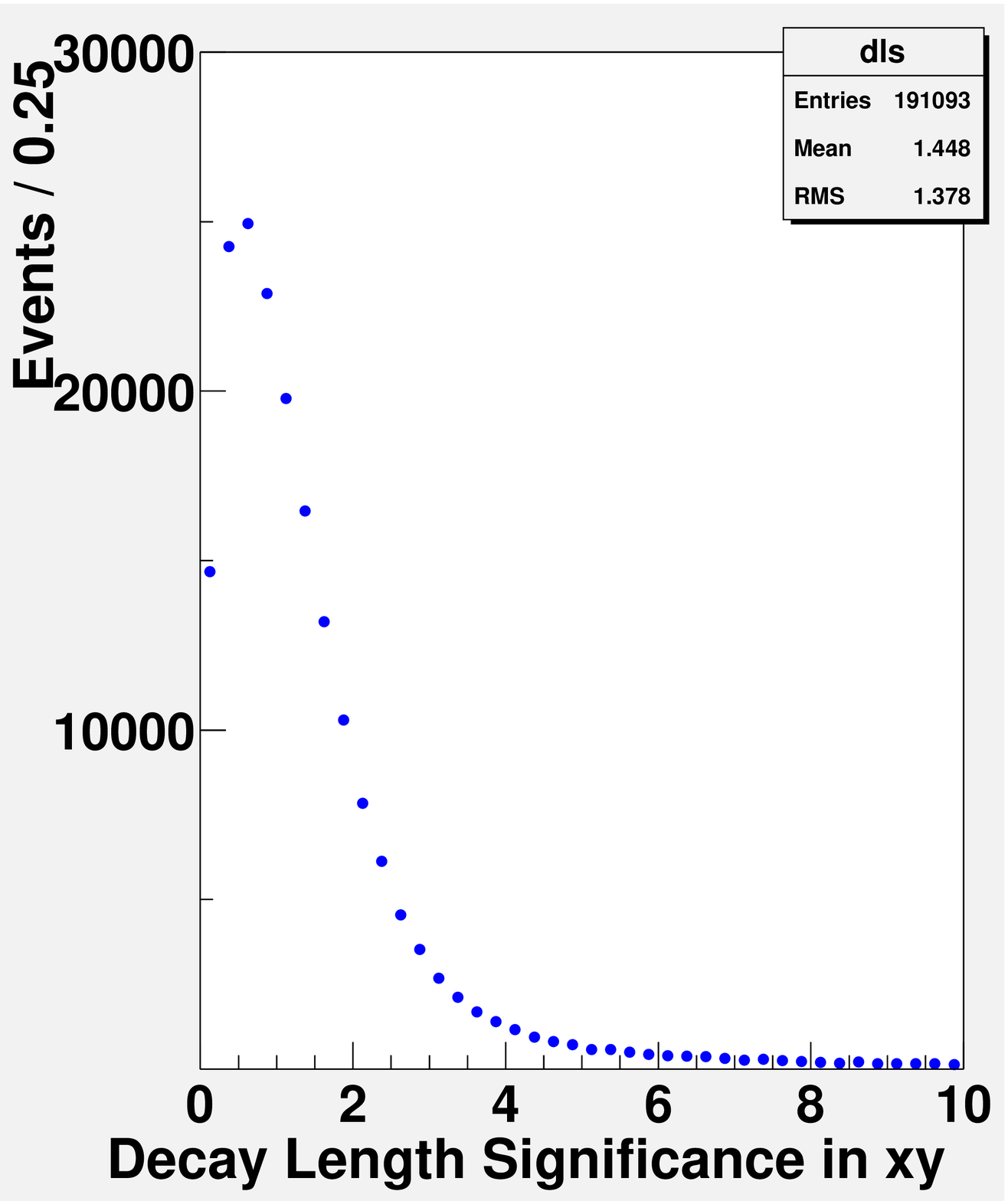}
\caption{Top: decay length distribution; and bottom: 
decay length significance distribution for data.}
\label{fig:decay_and_sig}
\end{center}
\end{figure}

To check the validity of the signal for the presence of 
long lived signal for lifetime measurement purposes, 
the decay length cut and decay length significance cut were varied.
Figure~\ref{fig:decay_and_sig} shows the decay length and decay length 
significance in data.

A transverse signed decay length is defined using the vector
between the $\Lambda_b^0$ vertex and the primary vertex,
$\vec{L}_{xy} = \vec{x}_{\Lambda_b} - \vec{x}_{PV}$, and forming:
$L_{xy} = (\vec{L}_{xy} \cdot \vec{p}_T(\Lambda_b))/|\vec{p}_T(\Lambda_b)|$.

\begin{figure}[h!tb]
\begin{center}
\includegraphics[height=6.0cm,width=60mm]{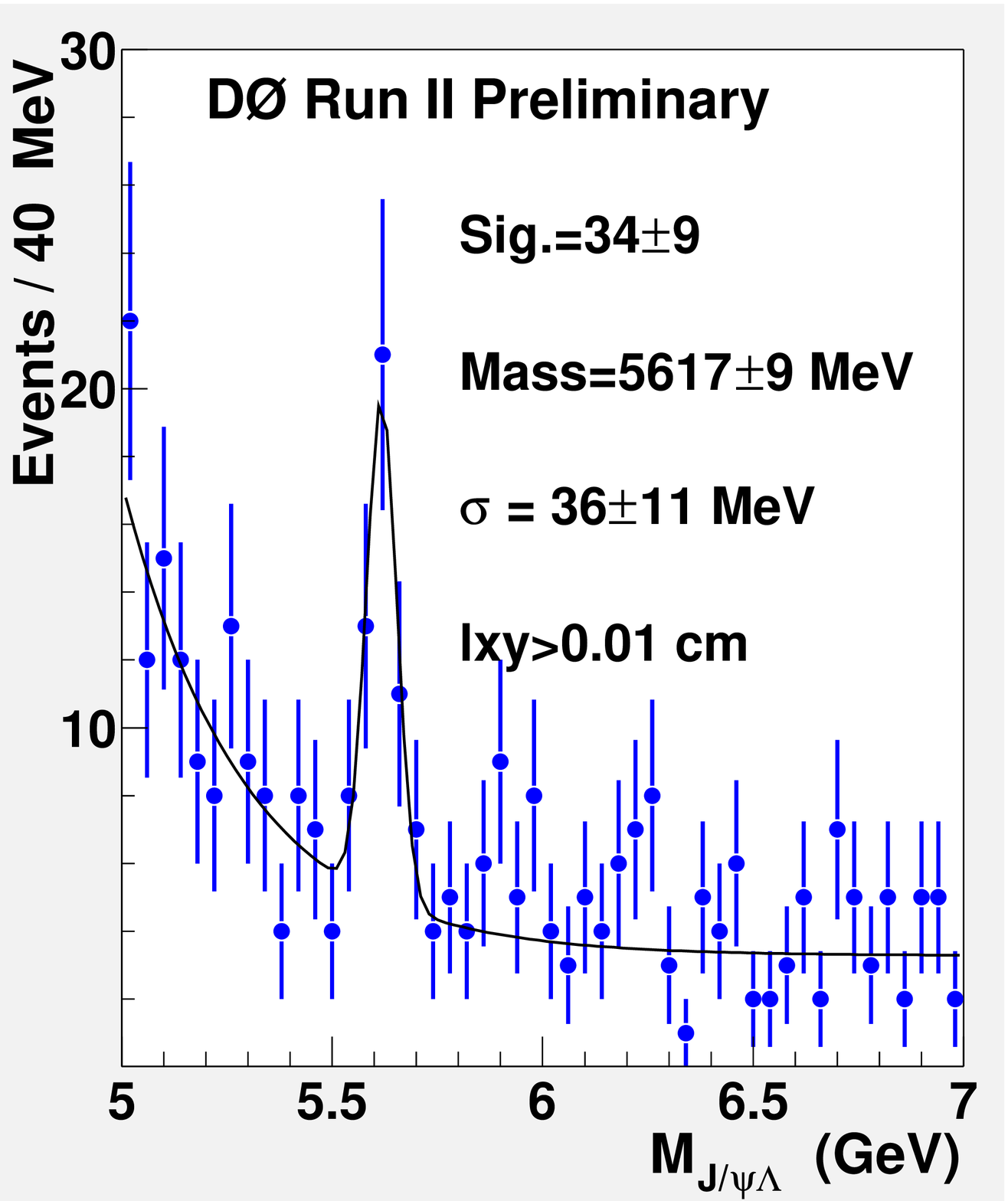}
\caption{Invariant mass distribution of the ($J/\psi,\Lambda$) system for
all $\Lambda^0_b$ candidates
plus the requirement $L_{xy} > 0.01$~cm. The signal is described by a 
Gaussian function and the background by a constant plus 
an exponential function.}
\label{fig:dl0.01}
\end{center}
\end{figure}

\begin{figure}[h!tb]
\begin{center}
\includegraphics[height=6.0cm,width=60mm]{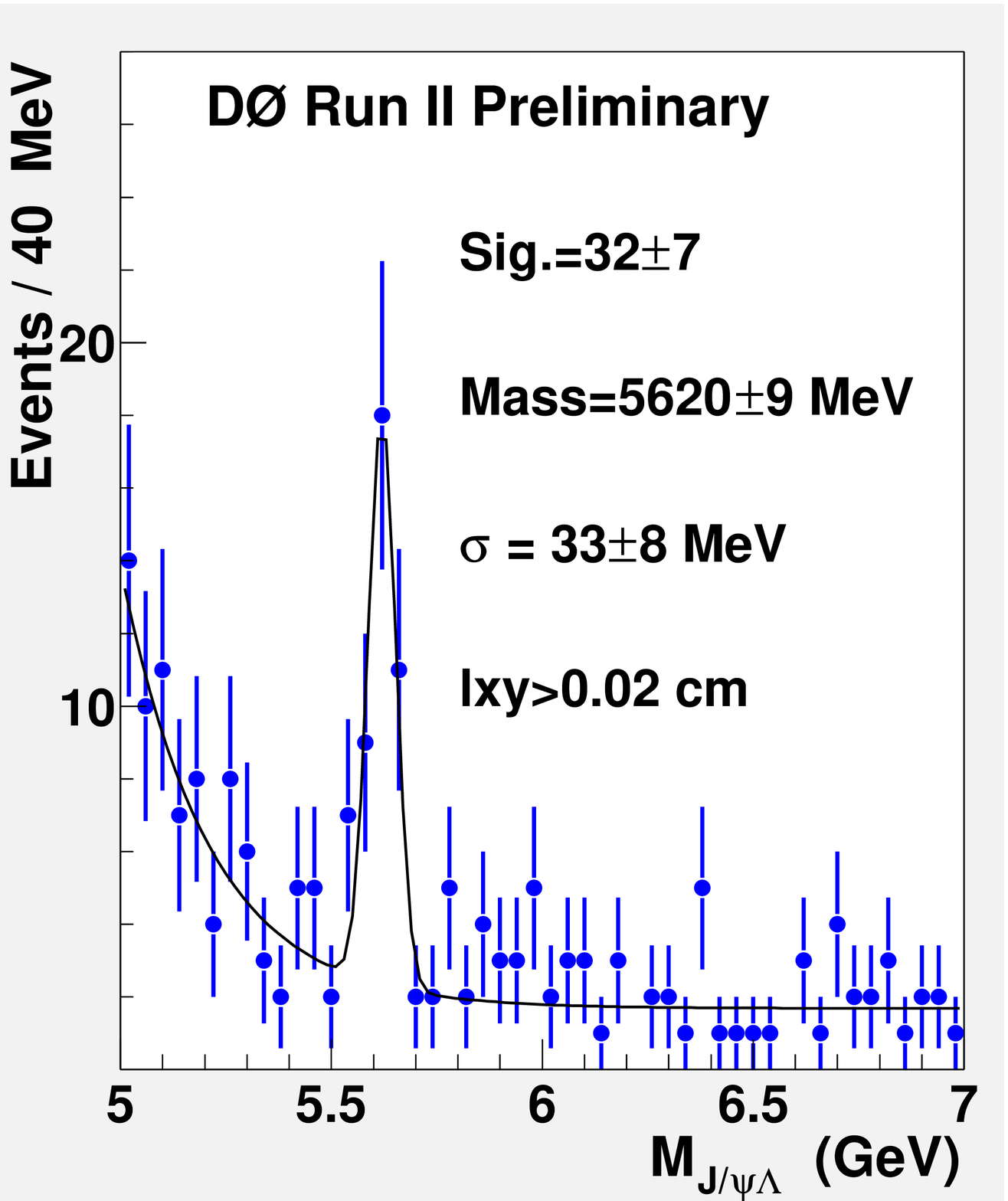}
\caption{Invariant mass distribution of the ($J/\psi,\Lambda$) system for
all $\Lambda^0_b$ candidates
plus the requirement $L_{xy} > 0.02$~cm. 
The signal is described by a Gaussian function 
and the background by a constant 
plus an exponential function.}
\label{fig:dl0.02}
\end{center}
\end{figure}

\begin{figure}[h!tb]
\begin{center}
\includegraphics[height=6.0cm,width=60mm]{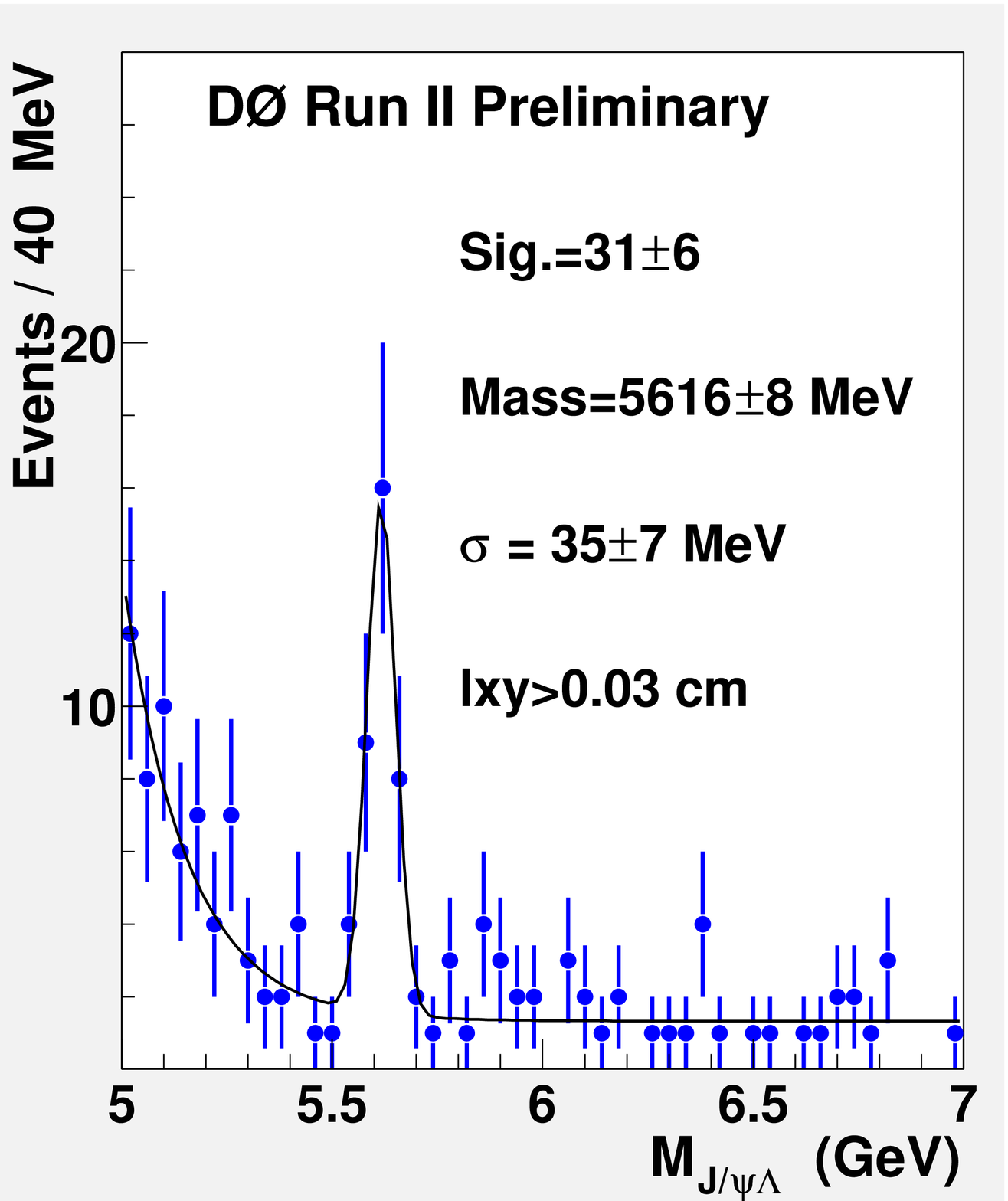}
\caption{Invariant mass distribution of the ($J/\psi,\Lambda$) system for
all $\Lambda^0_b$ candidates plus the requirement $L_{xy} > 0.03$~cm. 
ehe signal is described by a Gaussian function and the 
background by a constant plus an exponential function.}
\label{fig:dl0.03}
\end{center}
\end{figure}

Figures~\ref{fig:dl0.01}, \ref{fig:dl0.02}, and
\ref{fig:dl0.03} show the invariant mass of 
the $\mu^+\mu^-\Lambda$ system
subject to the above cuts and adding $L_{xy} > 0.01$~cm,
$L_{xy} > 0.02$~cm, and $L_{xy} > 0.03$~cm requirements, respectively.
The number of signal events extracted from the fit 
in each case is $33 \pm 8$, $33 \pm 7$, and $31 \pm 6$, respectively,
with the signal-to-noise increase in each case as expected
for a signal with lifetime.

\begin{figure}[h!tb]
\begin{center}
\includegraphics[height=12.0cm,width=100mm]{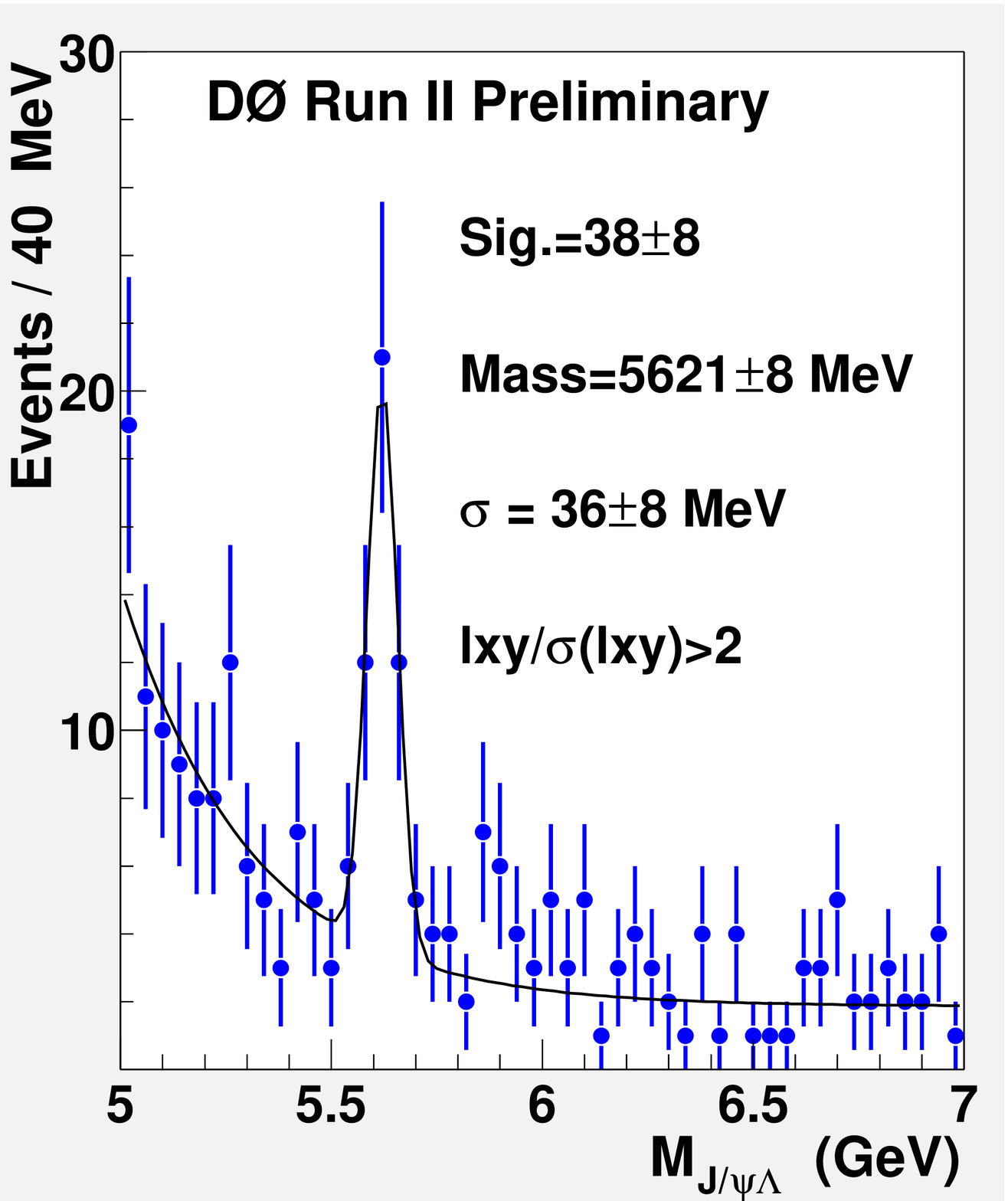}
\caption{Invariant mass distribution of the ($J/\psi,\Lambda$) system for
all $\Lambda^0_b$ candidates plus the requirement $L_{xy}/\sigma(L_{xy})>2$.
The signal is described by a Gaussian function and the background by a 
constant plus an exponential function.}
\label{fig:dls2}
\end{center}
\end{figure}

\begin{figure}[h!tb]
\begin{center}
\includegraphics[height=12.0cm,width=100mm]{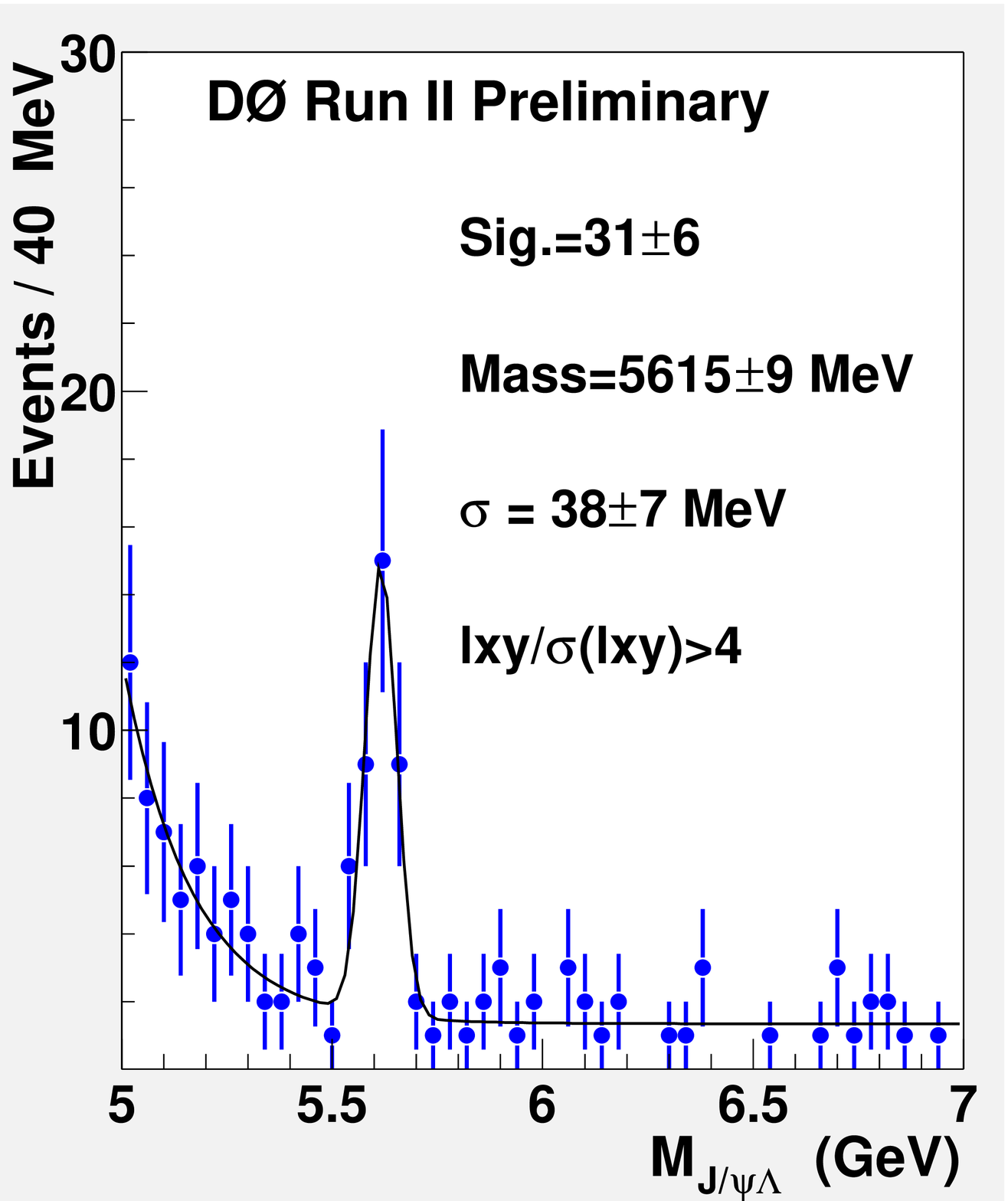}
\caption{Invariant mass distribution of the ($J/\psi,\Lambda$) system for
all $\Lambda^0_b$ candidates plus the requirement $L_{xy}/\sigma(L_{xy})>4$.
The signal is described by a Gaussian function and the 
background by a constant plus an exponential function.}
\label{fig:dls4}
\end{center}
\end{figure}

Figures~\ref{fig:dls2} and \ref{fig:dls4} show the invariant 
mass of the $\mu^+\mu^-\Lambda$ system
subject to the above cuts and adding 
$L_{xy}/\sigma(L_{xy})>2$ and $>4$, respectively.
The number of signal events extracted from each fit is $37 \pm 8$ and
$31 \pm 6$, respectively, with increased signal-to-noise, again
indicating lifetime information in the signal.

\begin{figure}[h!tb]
\begin{center}
\includegraphics[height=12.0cm,width=100mm]{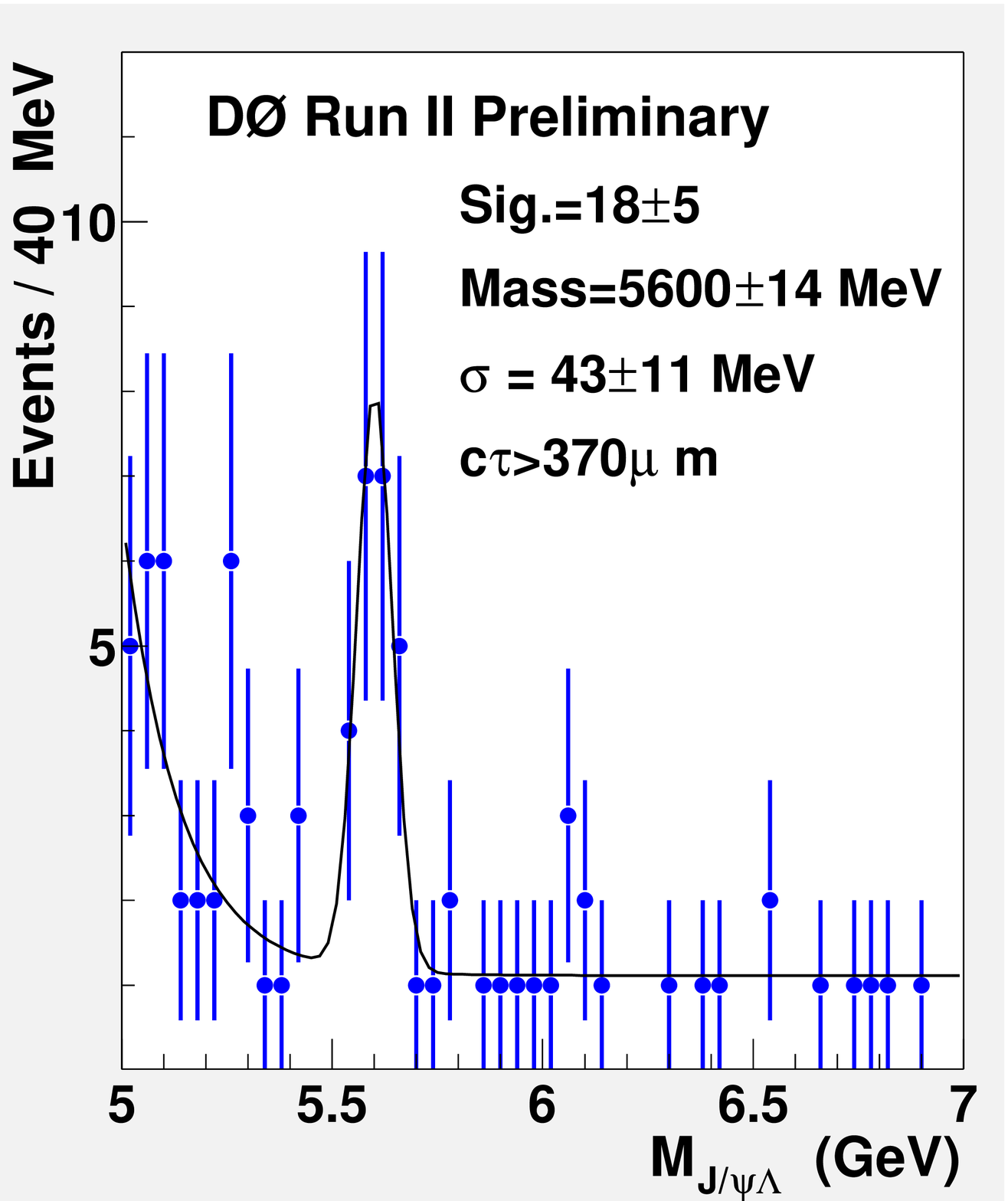}
\caption{Invariant mass distribution of the ($J/\psi,\Lambda$) system for
all $\Lambda^0_b$ candidates
plus the requirement $c \tau > 370$~$\mu$m. The signal is described by a 
Gaussian function and the background by a constant plus 
an exponential function.}
\label{fig:ctau370}
\end{center}
\end{figure}

\begin{figure}[h!tb]
\begin{center}
\includegraphics[height=12.0cm,width=100mm]{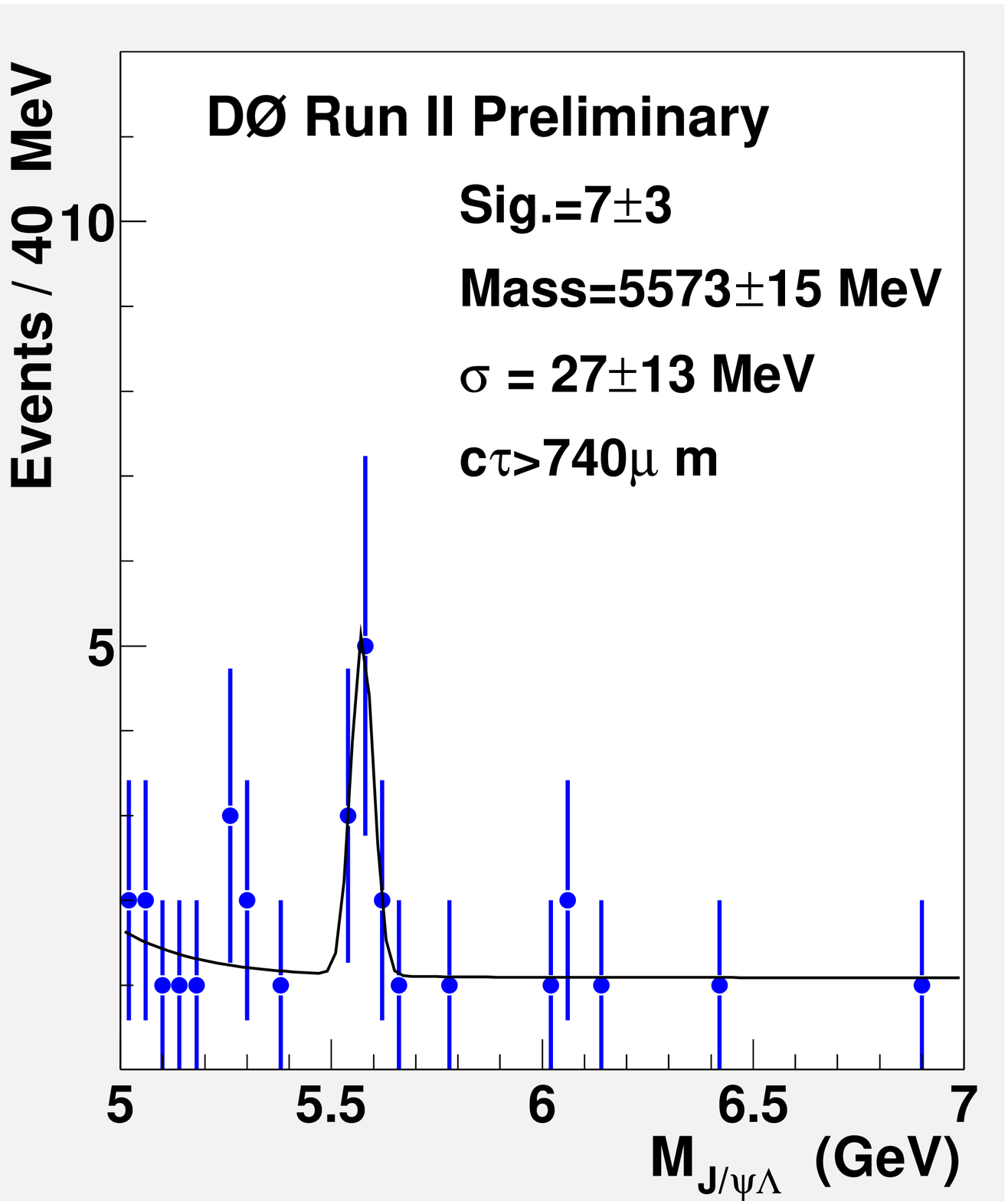}
\caption{Invariant mass distribution of the ($J/\psi,\Lambda$) system for
all $\Lambda^0_b$ candidates
plus the requirement $c \tau > 740$~$\mu$m. 
The signal is described by a Gaussian function 
and the background by a constant 
plus an exponential function.}
\label{fig:ctau740}
\end{center}
\end{figure}

\begin{figure}[h!tb]
\begin{center}
\includegraphics[height=12.0cm,width=100mm]{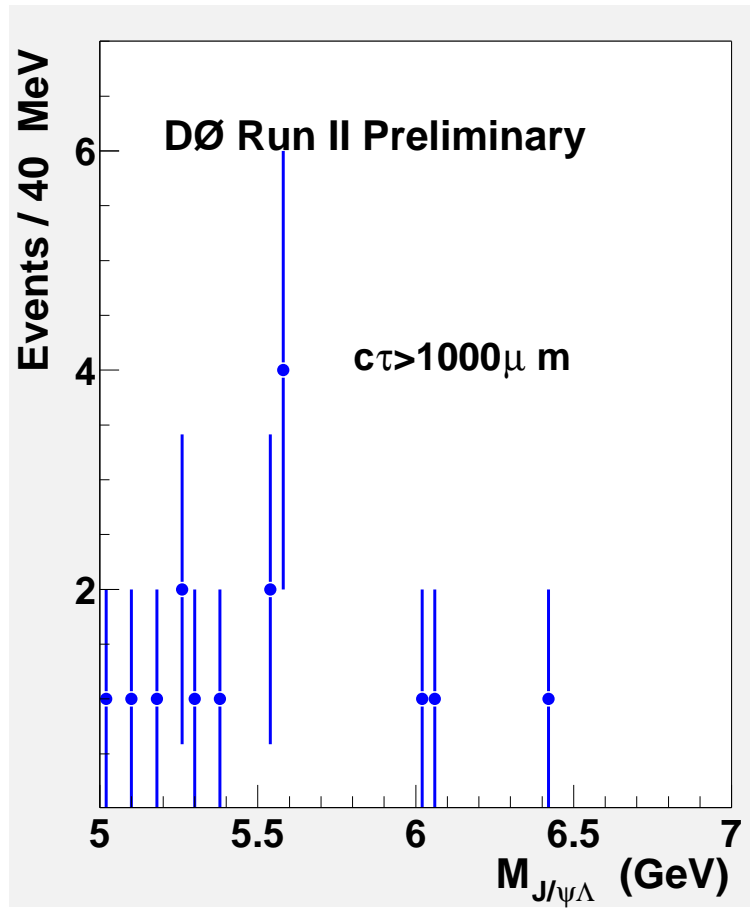}
\caption{Invariant mass distribution of the ($J/\psi,\Lambda$) system for
all $\Lambda^0_b$ candidates plus the requirement $c \tau > 1000$~$\mu$m. 
The signal is described by a Gaussian function and the 
background by a constant plus an exponential function.}
\label{fig:ctau1000}
\end{center}
\end{figure}

Figures~\ref{fig:ctau370}, \ref{fig:ctau740}, and
\ref{fig:ctau1000} show the invariant mass of 
the $\mu^+\mu^-\Lambda$ system
subject to the above cuts and adding 
proper lifetime cuts $c \tau > 370$~$\mu$m,
$c \tau > 740$~$\mu$m, and $c \tau > 1000$~$\mu$m requirements, respectively.
The number of signal events extracted from the fit 
in each case is $18 \pm 5$, $7 \pm 3$, 
and too small to allow a fit, respectively.
This fractional event loss is consistent with 
the signal having a lifetime of the order of a $B$ hadron. 

\clearpage
\subsection{Monte Carlo event samples}

\subsection{MC Signal \lbdec}

To simulate  the decay chain \lbdec, $J/\psi \rightarrow \mu ^+ \mu ^-$,
$\Lambda \rightarrow p \pi^-$, we use the {\tt evtgen} decay model
and the pythia generation program~\cite{pythia}.
Below we show the decay file used.
\begin{verbatim}
;
DECAY  LAMB
CHANNEL  0 1.000 PSI   LAM
ENDDECAY
;
DECAY  PSI
ANGULAR_HELICITY  -1   1.  0.  1.
ANGULAR_HELICITY   0   1.  0. -1.
ANGULAR_HELICITY   1   1.  0.  1.
CHANNEL  40 1.000 MU+   MU-
ENDDECAY
;
DECAY  LAM
CHANNEL  0 1.000  P+    PI-
ENDDECAY
;
\end{verbatim}



Before passing the generated events through 
the suite of programs for the detector simulation, 
hit simulation, trigger simulation, and track and particle reconstruction,
we apply the following ``pre-GEANT'' selection cuts using the d0mess
package~\cite{d0mess}:

\begin{itemize}
\item presence of the decay chain \lbbardec.
\item $p_T{(\mu)}> 1.5$~ GeV and $0.8 < |\eta| < 2.0 $ or  
      $p_T{(\mu)}> 3$~GeV and $0.8 < |\eta|$.

\end{itemize}              
``GEANT'' is a detector simulation tool, and is described in detail in Ref.~\cite{geant}.

\begin{figure}[h!tb]
\begin{center}
\includegraphics[height=12.0cm,width=100mm]{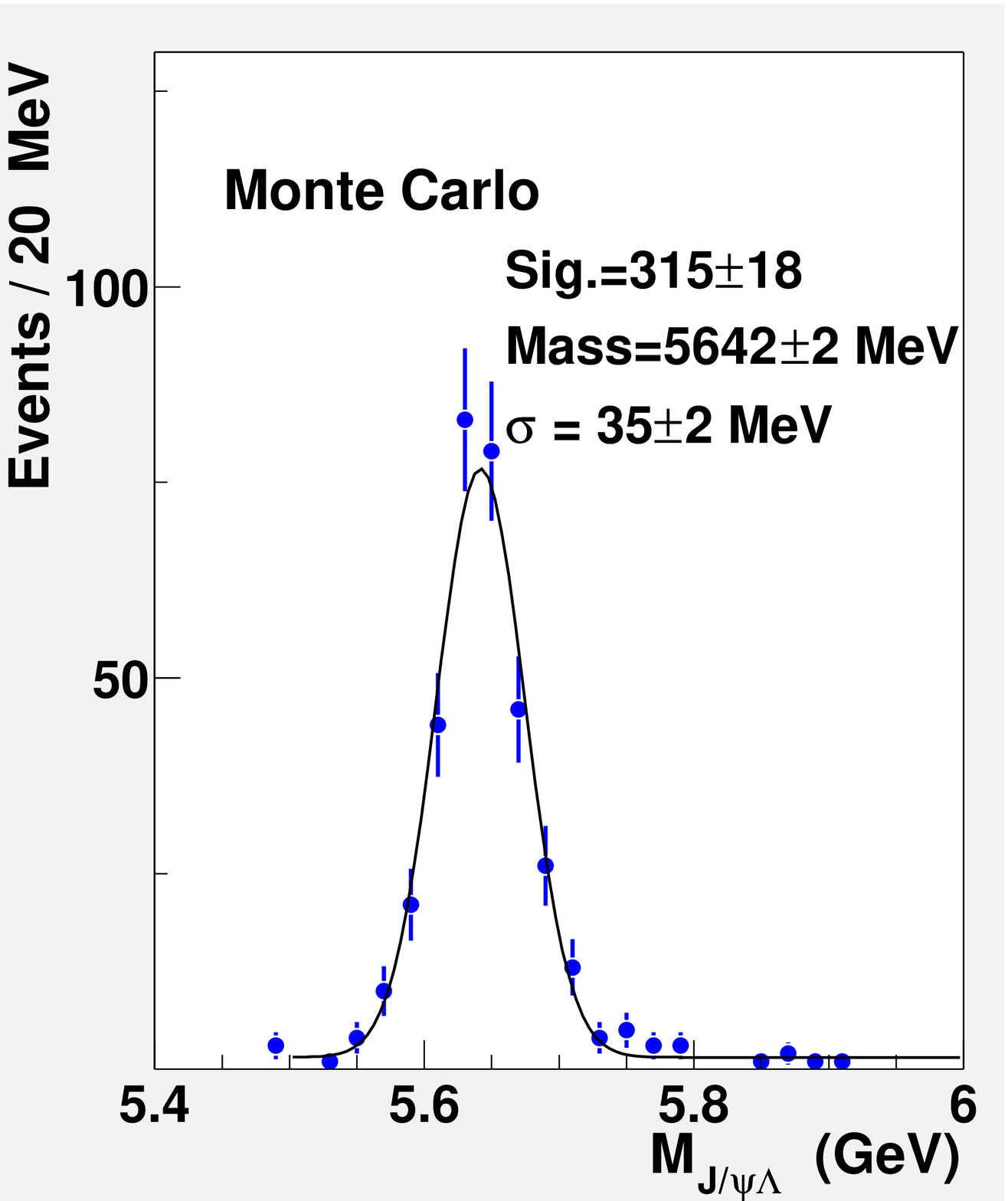}
\caption{Invariant mass distribution of   $\Lambda^0_b$ candidates
in MC simulated events.
A double Gaussian function was used to model the signal,
and a constant plus an exponential function was used for the background.}
\label{fig:lb_mc_mass}
\end{center}
\end{figure}

We apply the kinematic and quality cuts as described in
the previous section. 
After kinematic cuts, we end up with $50000$ events. 
Since the $\Lambda$ is a long-lived particle, it decays at the GEANT
processing stage, so out of $50000$ events, $31500$ events contain
the decay $\Lambda\rightarrow p \pi^-$. 
The same analysis cuts as data are applied.
The number of events passing all event selection criteria is 315.
The reconstruction efficiency after the kinematic cuts is 
$\epsilon=(1.0 \pm 0.06)\%$. 
Figure~\ref{fig:lb_mc_mass} shows the invariant mass of the combination ($\Lambda,J/\psi$) for the Monte Carlo sample. The mass and the width observed in MC 
are  M($\Lambda,J/\psi$) =5642~$\pm$~2~MeV and $\sigma =35 \pm 2$~MeV respectively. The mass is slightly lower in data compared to MC due to uncertainty in D\O\ momentum scale, 
and the fitted width observed in data is consistent with 
that obtained in the MC.

\vspace{5mm}

\begin{figure}[h!tb]
\begin{center}
\includegraphics[height=8.0cm,width=90mm]{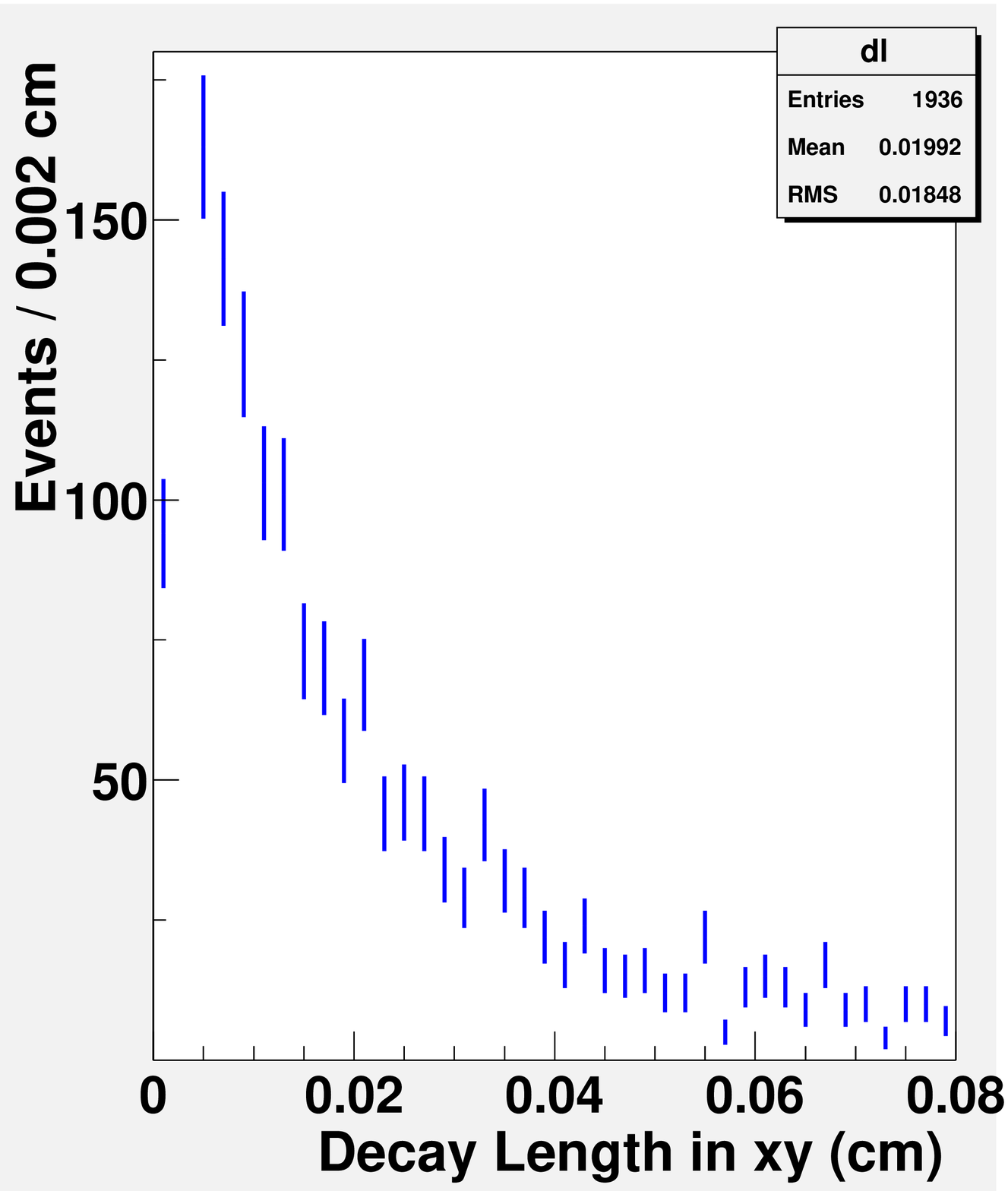}
\includegraphics[height=8.0cm,width=90mm]{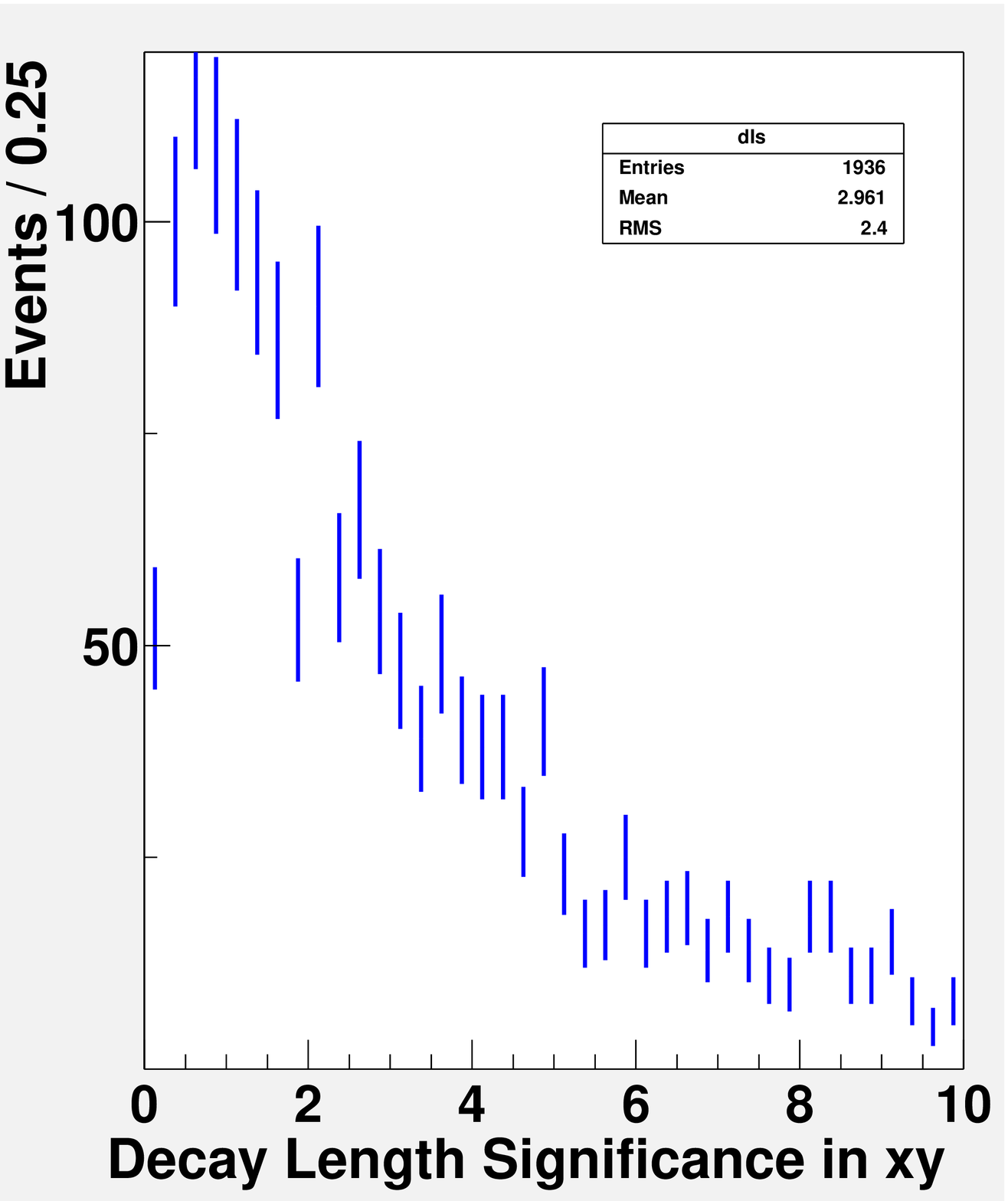}
\caption{Top: decay length distribution; bottom: decay length significance 
distribution for MC simulated events.}

\label{fig:decay_length_mc}
\end{center}
\end{figure}

Figure~\ref{fig:decay_length_mc} shows the decay length and 
decay length significance distribution from the MC 
within the mass window of \lb.
These distributions clearly indicate that the signal candidates have
a long lifetime component, as expected. 


\section{\bd\ Event Selections}

The decay \bddecks\ has similar topology to \lbdec\ decay and has higher statistics, making it a good test sample. 
Both decays are reconstructed the same way, 
the only difference is $K^0_{S}$ is that a reconstructed
$K^0_S \rightarrow \pi^+ \pi^-$ is combined with the
$J/\psi$ instead of a $\Lambda$.
To select $K^0_{S} \rightarrow \pi^+ \pi^-$ candidates, we require:
\begin{itemize}
\item two oppositely charged tracks forming a vertex;
\item limited number of track measurements downstream ($<1$) of the vertex;
\item number of CFT hits $\ge 1$ for each track;
\item 0.465~GeV~$<$~Mass$(\pi^+ \pi^-) < 0.52$~GeV;
\item $p_{T}(K^0_{S})>0.4$~GeV.
\end{itemize}

\begin{figure}[h!tb]
\begin{center}
\includegraphics[height=12.0cm,width=100mm]{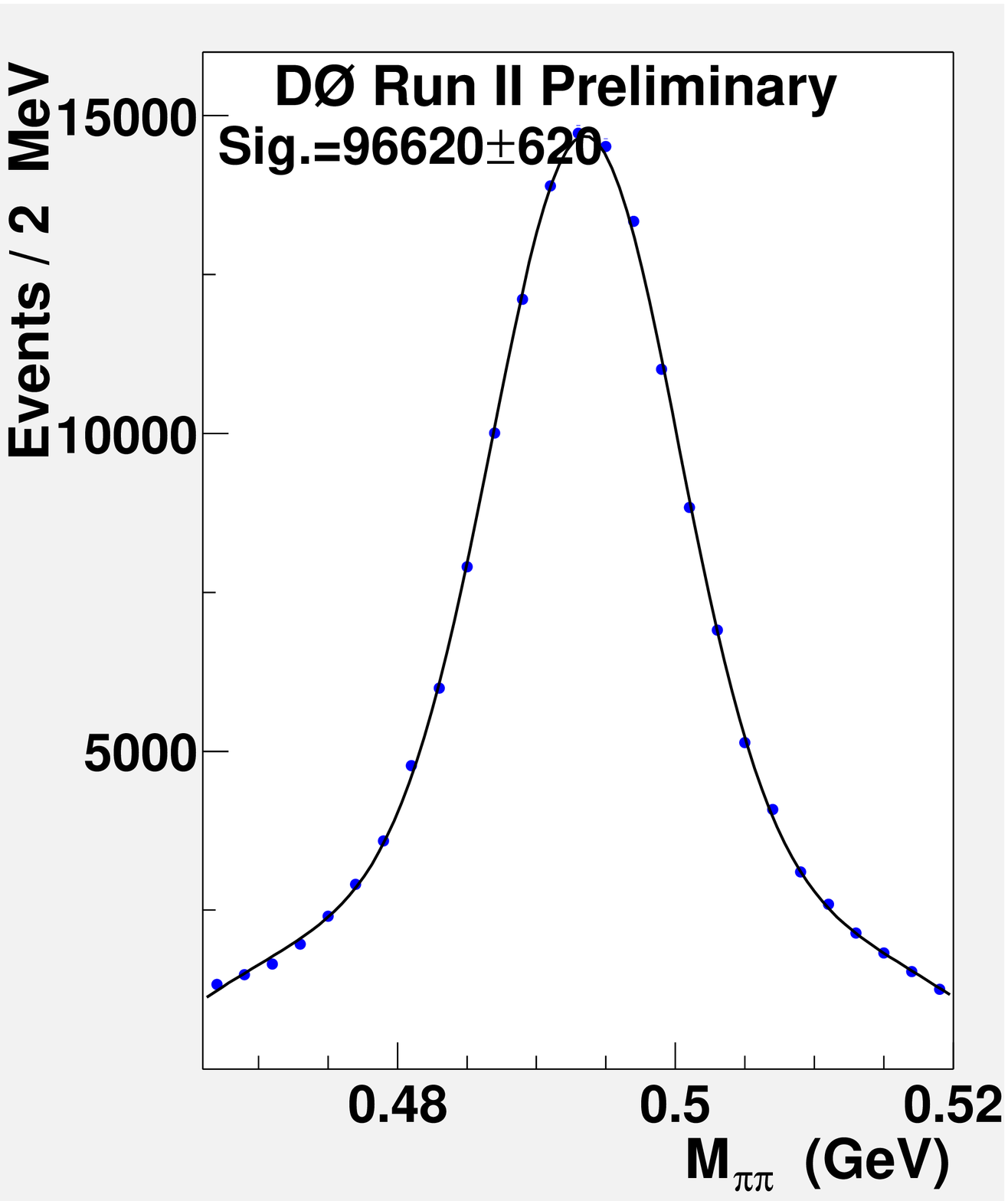}
\caption{Invariant mass distribution of the ($\pi^+\pi^-$) 
system for
all the events with a $J/\psi$ candidate. The signal is described by a double 
Gaussian function and the background by a second order polynomial. 
The signal mean is $493.5 \pm 0.1$~MeV and the fitted 
widths are $\sigma_{1}=5.9 \pm 0.1$~MeV and $\sigma_{2}=11.7 \pm 0.1$~MeV.}
\label{fig:ks_excl}
\end{center}
\end{figure}

Figure~\ref{fig:ks_excl} shows the invariant mass of $(\pi \pi)$  for all events containing a $J/\psi$ candidate. 

Once good $J/\psi$ and $K^0_{S}$ candidates have
been found in the same event,  the two muons 
from $J/\psi$ are combined with the reconstructed neutral $K^0_{S}$ track
 to form a \bd\ vertex. Further requirements are made on the vertex:

\begin{itemize}
\item vertex $\chi^2(B_{d})<25$;
\item $p(B_d^0) > 5$~GeV;
\item collinearity of \bd\ $>$~0.99, where collinearity is 
defined as the cosine of the angle between the momentum of \bd\ in the $xy$ 
plane and the direction from primary to secondary vertex 
in the $xy$ plane.
\end{itemize}

\begin{figure}[h!tb]
\begin{center}
\includegraphics[height=12.0cm,width=100mm]{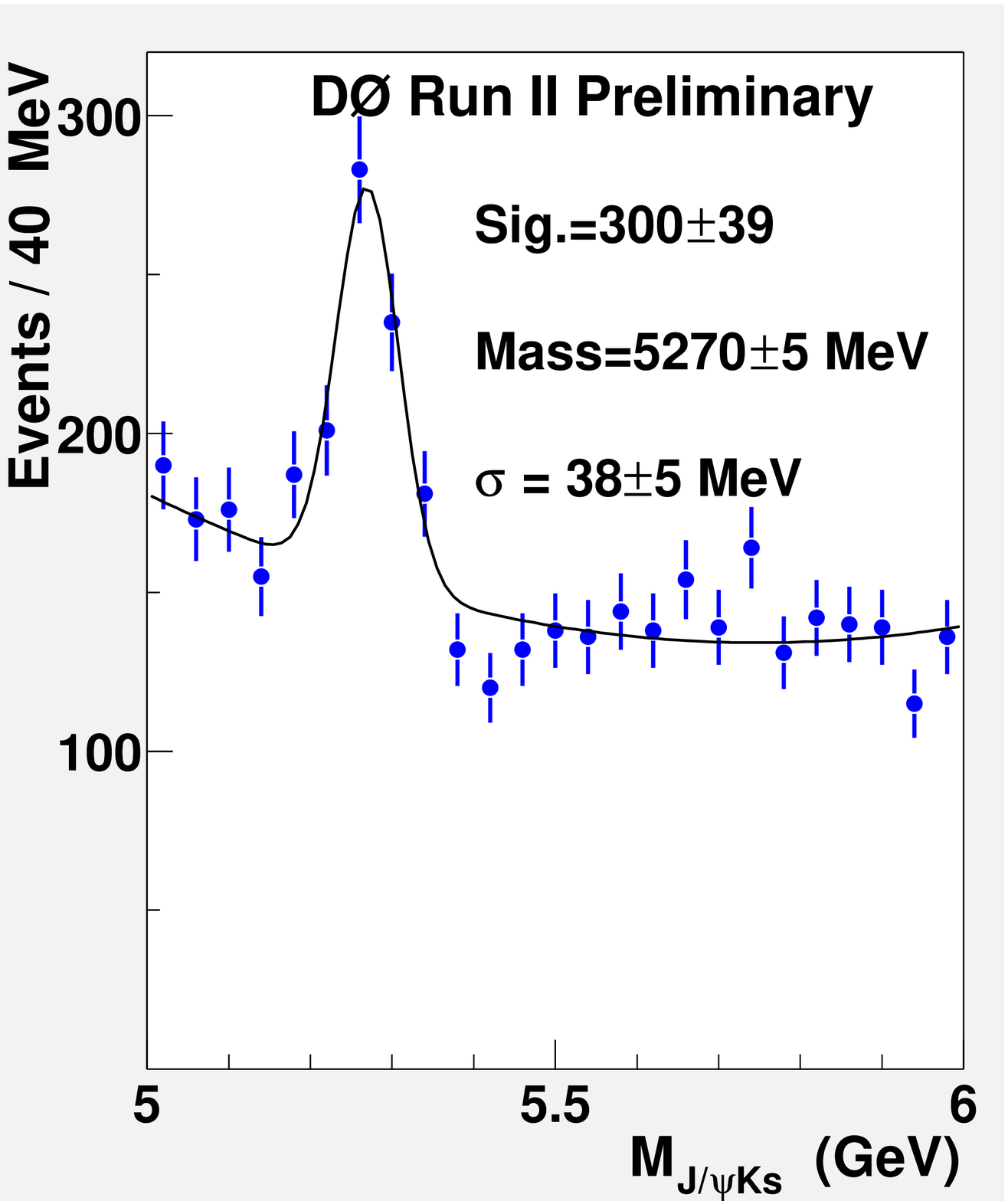}
\caption{Invariant mass distribution of the ($J/\psi,K^0_{S}$) system for
all $B^0_d$ candidates. The signal is described by a Gaussian function and the background by a second order polynomial.}
\label{fig:mass_jpsiks}
\end{center}
\end{figure}

Figure~\ref{fig:mass_jpsiks} shows the invariant mass 
of $\mu^+\mu^-K^0_{S}$
subject to the above cuts. 
The signal is modeled with a Gaussian function, and 
the background by a second-order polynomial.
The number of signal events 
extracted from the fit is $300 \pm 39$ with a mean
of $5270 \pm 5$~MeV and a width of $\sigma = 38 \pm 5$~MeV.

\begin{figure}[h!tb]
\begin{center}
\includegraphics[height=6.0cm,width=60mm]{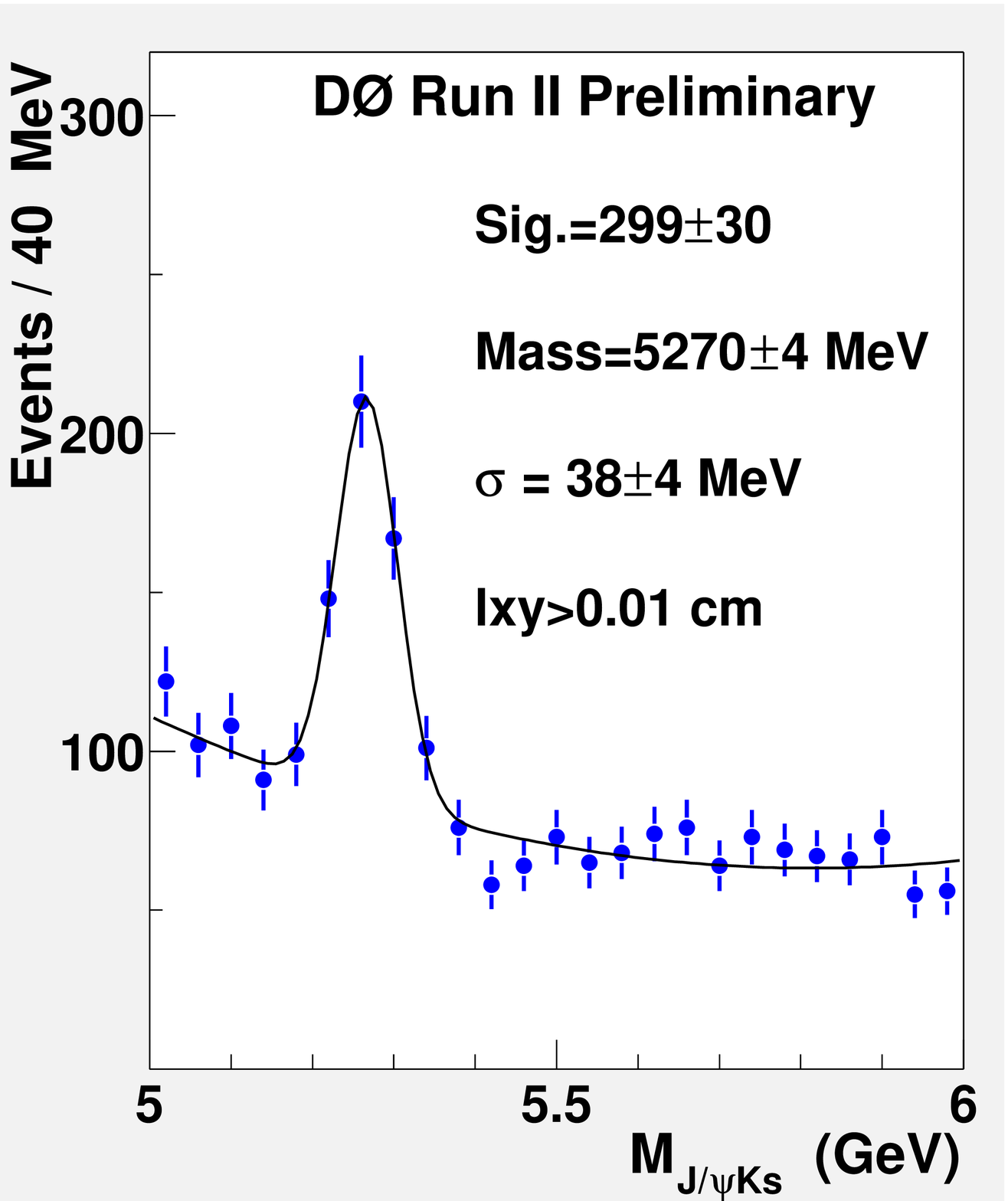}
\caption{Invariant mass distribution of the ($J/\psi,K^0_{S}$) system for
all $B^0_d$ candidates
plus the requirement $L_{xy} > 0.01$~cm. The signal is described by a 
Gaussian function and the background by a second order polynomial.}
\label{fig:ksdl0.01}
\end{center}
\end{figure}

\begin{figure}[h!tb]
\begin{center}
\includegraphics[height=6.0cm,width=60mm]{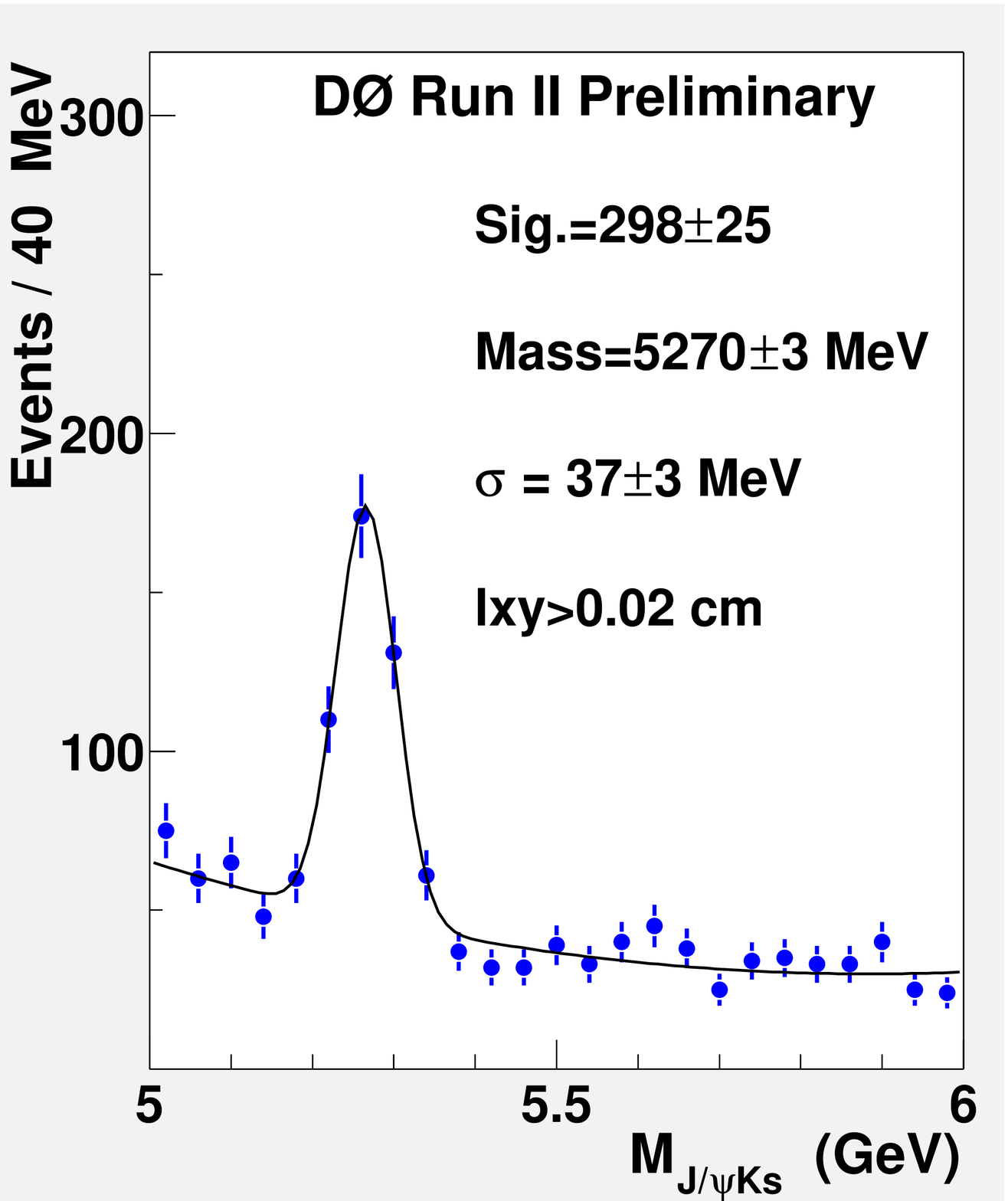}
\caption{Invariant mass distribution of the ($J/\psi,K^0_{S}$) system for
all $B^0_d$ candidates
plus the requirement $L_{xy} > 0.02$~cm. 
The signal is described by a Gaussian function 
and the background by a second order polynomial.}
\label{fig:ksdl0.02}
\end{center}
\end{figure}

\begin{figure}[h!tb]
\begin{center}
\includegraphics[height=6.0cm,width=60mm]{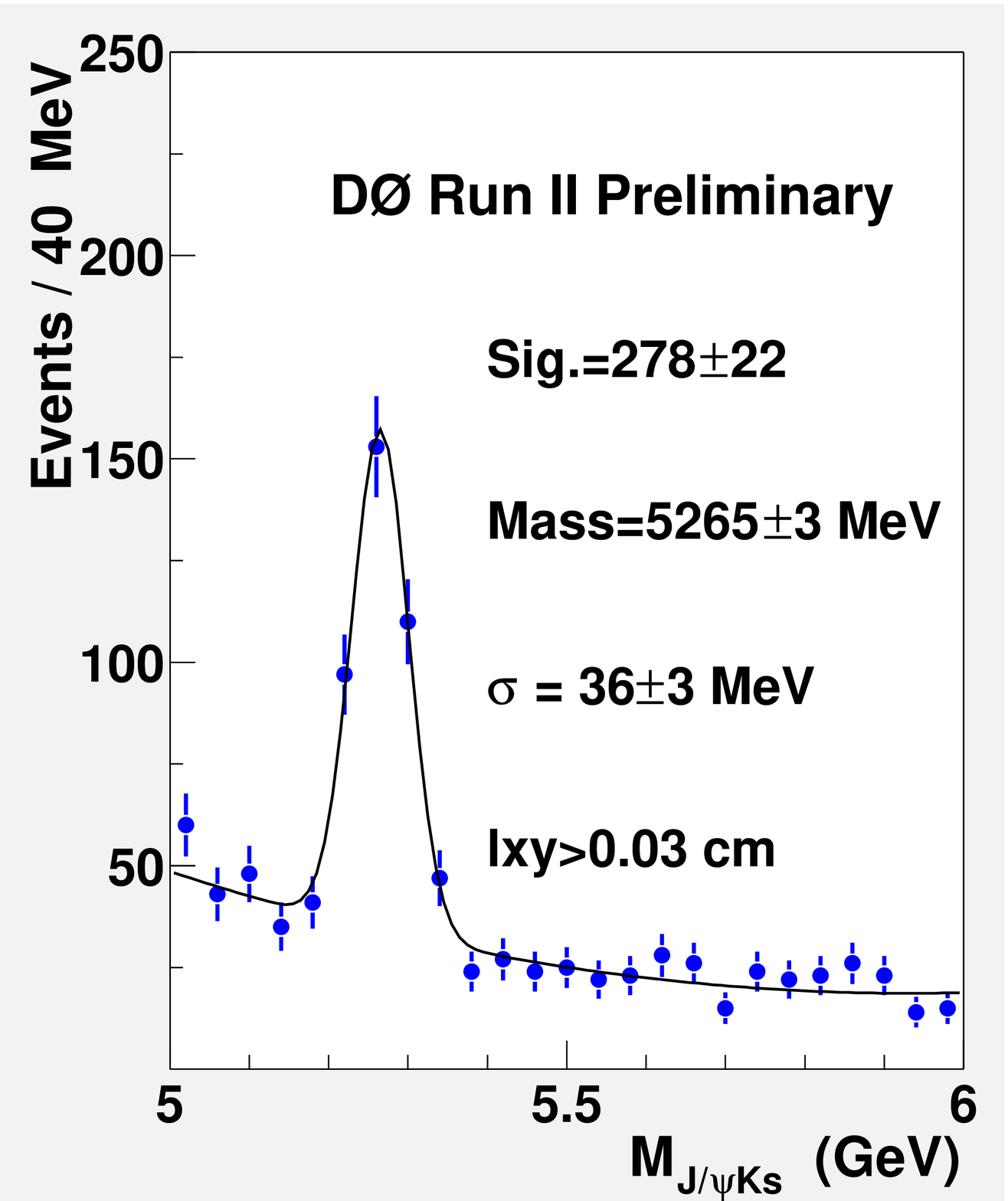}
\caption{Invariant mass distribution of the ($J/\psi,K^0_{S}$) system for
all $B^0_d$ candidates plus the requirement $L_{xy} > 0.03$~cm. 
The signal is described by a Gaussian function and the 
background by a second order polynomial.}
\label{fig:ksdl0.03}
\end{center}
\end{figure}

Figures~\ref{fig:ksdl0.01}, \ref{fig:ksdl0.02}, and
\ref{fig:ksdl0.03} show the invariant mass of 
the $\mu^+,\mu^-,K^0_{S}$ system
subject to the above cuts and adding $L_{xy} > 0.01$~cm,
$L_{xy} > 0.02$~cm, and $L_{xy} > 0.03$~cm requirements, respectively.
The number of signal events extracted from the fit 
in each case is $299 \pm 30$, $298 \pm 25$, and $278 \pm 22$, respectively. 

\begin{figure}[h!tb]
\begin{center}
\includegraphics[height=6.0cm,width=60mm]{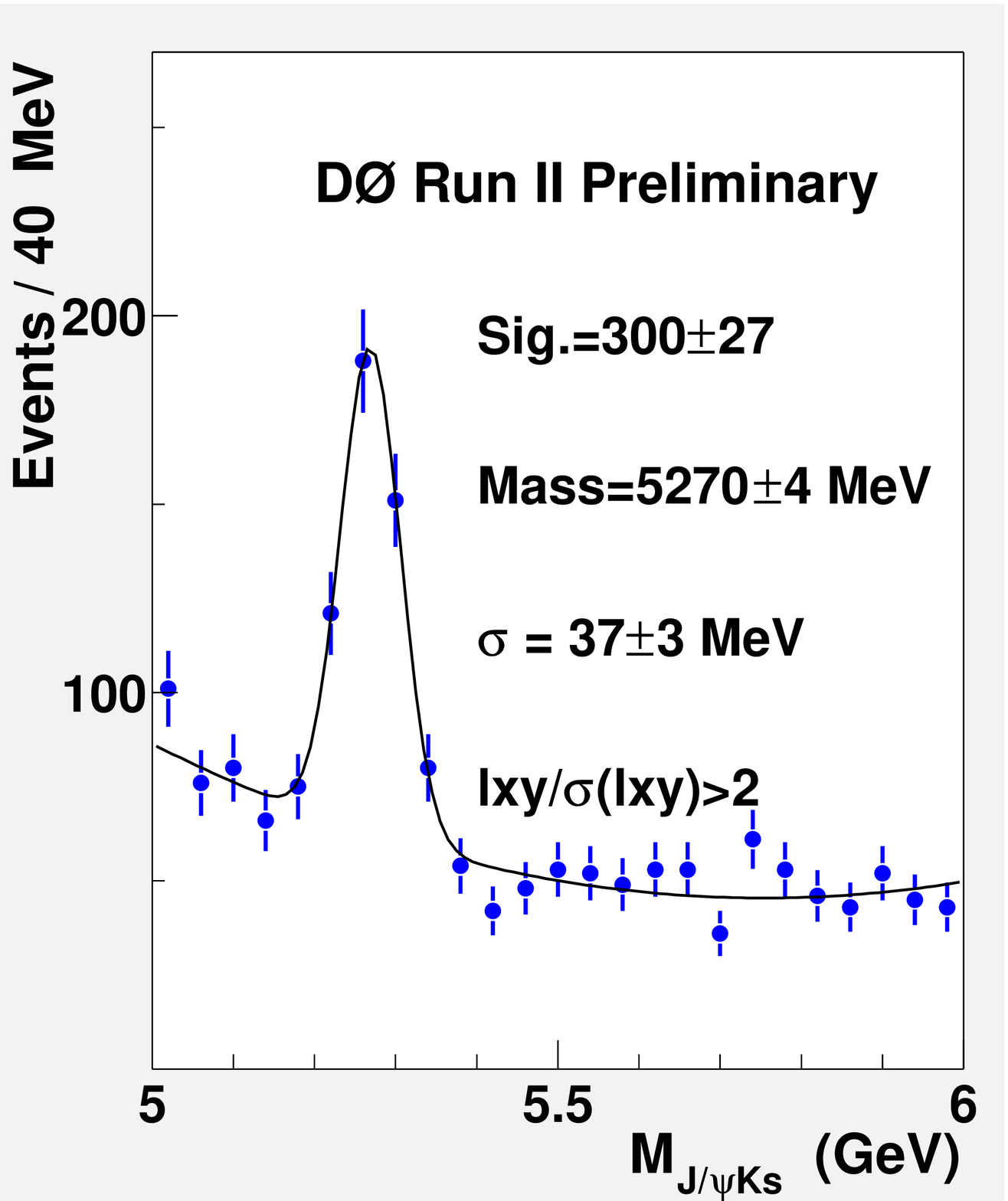}
\caption{Invariant mass distribution of the ($J/\psi,K^0_{S}$) system for
all $B^0_d$ candidates plus the requirement $L_{xy}/\sigma(L_{xy})>2$.
The signal is described by a Gaussian function and the background by a 
second order polynomial.}
\label{fig:ksdls2}
\end{center}
\end{figure}

\begin{figure}[h!tb]
\begin{center}
\includegraphics[height=6.0cm,width=60mm]{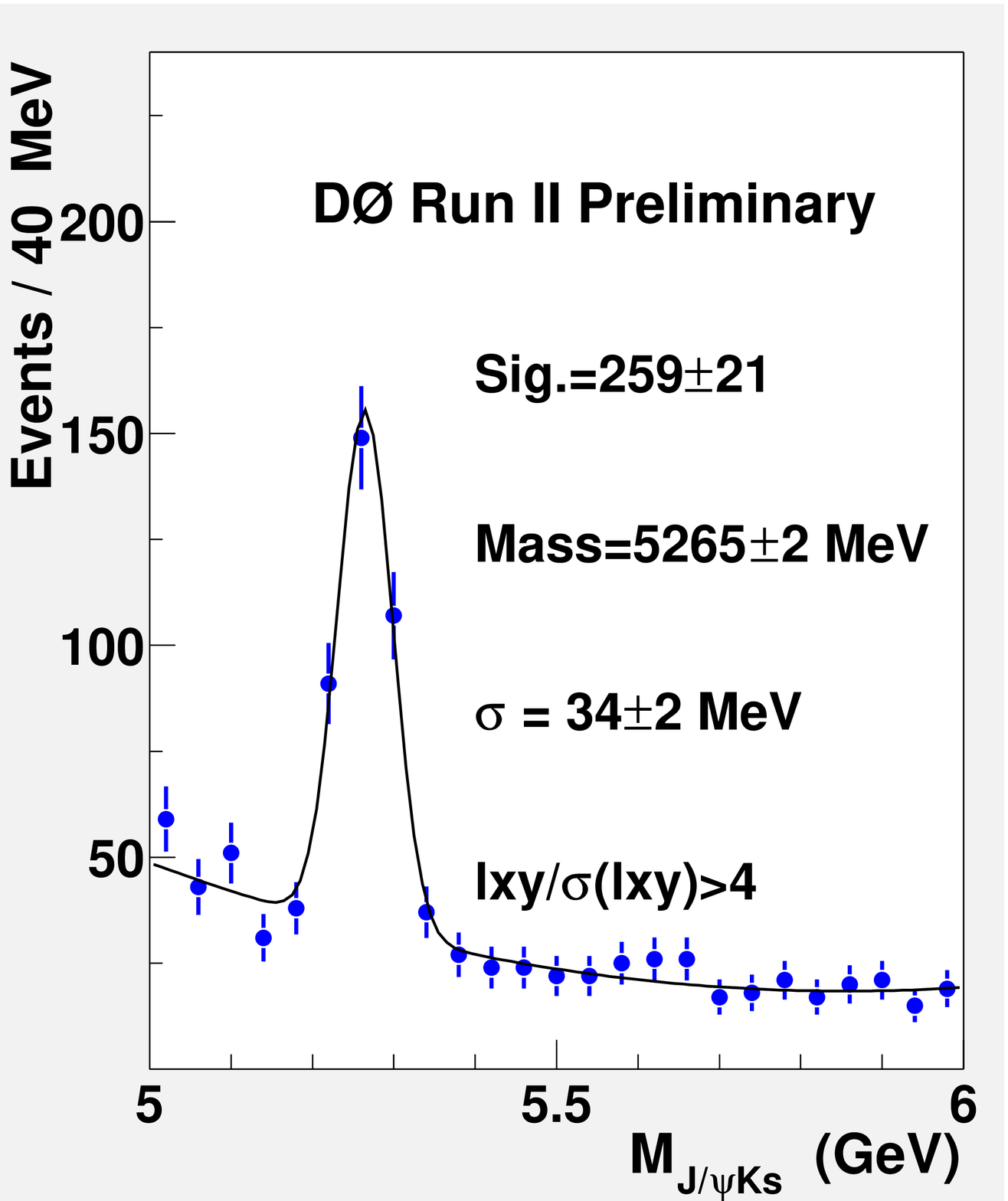}
\caption{Invariant mass distribution of the ($J/\psi,K^0_{S}$) system for
all $B^0_b$ candidates plus the requirement $L_{xy}/\sigma(L_{xy})>4$.
The signal is described by a Gaussian function and the 
background by a second order polynomial.}
\label{fig:ksdls4}
\end{center}
\end{figure}

Figures~\ref{fig:ksdls2} and \ref{fig:ksdls4} show the invariant 
mass of the $\mu^+,\mu^-,K^0_{S}$ system
subject to the above cuts and adding 
$L_{xy}/\sigma(L_{xy})>2$ and $>4$, respectively.
The number of signal events extracted from each fit is 300$ \pm $~27 and
$259 \pm 21$, respectively.

\begin{figure}[h!tb]
\begin{center}
\includegraphics[height=6.0cm,width=60mm]{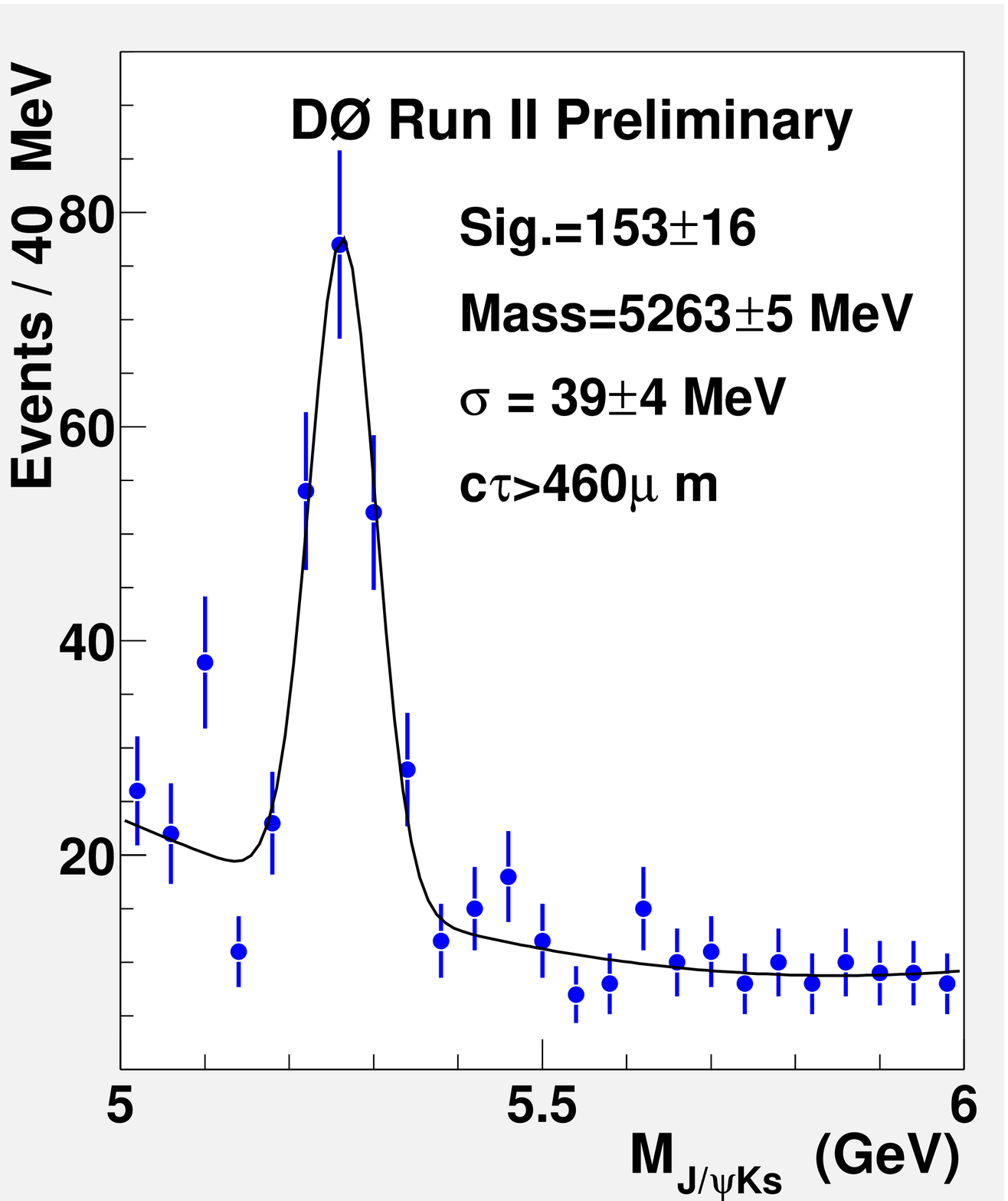}
\caption{Invariant mass distribution of the ($J/\psi,K^0_{S}$) system for
all $B^0_d$ candidates
plus the requirement $c \tau > 460$~$\mu$m. The signal is described by a 
Gaussian function and the background by a second order polynomial.}
\label{fig:ctau460}
\end{center}
\end{figure}

\begin{figure}[h!tb]
\begin{center}
\includegraphics[height=6.0cm,width=60mm]{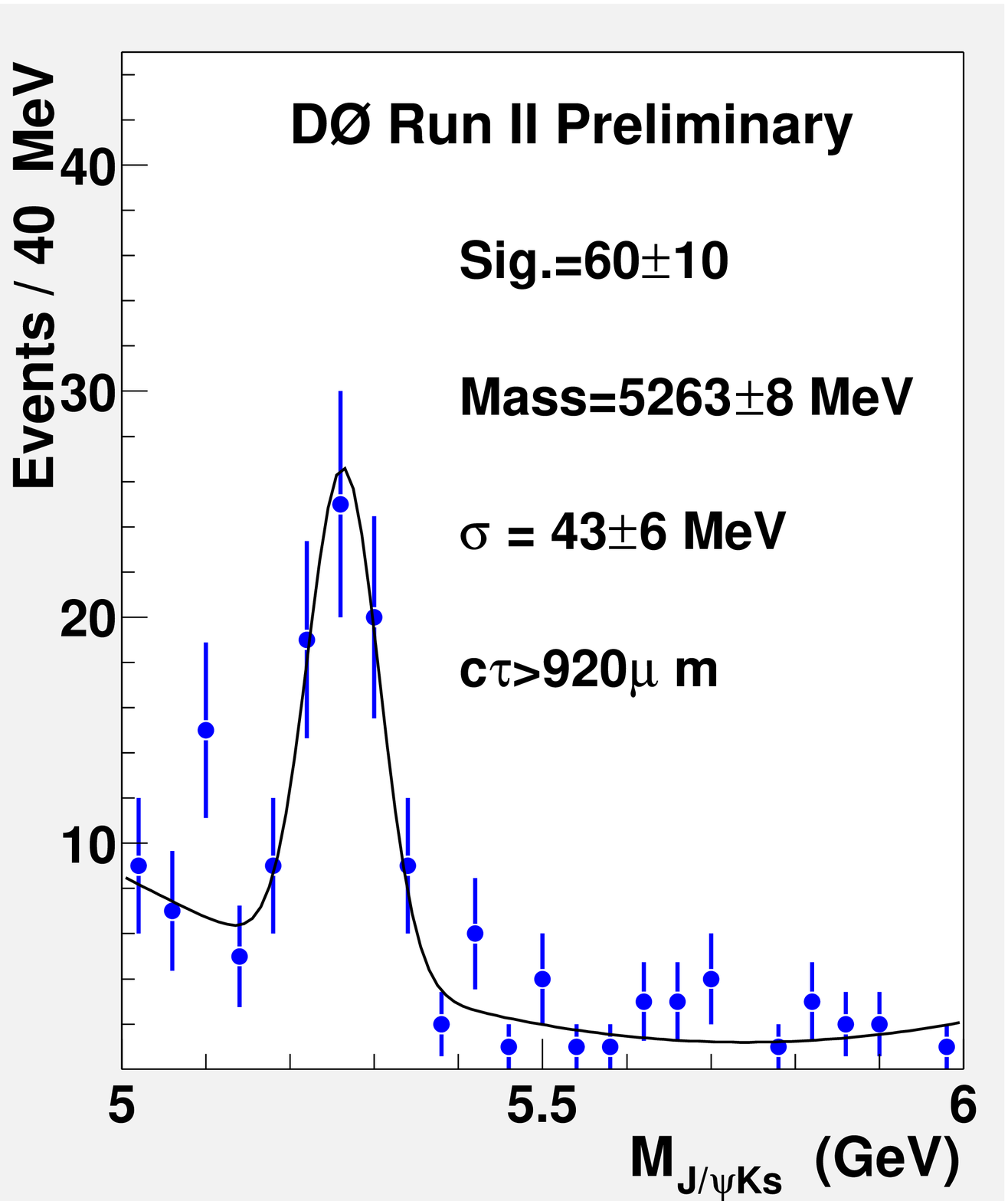}
\caption{Invariant mass distribution of the ($J/\psi,K^0_{S}$) system for
all $B^0_d$ candidates
plus the requirement $c \tau > 920$~$\mu$m. The signal is described by a 
Gaussian function and the background by a second order polynomial.}
\label{fig:ctau920}
\end{center}
\end{figure}

\begin{figure}[h!tb]
\begin{center}
\includegraphics[height=6.0cm,width=60mm]{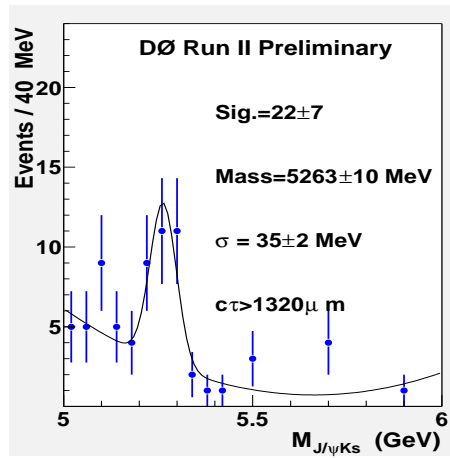}
\caption{Invariant mass distribution of the ($J/\psi,K^0_{S}$) system for
all $B^0_d$ candidates
plus the requirement $c \tau > 1320$~$\mu$m. The signal is described by a 
Gaussian function and the background by a second order polynomial.}
\label{fig:ctau1320}
\end{center}
\end{figure}

Figures~\ref{fig:ctau460}, \ref{fig:ctau920}, and
\ref{fig:ctau1320} show the invariant mass of 
the $\mu^+,\mu^-,K^0_{S}$ system
subject to the above cuts and adding $c \tau > 460$~$\mu$m,
$c \tau > 920$~$\mu$m, and $c \tau > 1320$~$\mu$m requirements, respectively.
The number of signal events extracted from the fit 
in each case is $153 \pm 16$, $60 \pm 10$, and $22 \pm 7$, respectively. These numbers are consistent with an exponentially falling $B$ lifetime. 

\chapter{$\Lambda^0_b$ and $B_d$ Lifetime Measurement}
\label{ch:lifelb}
\section{Proper Decay Length}

We define the signed  decay length of a $\Lambda^0_b$ baryon as the vector 
pointing
from the primary vertex to the decay vertex projected on the
$\Lambda^0_b$ momentum in the transverse plane:

\begin{equation} L^{\Lambda^0_b}_{xy} = (\vec{x}_{\Lambda^0_b} - \vec{x}_{prim})\cdot \vec{p}_T/p_T,
\end{equation}
where $\vec{p}_T$ is the measured transverse momentum vector and $p_T$ 
is its magnitude.
The primary vertex is reconstructed individually for each event as described earlier.

The proper lifetime, $\tau$, and the proper decay length, $c\tau$, are then 
defined by the relation:

\begin{equation} c\tau = L^{\Lambda^0_b}_{xy}\cdot M_{\Lambda^0_b}/p_T,
\end{equation}
where $M_{\Lambda^0_b} = 5624 \pm 9$~MeV is the world average mass of the $\Lambda^0_b$ baryon~\cite{pdg2004}.

\subsection{Proper Decay Time Distributions}

\begin{figure}[h!tb]
\begin{center}
\psfig{file=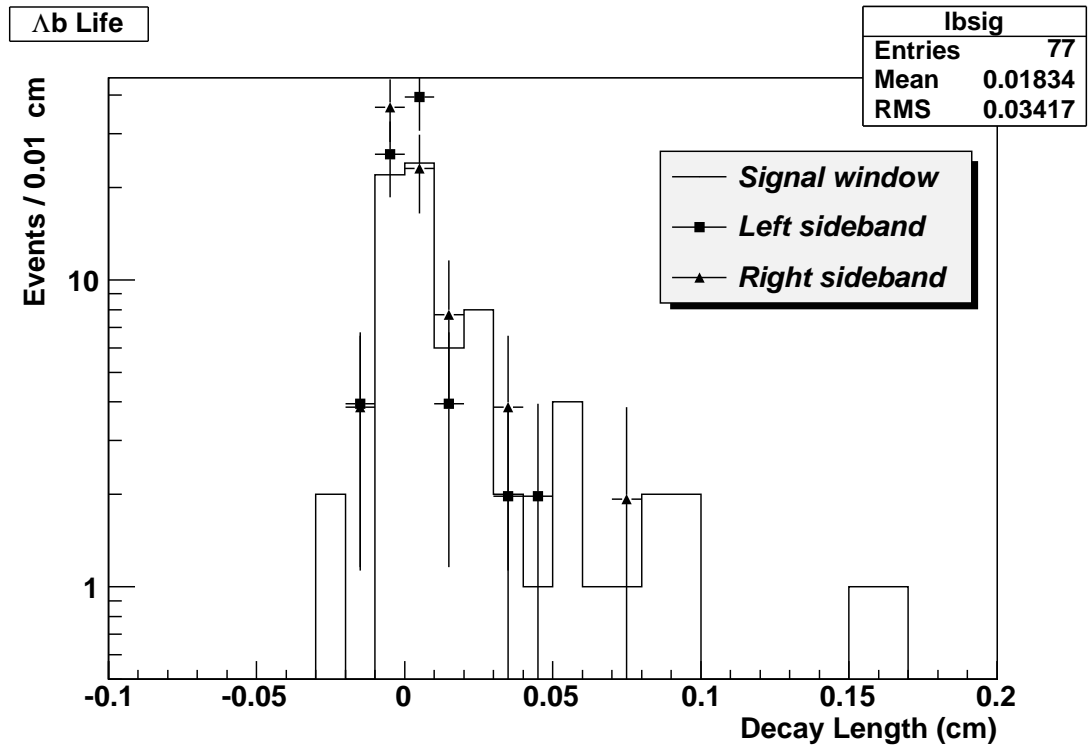,height=10.0cm}
\caption{The proper lifetime distribution in three mass bands (see text).}
\label{fig:lb_lifelh}
\end{center}
\end{figure}

We divide the $\Lambda^0_b$ mass range into three bands: ``low side'',
$5.37<M(\Lambda^0_b)<5.5$~GeV;  
``middle'', $5.5<M(\Lambda^0_b)<5.73$~GeV;
and ``high side'', $5.73<M(\Lambda^0_b)<5.87$~GeV. The middle mass band includes
the signal; all three contain comparable numbers of background
events. The proper lifetime distributions in the three
mass bands are compared in Fig.~\ref{fig:lb_lifelh}. The distributions are normalized to the number of events in the signal region.
The low and high sidebands are dominated by physics background, such 
as $B$ meson
background.   
We assume the same parametrization
of the  background shape in the entire mass region. 

\vspace{5mm}

\section{Fitting Procedure for \lb\ }

\vspace{5mm}

We use the MINUIT minimization program in root (TMinuit class).
The source code and user tips are available at~\cite{bs:chunhui_fit}.

\vspace{5mm}

The proper decay length and invariant mass distribution of the 
$\Lambda^0_b$ candidates are fit 
simultaneously using an unbinned maximum log-likelihood method. 
The likelihood function ${\cal L}$ is given by:
\begin{eqnarray}
{\cal L} & = & \prod^{N}_{i=1}[ f_{sig}{\cal F}^i_{sig} + (1-f_{sig}){\cal F}^i_{bck}],
\end{eqnarray}
where $N$=353 is the total number of events, 
 ${\cal F}^i_{sig}$ is the product of the signal mass and proper
decay-length probability density functions, ${\cal F}^i_{bck}$ is the
product of the background mass and proper decay-length probability
density functions, and $f_{sig}$ is the fraction of signal.

\vspace{5mm}

\begin{figure}[h!tb]
\begin{center}
\includegraphics[height=12.0cm,width=100mm]{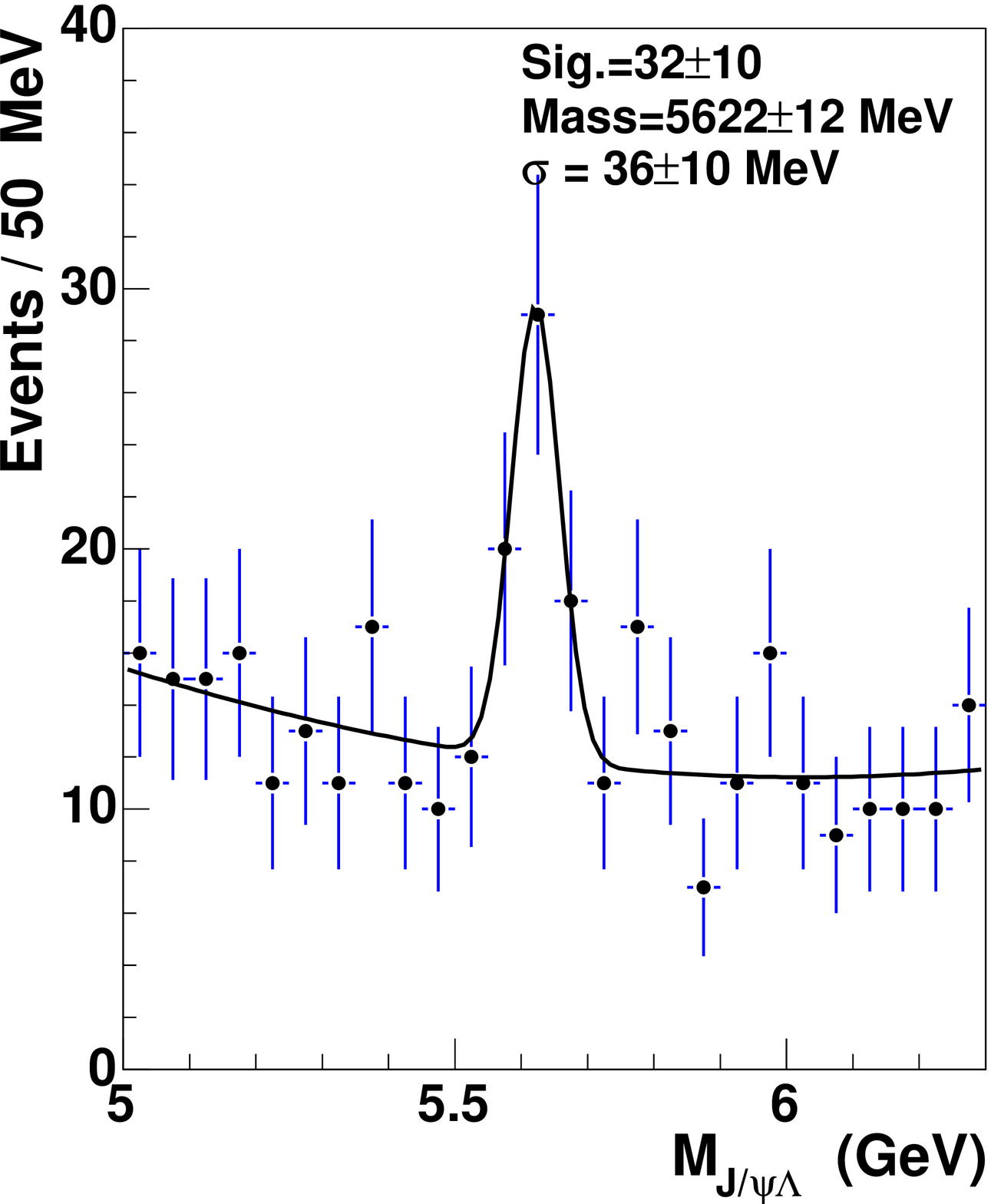}
\caption{Invariant mass distribution of the ($J/\psi,\Lambda$) system for
all $\Lambda^0_b$ candidates without any lifetime cuts. The signal is described by a Gaussian function and the background by a second order polynomial.}
\label{fig:mass_no_life}
\end{center}
\end{figure}

We use a range of 5.0~GeV~$<$~M($\Lambda^0_b$)~$<$~6.3~GeV 
for the $\Lambda^0_b$ mass window.
Previously we implemented lifetime related cuts~(collinearity and decay length related cuts), which were used to solidly establish the signal. However, for lifetime purposes we have to remove these cuts. We will use the signal with no lifetime cuts as shown in Fig~\ref{fig:mass_no_life}. The cuts used for selecting these events are:

\begin{itemize}
\item vertex $\chi^2(\Lambda_{b})<12$;
\item $p(\Lambda_b^0) > 5$~GeV;
\item $xy$ distance from $J/\psi$ vertex to 
$\Lambda$ vertex $>$~0.9~cm;
\item $p_T(\Lambda)>2.4$~GeV; and 
\item $\sigma(c\tau)<100$~fs.
\end{itemize}
These are the events used to extract the lifetime. 
Figure~\ref{fig:mass_no_life} shows the invariant mass 
of $\mu^+\mu^-\Lambda$
subject to the above cuts. 
The signal is modeled with a Gaussian function, and 
the background by a constant plus a second order polynomial.
The number of signal events 
extracted from the fit is $ 32\pm 10$ with a mean
of $5622 \pm 12$~MeV and a width of $\sigma = 36 \pm 10$~MeV.

The lifetime distribution of the signal is parameterized by an
exponential convoluted with a Gaussian function.
The lifetime resolution of background is approximated by a superposition of
a Gaussian function centered at zero, one exponential for the
negative $c\tau$ region 
and one exponential
for the positive $c\tau$ region, with free slopes. The width of the Gaussian 
functions is taken from  the  event-by-event measurement of decay length error, $\sigma_L$.
To allow for the possibility of the lifetime uncertainty to be systematically
mis-estimated, we introduce a free scale factor $s$ that multiplies the decay length error, i.e., $s \cdot \sigma_L$.    
The fraction of the signal is set to the central value obtained in the 
separate mass
fit and allowed to vary within its statistical errors. This fraction is
$f_{sig} =  N_{signal}/N_{total}$ in the defined mass window.

\subsubsection{Fit to the data}

The fit results are summarized in Table~\ref{likelihood-fitting}. 

\begin{table}
\begin{center}
\begin{tabular}{|c|c|}
\hline
Parameter   &  Central value $\pm$ error  \\
\hline \hline
{\bf Proper decay length, $c\tau$}          & {\bf $408 \pm 89\mu$m} \\
\hline
Signal fraction, $f_{sig}$  &  0.091 $\pm$ 0.028 \\
\hline
Mass of $\Lambda_b$, {\it M}($\Lambda_b$)  & 5622$\pm$12~MeV \\
\hline
Width of $\Lambda_b$, $\sigma(\Lambda_b)$  & 36$\pm$ 10~MeV \\
\hline
Mass Slope(0), a0             & $-0.31\pm 0.031$\\
\hline
Mass Slope(1), a1             & $0.026 \pm 0.028$\\
\hline
Error scale factor, $s$       &  1.7~$\pm$~0.1    \\
\hline
$slope-bkg-neg$  &  88~$\pm$~52~$\mu$m\\
\hline
$slope-bkg-pos$  &  423~$\pm$~114 $\mu$m\\
\hline
$norm-bkg-neg$   &  0.02~$\pm$~0.02  \\
\hline
$norm-bkg-pos$   &  0.1~$\pm$~0.02  \\
\hline
\end{tabular}
\caption {Unbinned maximum likelihood fitting results for the \lb\ signal.}
\label{likelihood-fitting}
\end{center}
\end{table}

\vspace{5mm}

\begin{figure}[h!tb]
\begin{center}
\psfig{file=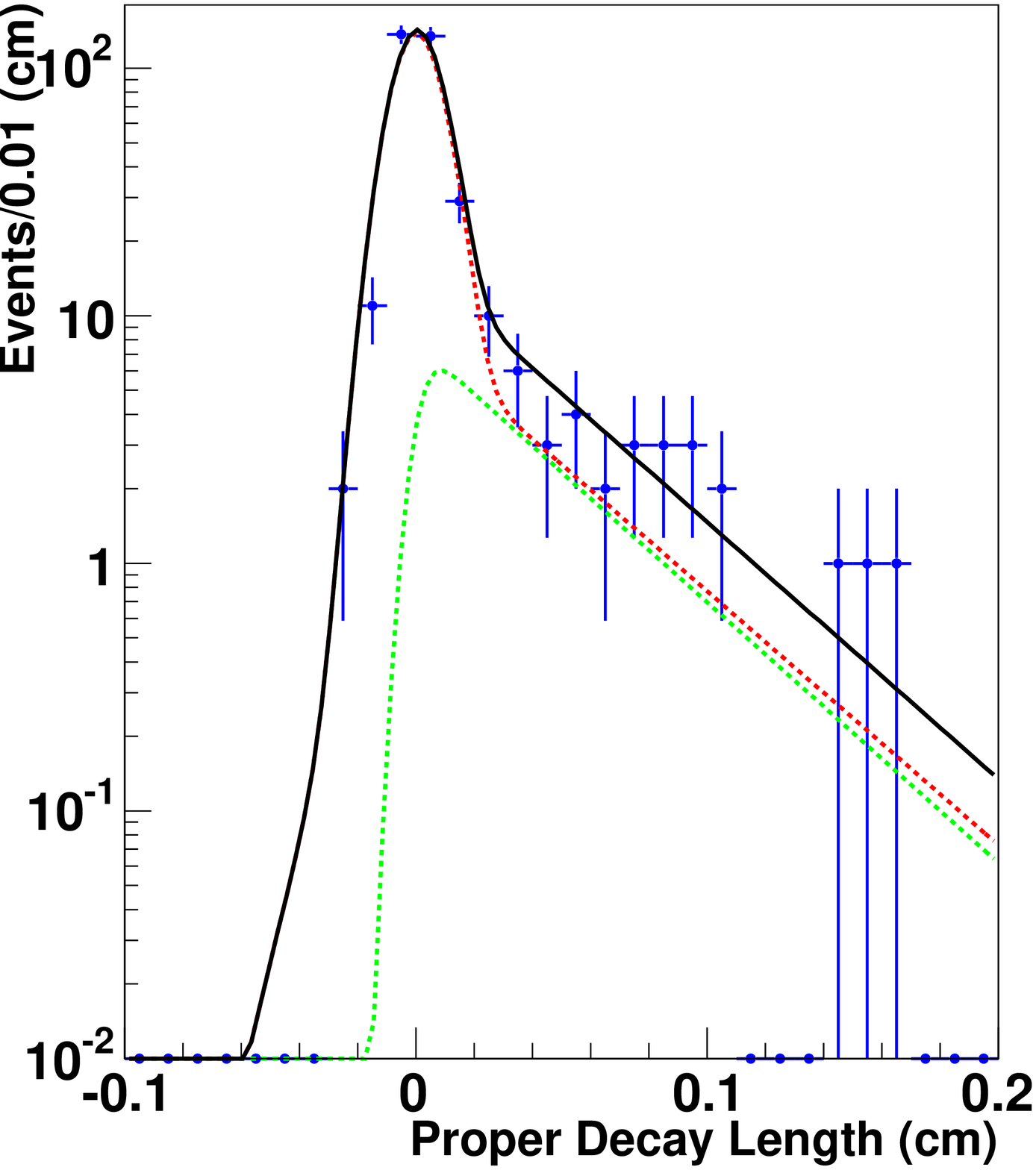,height=8.0cm,width=100mm}
\caption{The proper decay length, $c\tau$,  of the $\Lambda^0_b$ candidates. 
The curves show:  the signal contribution,  dotted (green); 
the background, dashed (red); and total, solid (black).}
\label{fig:lb_lifetime_fit}
\end{center}
\end{figure}

The lifetime distribution with the fit results overlaid is shown 
in Fig.~\ref{fig:lb_lifetime_fit}. In the figure, for illustration purposes, we assume a constant decay length resolution of 50~$\mu$m, which is the average value observed in the 
data. 

\begin{figure}[h!tb]
\begin{center}
\psfig{file=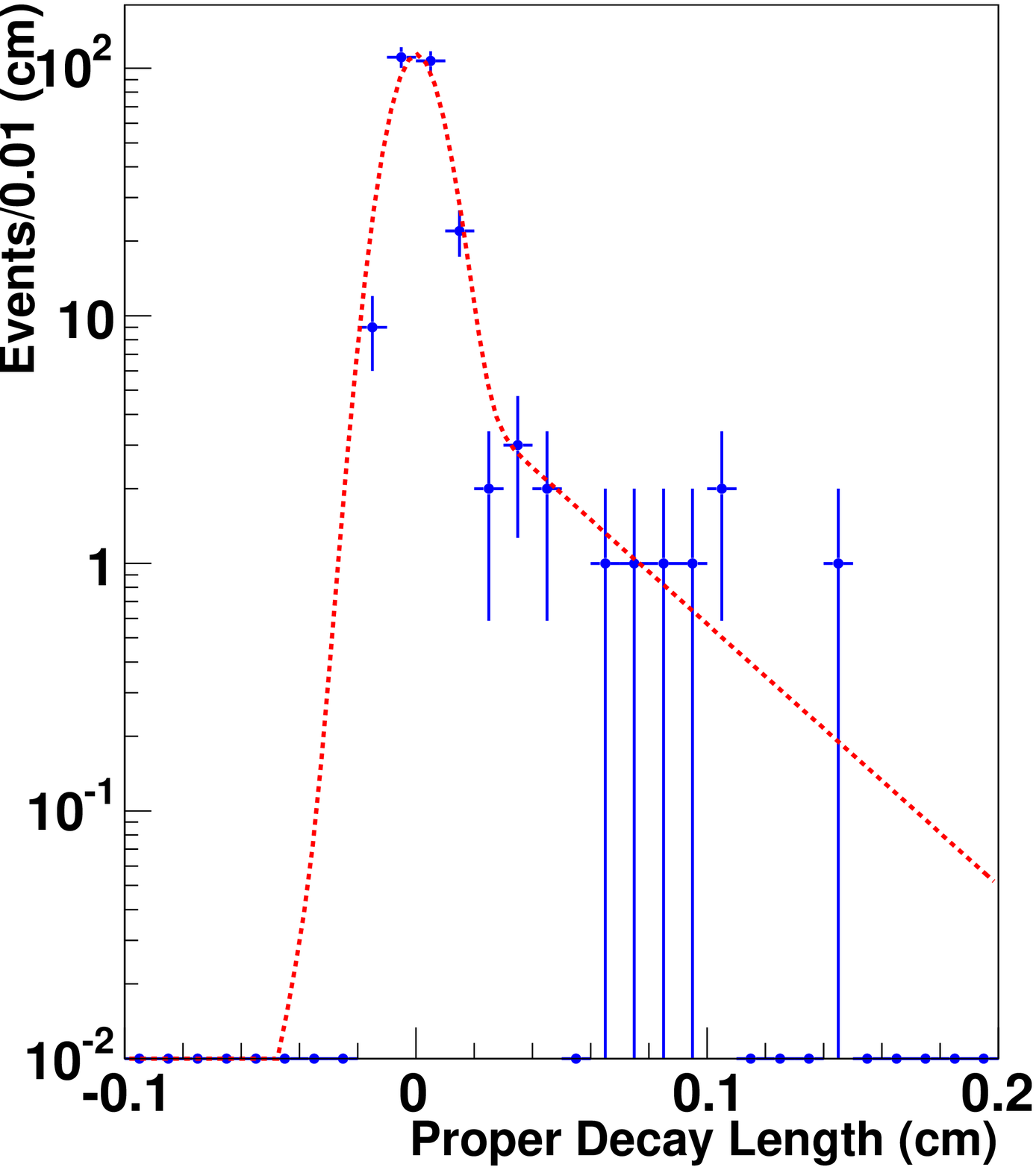,height=8.0cm,width=100mm}
\caption{The proper decay length, $c\tau$, of the sidebands.}
\label{fig:sideband2}
\end{center}
\end{figure}

In Fig.~\ref{fig:sideband2} we fit only the 
sidebands with the background functional form showing a good fit as expected.

For the $\Lambda^0_b$
mean proper decay length we obtain:

\begin{equation} 
c\tau({\Lambda^0_b}) = 408 \pm 89~({\rm stat})~\mu{\rm m}.
\end{equation}
The central value of \lb\ mean proper decay length is larger than what has been  reported previously~\cite{pdg2004} ($367 \pm 24~\mu{\rm m}$); however, it is consistent within statistical errors.

The fraction of signal in the sample is
$f_{sig} =  0.091 \pm 0.028$. 
The fitted value of the lifetime resolution scale factor,  
$s = 1.7 \pm 0.1$, is inconsistent with unity which means that our errors in the data have been underestimated. This has been shown also in other lifetime analyses~\cite{eduardo}. 


\section{Fitting Procedure for $B_d$ }

We used the same fitting procedure for $B_d$ as for \lb.  
~The proper decay length and invariant mass distribution of the 
$B_d$ candidates are fit 
simultaneously using an unbinned maximum log-likelihood method. 
The likelihood function ${\cal L}$ is given by:
\begin{eqnarray}
{\cal L} & = & \prod^{N}_{i=1}[ f_{sig}{\cal F}^i_{sig} + (1-f_{sig}){\cal F}^i_{bck}],
\end{eqnarray}
where $N$=2409 is the total number of events, 
 ${\cal F}^i_{sig}$ is the product of the signal mass and proper
decay-length probability density functions, ${\cal F}^i_{bck}$ is the
product of the background mass and proper decay-length probability
density functions, and $f_{sig}$ is the fraction of signal.

\vspace{5mm}

\begin{figure}[h!tb]
\begin{center}
\includegraphics[height=12.0cm,width=100mm]{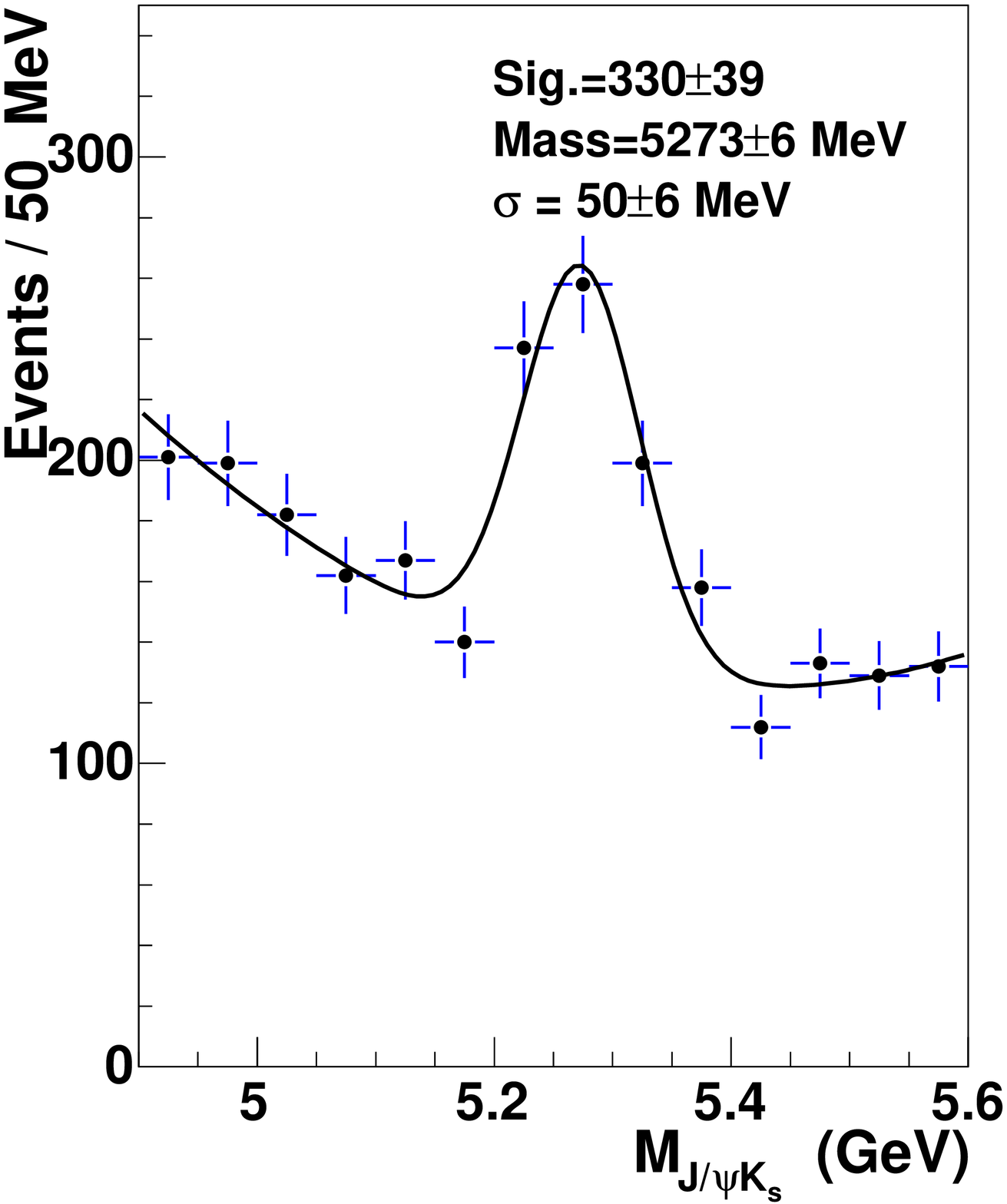}
\caption{Invariant mass distribution of the ($J/\psi,K_S$) system for
all $B_d$ candidates without lifetime cuts. The signal is described by a Gaussian function and the background by a second order polynomial.}
\label{fig:mass_no_life_bd}
\end{center}
\end{figure}

We use a range of 4.9~GeV~$<$~M($B_d$)~$<$~5.6~GeV 
for the $B_d$ mass window.
 We will use the signal with no lifetime cuts as shown in Fig~\ref{fig:mass_no_life_bd}. The cuts used for selecting these events are:

\begin{itemize}
\item vertex $\chi^2(B_d)<25$;
\item $p(B_d) > 5$~GeV;
\item $xy$ distance from $J/\psi$ vertex to 
$K_s$ vertex $>$~0.3~cm; and 
\item $p_T(K^0_S)>1.8$~GeV.
\end{itemize}
These are the events used to extract the lifetime. 
Figure~\ref{fig:mass_no_life_bd} shows the invariant mass 
of ($\mu^+\mu^-\Lambda^0$)
subject to the above cuts. 
The signal is modeled with a Gaussian function, and 
the background by a constant plus a second order polynomial.
The number of signal events 
extracted from the fit is $ 330 \pm 39$ with a mean
of $5622 \pm 12$~MeV and a width of $\sigma = 50 \pm 6$~MeV.

Signal and background are parametrized identically to \lb. 

\subsubsection{Fit to the data for $B_d$}

The fit results are summarized in Table~\ref{likelihood-fitting_bd}. 

\begin{table}
\begin{center}
\begin{tabular}{|c|c|}
\hline
Parameter   &  Central value $\pm$ error  \\
\hline \hline
{\bf Proper decay length, $c\tau$}          &  {\bf 428 $\pm$35  $\mu$m} \\
\hline
Signal fraction, $f_{sig}$  &  0.135 $\pm$ 0.016 \\
\hline
Mass of $B_d$, {\it M}($B_d$)  & 5273 $\pm$ 6~MeV \\
\hline
Width of $B_d$, $\sigma(B_d)$  & 50 $\pm$ 6~MeV \\
\hline
Mass Slope~(0), a0             & $-0.37 \pm 0.003$\\
\hline
Mass Slope~(1), a1             & 0.034~$\pm$~0.003\\
\hline
Error scale factor, $s$       &  1.56~$\pm$~0.05    \\
\hline
$slope-bkg-neg$  &  131~$\pm$~19 $\mu$m\\
\hline
$slope-bkg-pos$  &  466~$\pm$~35 $\mu$m\\
\hline
$norm-bkg-neg$   &  0.040~$\pm$~0.008  \\
\hline
$norm-bkg-pos$   &  0.18~$\pm$~0.01  \\
\hline
\end{tabular}
\caption {Unbinned maximum likelihood fitting results for the $B^0_d$ signal.}
\label{likelihood-fitting_bd}
\end{center}
\end{table}

\vspace{5mm}

\begin{figure}[h!tb]
\begin{center}
\psfig{file=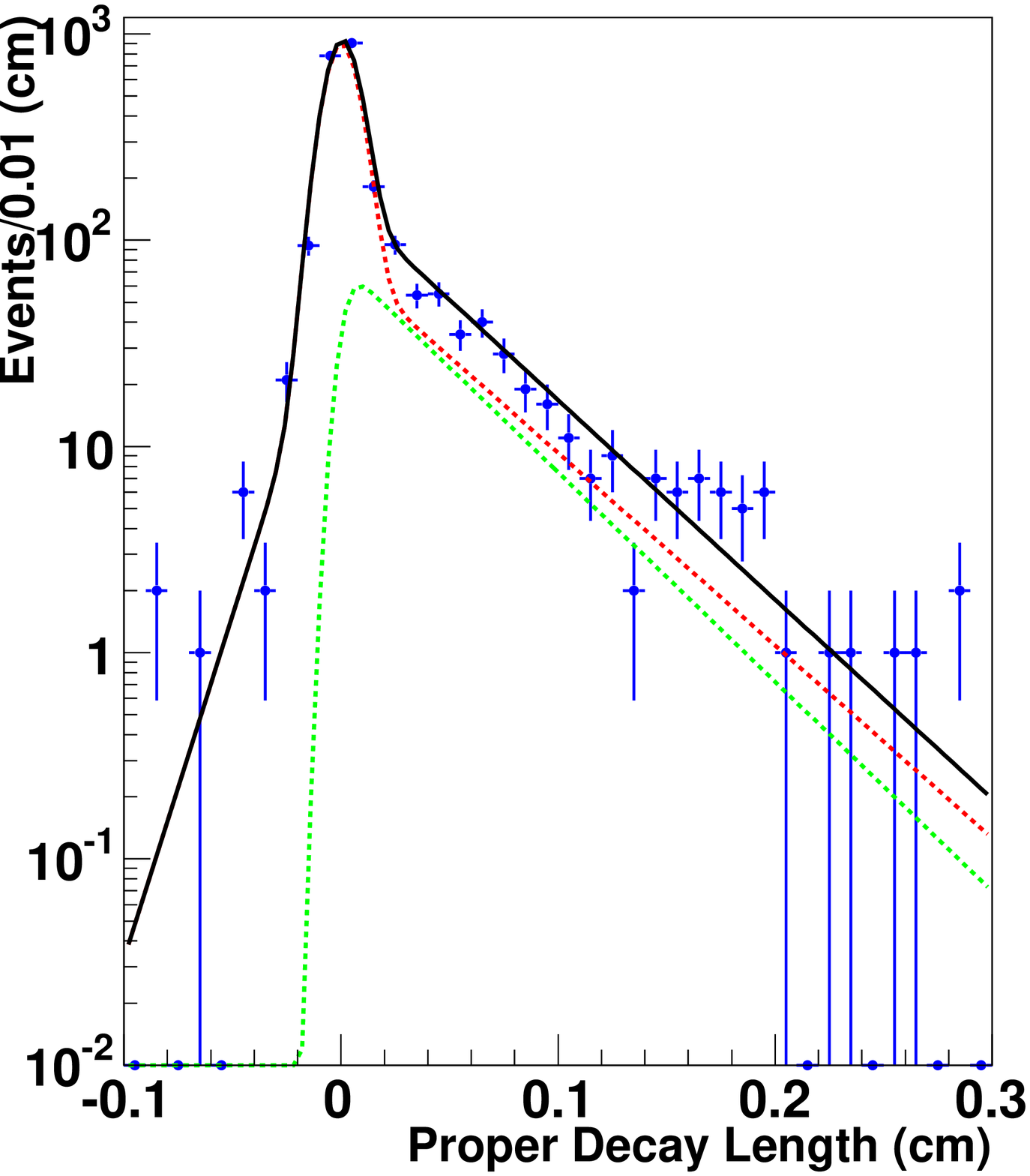,height=8.0cm,width=100mm}
\caption{The proper decay length, $c\tau$,  of the $B_d$ candidates. 
The curves show:  the signal contribution,  dotted (green); 
the background, dashed (red); and total, solid (black).}
\label{fig:lb_lifetime_fit_bd}
\end{center}
\end{figure}

The lifetime distribution with the fit results overlaid is shown 
in Fig.~\ref{fig:lb_lifetime_fit_bd}. In the figure for illustration purposes, we again assume a constant decay length resolution of 50~$\mu$m, which is the average value observed in the data. 

For the $B_d$
mean proper decay length we obtain:

\begin{equation} 
c\tau({B_d}) = 428 \pm 35~({\rm stat})~\mu{\rm m}.
\end{equation}
The central value of $B_d$ mean proper decay length is in a good agreement with previous measurements of this value~\cite{pdg2004} and the world average of $c\tau({B_d}) = 460 \pm 4~({\rm stat})~\mu{\rm m}$.

The fraction of signal in the sample is
$f_{sig} =  0.135 \pm 0.016$. 
The fitted value of the lifetime resolution scale factor,  
$s = 1.56 \pm 0.05$, is consistent with that observed for the \lb\ channel. 



\section{Systematic Errors for $\Lambda_b$ and $B_d$ Lifetime}

We summarize the systematic errors for our measurements for \lb\ and $B_d$ lifetime in Table~\ref{tablesys}.

The contribution from the uncertainty in detector alignment is estimated by reconstructing the $B^0$ sample with the positions of the SMT sensors shifted outwards radially by the alignment error of 30~$\mu {\rm m}$ in the radial position of the sensors and then re-fitting for the lifetime.

To check the stability of the fits, we varied the fitted parameters other than $c\tau$ within $\pm \sigma$ of their nominal value, one at a time. This systematic turned out to be the dominant error for \lb\ but, comparable with the other ones for $B_d$.

To estimate the uncertainity in the modeling, we used different background shapes to describe the background. We used one and two exponential PDF for the lifetime background. We also used an exponential and second order polynomial PDF for the mass. 

We also fitted the background only separately and the fitted parameters from this fit are used as fixed parameters when we fit background plus signal. 
The difference in the results between the lifetime we get when using fixed background and the unfixed background gives rise to only a small uncertainity.

\begin{table}
\begin{center}
\begin{tabular}{|c|c|c|c|}
\hline
Source   &  $\Lambda_b$ ($\mu$~m) & $B_d$ ($\mu$m) & Ratio   \\
\hline \hline
Alignment  & 5  & 5 & 0.016\\
\hline
Variation of Fitting Parameters  & 16 & 5 & 0.039 \\
\hline 
Modeling & 11 & 3 & 0.026\\
\hline
Fixing the parameters from sideband fit & 1 & 9 & 0.020 \\
\hline
{\bf Total}  & {\bf 20} & {\bf 12} & {\bf 0.050} \\
\hline
\end{tabular}
\caption {Summary of systematic uncertainties in the measurement of $c\tau$ for \lb\ and $B_d$ and their ratio. The total uncertainties are also given combining the individual uncertainties in quadrature.}
\label{tablesys}
\end{center}
\end{table}

\chapter{Discussion and Conclusion}
\label{ch:lbconc}
\section{Summary on Establishing the \lb\ Signal}

We observe evidence for the \lb\ baryon using the
exclusive decay \lbdec\
followed by $J/\psi \rightarrow \mu^+ \mu^-$ and
$\Lambda \rightarrow p \pi^-$.
The data sample corresponds to an integrated luminosity of
approximately 225~pb$^{-1}$.
We find 48$\pm$10 \lb\ candidates and determine the mass of \lb\ baryon to be
$5630\pm10$~(stat)~MeV.
This result is consistent with previous $\Lambda^0_b$ baryon mass measurements~\cite{pdg2004}.
The \lb\ yield following lifetime-related cuts also indicates
that the signal contains a lifetime distribution different from background.
The decay \bddecks\ is used as a test sample,
so for completenness, we also show the results for \bd\ meson.
We find 300$\pm$39 \bd\ candidates and determine the mass
of the \bd\ meson to be $5270\pm5$~(stat) MeV.

\section{Summary of Preliminary Lifetimes}
We were able to extract the lifetimes of \lbdec\ and \bddecks. 
~The lifetime of the $\Lambda^0_b$ baryon has been measured to be
 
\begin{equation}
c\tau(\Lambda^0_b) = 408 \pm 89~{\rm (stat)}\pm 20~{\rm (syst)}~\mu{\rm m}.
\end{equation}
 
\begin{equation}
\tau(\Lambda^0_b) = 1.36 \pm 0.30~{\rm (stat)} \pm 0.07~{\rm (syst)}~{\rm ps}.
\end{equation}
 
This result is consistent with the world average~\cite{pdg2004} of
$\tau(\Lambda^0_b) = 1.229 \pm 0.080$~ps.
 
The lifetime of the $B_d$ baryon has been measured for completeness:
 
\begin{equation}
c\tau(B_d) = 428 \pm 35~{\rm (stat)} \pm 12~{\rm (syst)}~\mu{\rm m}.
\end{equation}
 
\begin{equation}
\tau(B_d) = 1.43 \pm 0.12~{\rm (stat)} \pm 0.04~{\rm (syst)}~{\rm ps}.
\end{equation}
 
This result is in agreement with the world average~\cite{pdg2004} of
$\tau(B_d) = 1.536 \pm 0.014$~ps.
Finally we report the ratio $\tau(\Lambda_b)/\tau(B_d)$: 
\begin{eqnarray}
\frac{\tau(\Lambda_b)}{\tau(B_d)}|_{exp}=0.95 \pm 0.22~{\rm (stat)} \pm 0.05~{\rm (syst)},
\label{eq:lbratioexp}
\end{eqnarray}
which is consistent with previous world measurements and with the theoretical predictions.
Although the data used in this study are limited we look forward to adding more
data and should be able to
carry out a competitive measurement of the $\Lambda^0_b$ lifetime via this exclusive decay channel. 



\appendix
\chapter{Further Details on Systematics Error Determination}
\label{app:systematics}
\begin{figure}
\includegraphics[height=2.5in]{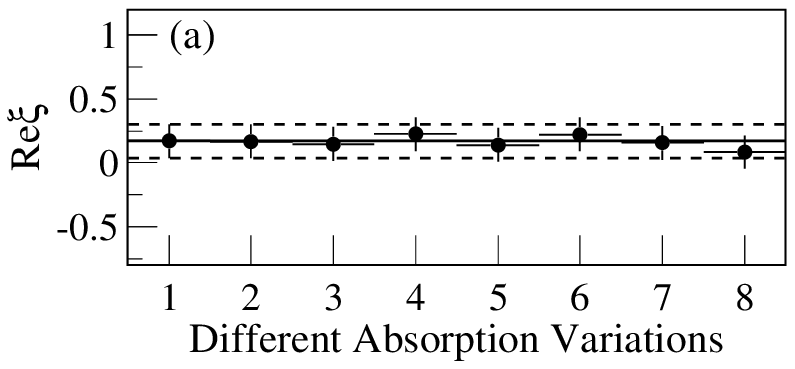}
\includegraphics[height=2.5in]{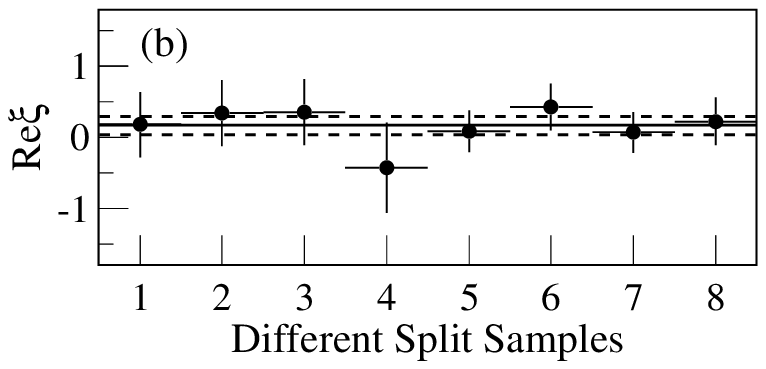}\\
\caption{Re~$\xi$ for different MC absorption models. The order of
each point from left to right is 1) uncorrected 2) standard FOCUS
(baseline): 3) baseline and byproducts 4) baseline and charm absorption
zero 5) baseline, charm absorption zero and byproducts, 6)
baseline, $\sigma(\d0b)=1/2\times\sigma(\pi^+)=11.5mb$ and
$\sigma(D^0)=1/2\times\sigma(\pi^-)=13mb$, 7)
baseline and $\sigma(D^0)=3/2\times\sigma(\d0b)$, 8)
baseline and $\sigma(D^0)=3\times\sigma(\d0b)$ (a). Different split 
samples (b).}  
\label{fig:multi_mc_data}
\end{figure}

We will show here a case study for systematics: we assume the mixing parameters  $x$ and $y$ are 5\% and Im~$\xi=0$, and investigate Re~$\xi$.
We previously observed that MC corrections were very small in what we
call the standard FOCUS absorption model. It includes $\d0b$ and $D^0$
cross sections half the pion cross section, Moliere scattering and the 
absorbed particles having no byproducts. While interactions of pions and
kaons with matter have been measured, no such measurement exist for
intereactions of charm particles with matter. We assumed that
$\sigma(D^0)=\sigma(\d0b)=1/2\times\sigma(\pi)$ and even though this is a
good assumption we want to examine the sensitivity of 
CPT parameters when cross sections are not equal.
In Fig.~\ref{fig:multi_mc_data}(a) several variations from standard
absorption are shown. We observe that we are not very sensitive to
the difference between the cross sections of $\d0b$ versus $D^0$. For
instance, increasing the $\d0b$ cross section by 20\% relative to $D^0$,
 results in a 4\% decrease in Re~$\xi$. In reality these cross
sections should be closely equal which results in
systematic effect of less than 1\% associated with typical uncertainties in charm cross sections.  


Fig.~\ref{fig:multi_mc_data}(b) shows the stability of Re~$\xi$ versus
split samples. Some momentum dependence is observed but the variations are not very significant.

\begin{figure}
\includegraphics[height=2.5in]{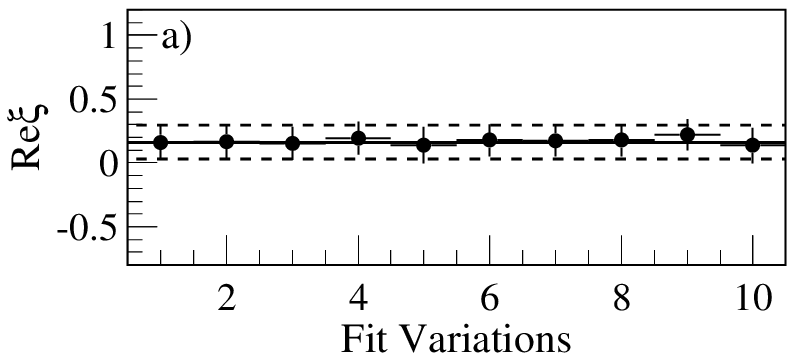}
\includegraphics[height=2.5in]{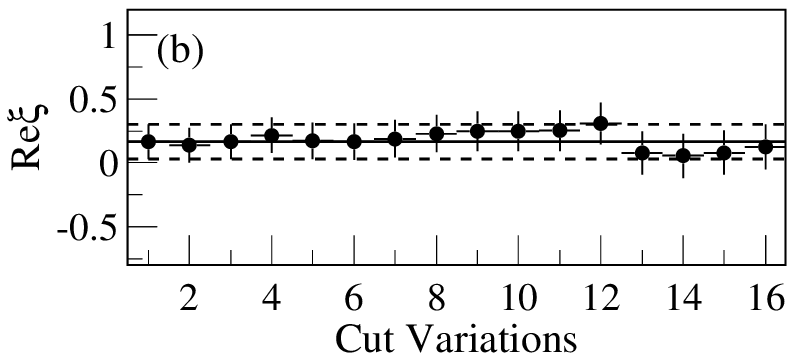}\\
\caption{Stability of $Re\xi$ for (a) different fit variants and
(b) cut variants ($L\over{\sigma}$,Kaonicity) }  
\label{fig:rexi_vs_cuts}
\end{figure}

We vary $L\over{\sigma}$ and Kaonicity $(W_{\pi}-W_{K})$ requirements.
We observe in Fig.\ref{fig:rexi_vs_cuts}(b) the fitted ${\rm Re}\xi$ versus
16 combinations of $L\over{\sigma}$ and $(W_{\pi}-W_{K})$ for $x$ and $y$
mixing of 5\%. We saw that the corrections remain small independent of
the set of cuts we used.
From the way these cuts are applied there is a correlation of the data
from one set to the other one but the underlying systematic effect is
small. 

We also investigated bin and sideband variations to check for any
systematic effect. Fig.\ref{fig:rexi_vs_cuts}(a) shows five different
sideband methods and two different bin widths. The five different sidebands are 
($\pm2\sigma$,$\pm3\sigma$,$\pm7\sigma$),
($\pm2\sigma$,$\pm4\sigma$,$\pm8\sigma$),
($\pm2\sigma$,$\pm3\sigma$,$\pm9\sigma$),
($\pm2\sigma$,$3\sigma$,$7\sigma$) and 
($\pm2\sigma$,$-3\sigma$,$-7\sigma$) 
and two bins are $500~f$s and $300~f$s. In Fig.\ref{fig:rexi_vs_cuts}a), the
first 5 points correspond to 500fs bin width and the last five points
correspond to 300~fs bin width. 


\clearpage{\pagestyle{empty}\cleardoublepage}



\referencepage
\thispagestyle{empty}
\noindent{\Huge\bf\centerline {Curriculum Vitae}}
\vspace*{0.5cm}
\flushleft{\Large \bf \underline{Personal}}
\vspace*{-0.5cm}
\begin{tabbing}
mail\= address and all that \= \kill
\>{  Name:}\>         Abaz Kryemadhi\\
\>{  Date of birth}\> March 31, 1973\\
\>{  Citizenship}\> Albanian\\
\end{tabbing}

\vspace*{-1.0cm}
\flushleft{\Large \bf \underline{Education}}
\vspace*{-0.5cm}
\begin{tabbing}
mail\= address and all that \= \kill
\>1998-2004 \hspace{0.3in} {\bf Ph.D.}, Indiana University, Bloomington, IN\\
\>1998-1999 \hspace{0.3in} {\bf M.Sc.} Indiana University, Bloomington, IN\\
\>1996-1998 \hspace{0.32in} {\bf Diplom Physik.} Siegen University, Siegen, Germany \\
\>1991-1996 \hspace{0.32in} {\bf B.Sc.} University of Tirana, Tirana, Albania \\

\end{tabbing}

\vspace*{-1.0cm}
\flushleft{\Large \bf \underline {Research Experience}}
\vspace*{-0.5cm}
\begin{tabbing}
mail\= address and all that \= \kill
Indiana University, \D0, FNAL \hspace{0.3in}  Measuring the Lifetime of \lb\ baryon\\
Indiana University, E831, FNAL \hspace{0.3in}  Tests of CPT in charm mesons\\
\end{tabbing}

\vspace*{-1.0cm}
\flushleft{\Large \bf \underline {Work Experience}}
\vspace*{-0.5cm}
\begin{tabbing}
mail\= address and all that \= \kill
\>1999-2004 \hspace{0.3in} Research Assistant -- Department of Physics, Indiana University\\
\>1998-1999 \hspace{0.3in} Teaching Assistant -- Department of Physics, Indiana University\\
\end{tabbing}

\vspace*{-1.0cm}
\flushleft{\Large \bf \underline {Honors}}
\vspace*{-0.5cm}
\begin{tabbing}
mail\= address and all that \= \kill
\>1997 \hspace{0.65in} Heinrich Hertz Foundation Fellowship
\end{tabbing}

\end{document}